\documentclass[prmaterials,showpacs,floats,preprintnumbers,amsmath,amssymb,nofootinbib]{revtex4}

\usepackage{graphicx,epsfig}
\usepackage{times,bbm}
\usepackage{graphics,dcolumn,bm,float}
\usepackage{amssymb,amsmath,rotate,color,amsfonts}
\usepackage[title,titletoc,toc]{appendix}
\usepackage{mathtools}
\usepackage{booktabs}
\usepackage{tcolorbox}
\usepackage[normalem]{ulem}
\usepackage[pagebackref=false,colorlinks,linkcolor=magenta,citecolor=blue,urlcolor=magenta]{hyperref}

\usepackage{wrapfig}
\usepackage{lipsum}
\usepackage{mwe}
\usepackage[mathlines]{lineno}
\usepackage[mathscr]{euscript}
\usepackage{hyperref}
\usepackage{breqn}
\usepackage{bbold}
\usepackage{pgfplots}
\usepackage{float}
\usepackage{tkz-euclide}
\usepackage{braket}
\usepackage{physics}
\usepackage[caption=false]{subfig}
\usepackage[export]{adjustbox}
\usepackage{tikz}
\usepackage{slashed}
\usepackage{bm}
\usetikzlibrary{through,calc}
\usetikzlibrary{positioning}

\usepackage{epstopdf}
\usepackage{color}
\definecolor{orange}{RGB}{255,127,0}
\definecolor{blue2}{RGB}{33,114,173}

\usepackage{amsmath}
\usepackage{dcolumn}
\usepackage{bm}
\usepackage{color}
\usepackage[normalem]{ulem}
\usepackage{subfig}
\usepackage{float}

\newcommand{\hlt}[1]{{#1}}

\newcommand{\be}{\begin{equation}}
\newcommand{\ee}{\end{equation}}

\newcommand{\bea}{\begin{eqnarray}}
\newcommand{\eea}{\end{eqnarray}}
\newcommand{\nn}{\nonumber}

\newcommand{\bk}{{\boldsymbol{k}}}

\newcommand{\bs}{\boldsymbol}
\newcommand{\bmt}{\left[\begin{matrix}}
\newcommand{\emt}{\end{matrix}\right]}

\begin{document}
\preprint{}
\title{Tunning the tilt of a Dirac cone by atomic manipulations: application to 8Pmmn borophene}

\author{Y. Yekta$^{1}$}
\email{yasinyekta1989@gmail.com}
\author{H. Hadipour$^{2}$}
\author{S. A. Jafari$^{1}$}
\email{jafari@sharif.edu}
\affiliation{$^{1}$Department of Physics$,$ Sharif University of  Technology$,$ Tehran 11155-9161$,$ Iran \\
$^{2}$Department of Physics$,$ Sharif University of  Technology$,$ Tehran 11155-9161$,$ Iran}

\date{\today}

\begin{abstract}
 We decipher the microscopic mechanism of the formation of tilt in the two-dimensional Dirac cone of $8Pmmn$ borphene. 
In our {\em ab initio} calculations, we identify relevant low-energy degrees of freedom on the $8Pmmn$ lattice
and find that these atomic orbitals reside on an effective honeycomb lattice (inner sites), while the high-energy degrees of freedom
reside on the rest of the $8Pmmn$ lattice (ridge sites). 
Integrating out the high-energy atomic orbitals, gives rise to remarkably
large {\em effective} further neighbor hoppings on the coarse grained "honeycomb graph" of inner sites that determine the location and tilt of the Dirac cone.
Molecular orbitals -- that can be modified by atomic manipulations -- are responsible for the creation of further neighbor edges on the honeycomb graph that 
controls the tilt of the resulting Dirac cone. 
This leads to an effective tight-binding model on a parent honeycomb graph 
that facilitates numerical modeling of various effects such as disorder/interactions/symmetry-breaking for tilted Dirac cone
fermions. Since the tilt is a proxy to spacetime metric, our result offers a robust 
perspective on the fabrication of desired emergent spacetime structure  and synthesis of geometric forces at ambient conditions
that are likely to enhance our control on the movement of electrons in electric/optical devices. 
\end{abstract}



\maketitle

\section{Introduction}
The marriage between the mathematical concept of "topology" and quantum materials lead to the birth of "topological materials"
and developments that followed afterwards. Can the (local) "geometry" play a similar role in combination with quantum materials?
In the same way that the appearance of Dirac fermions in graphene and other Dirac materials can be attributed to an
emergent Minkowski spacetime structure at distances much larger than the atomic distances, quantum materials with 
"tilted Dirac cone" can be naturally associated with an emergent spacetime (not merely the space) metric. In these class of materials
the tilt is a proxy to spacetime geometry. We use {\em Geometric Quantum Material} (GQM) to refer to them.

Every periodic structure in quantum condensed matter is mounted on a mathematical object called lattice.
Irrespective of which atoms one wishes to place on the sites of a given lattice, they come in 230 possible structures~\cite{Inui1990}.
The presence of lattice breaks translation and rotation, and some times parity and/or time reversal invariance
of the vacuum~\cite{Girvinbook}. The lattice therefore breaks the Poincar\'e group~\cite{RyderQFTbook1996}
thereby the connection between spin and statistics of the particles is lost and therefore one may have fermions
with integer spin that is enforced by the irreducible representations of its space group (SG)~\cite{Bradlyn2016}.

Is there any other interesting consequence that can be associated with the underlying SG?
To set the stage for answering this question, let us start by a simple, but profound observation~\cite{Girvinbook}:
Consider a simple quantum mechanical hopping process on a lattice and think about the wave equation
in the continuum limit. On the square lattice the Hamiltonian becomes $-\hbar^2{\nabla}^2/(2m^*)$,
while on the honeycomb lattice it becomes $i\hbar v_F{\bs\sigma}.{\nabla}$, namely the massless Dirac Hamiltonian~\cite{Ando2011}.
Therefore despite that in the continuum limit, the lattice spacing is immaterial, but still remnants of
the microscopic symmetries of the pertinent SG are manifested in the long-distance behavior and decide whether the structure of
the ensuing spacetime is Galilean or Minkowski.
This can be regarded as an example of metamorphosis of atomic scale symmetry (SG) to a long-distance geometrical structure. 
The duality between atomic scale SG symmetry and long distance geometry is a central to understand the geometric structure in GQMs. 

What is the microscopic mechanism of such a duality between the microscopic structure and
the long-distance (low-energy) characteristics in more complicated lattice structures?
The long-distance behavior of interest for us is the tilted Dirac cone band whose Hamiltonian is
$\propto({\bs\sigma}.\nabla)+\sigma_0{\bs\zeta}.\nabla$~\cite{Tajima2006,Goerbig2008,Kajita2014,Farajollahpour2019,SaharCovariance,Tarun2017}
where $\sigma_0$ is the $2\times 2$ unit matrix. A vector-looking quantity $\bs\zeta$ here determines the tilt of the
Dirac cone. But at a much deeper level, it can be encoded into an emergent spacetime metric
$ds^2=-(v_Fdt)^2+(d\bs x-\bs\zeta v_F dt)^2$~\cite{Farajollahpour2019,SaharCovariance,Jafari2019,Ojanen2017PRX,Ojanen2019PRR,Volovik2016,Volovik2018,Nissinen2017},
where $v_F$ is the Fermi velocity. {\em Therefore the tilt of a Dirac cone is actually a proxy for an emergent solid-state spacetime structure.}
Since the density of states is enhanced by a $1/\sqrt{1-\zeta^2}$, the tilted Dirac cone materials can generically
give stronger responses to external stimuli~\footnote{This enhancements nicely fits into spacetime description as a "redshift" factor~\cite{Mohajerani2021}.}.
Therefore it is crucial to understand the atomistic mechanism of the formation of the tilt in order to utilize it. 

\begin{figure}[]
	\centering
\subfloat[]{ \includegraphics[width=0.23\textwidth]{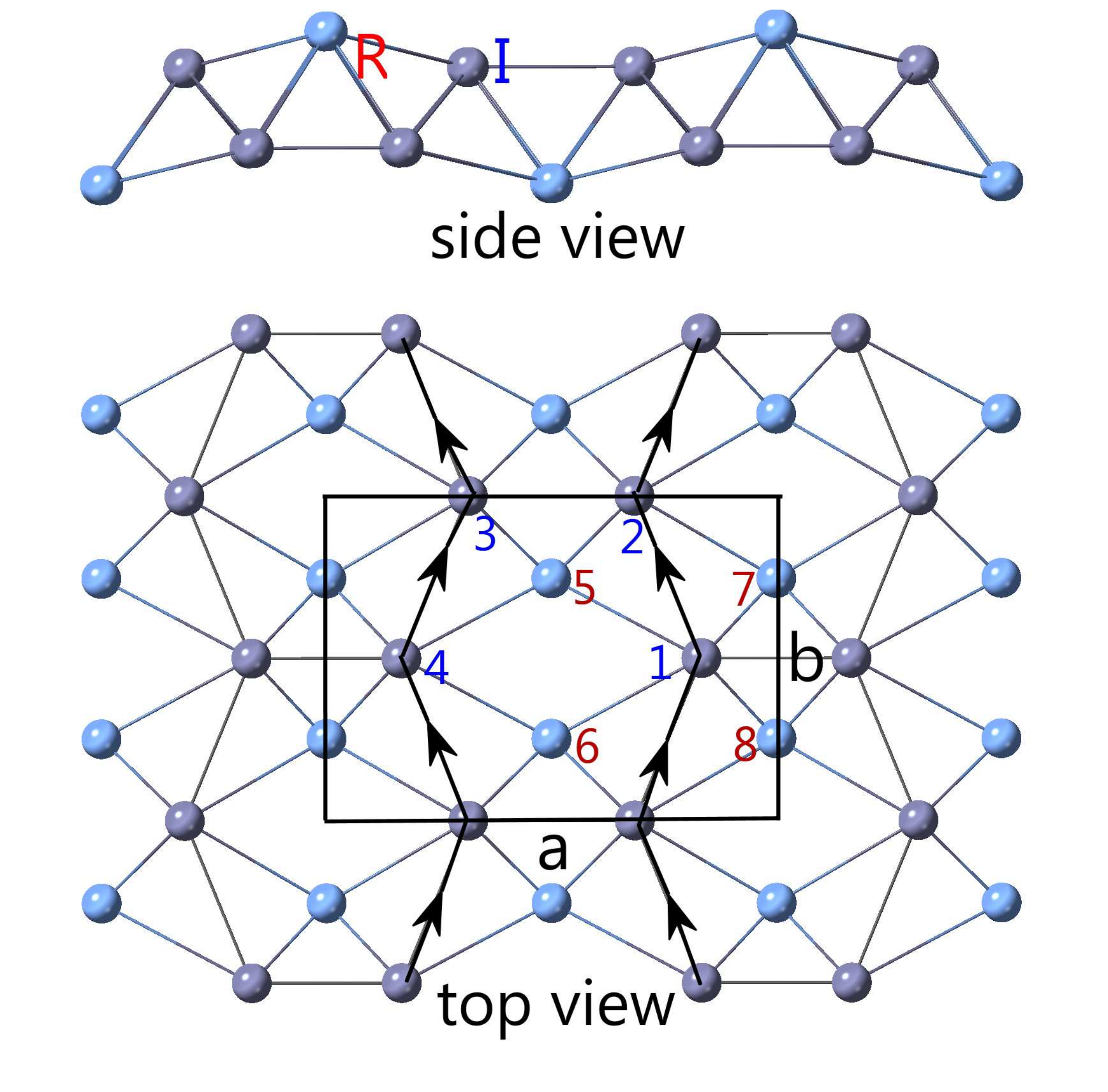}}
  \subfloat[]{    \includegraphics[width=0.21\textwidth]{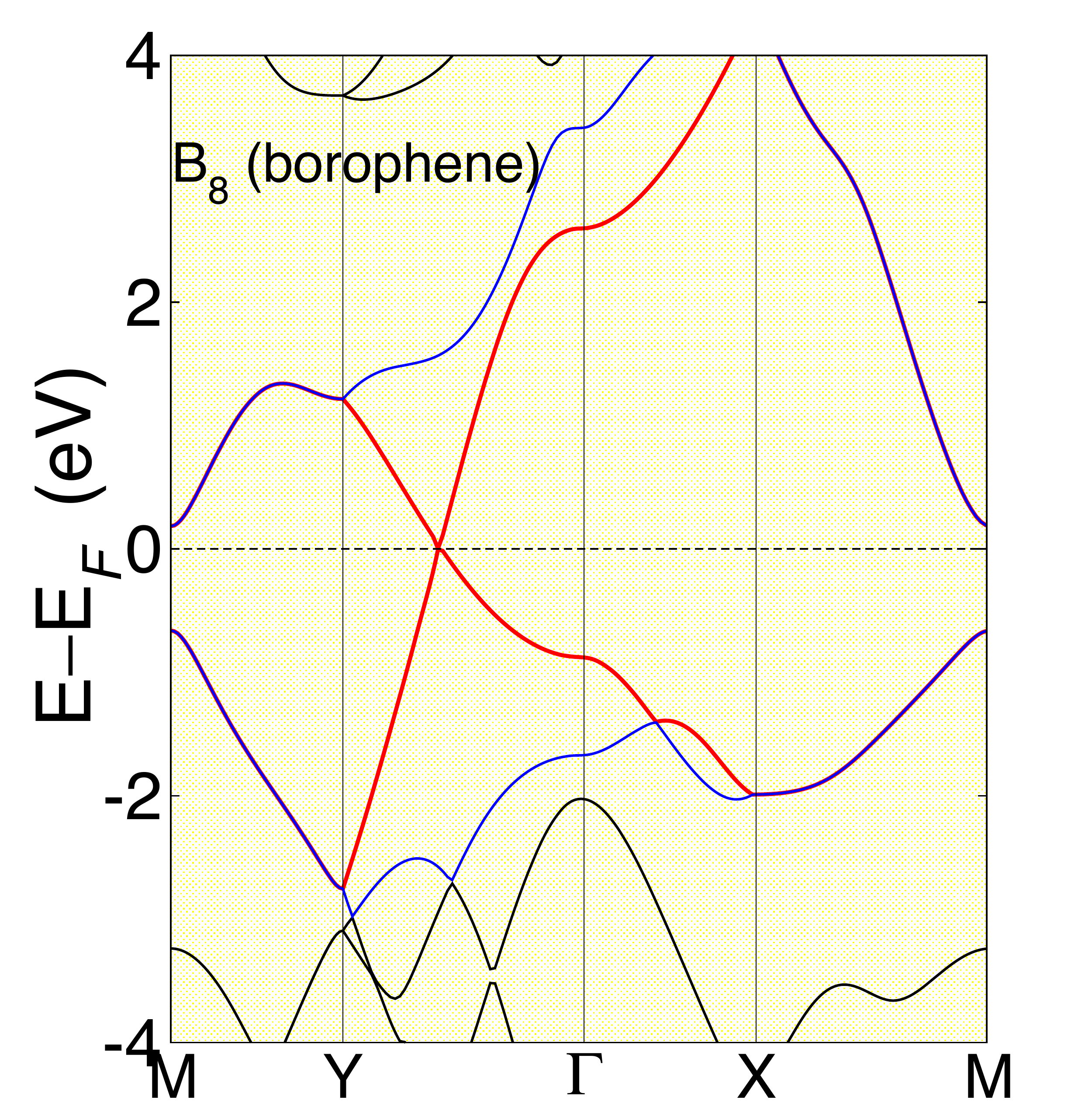}}
  \subfloat[]{    \includegraphics[width=0.27\textwidth]{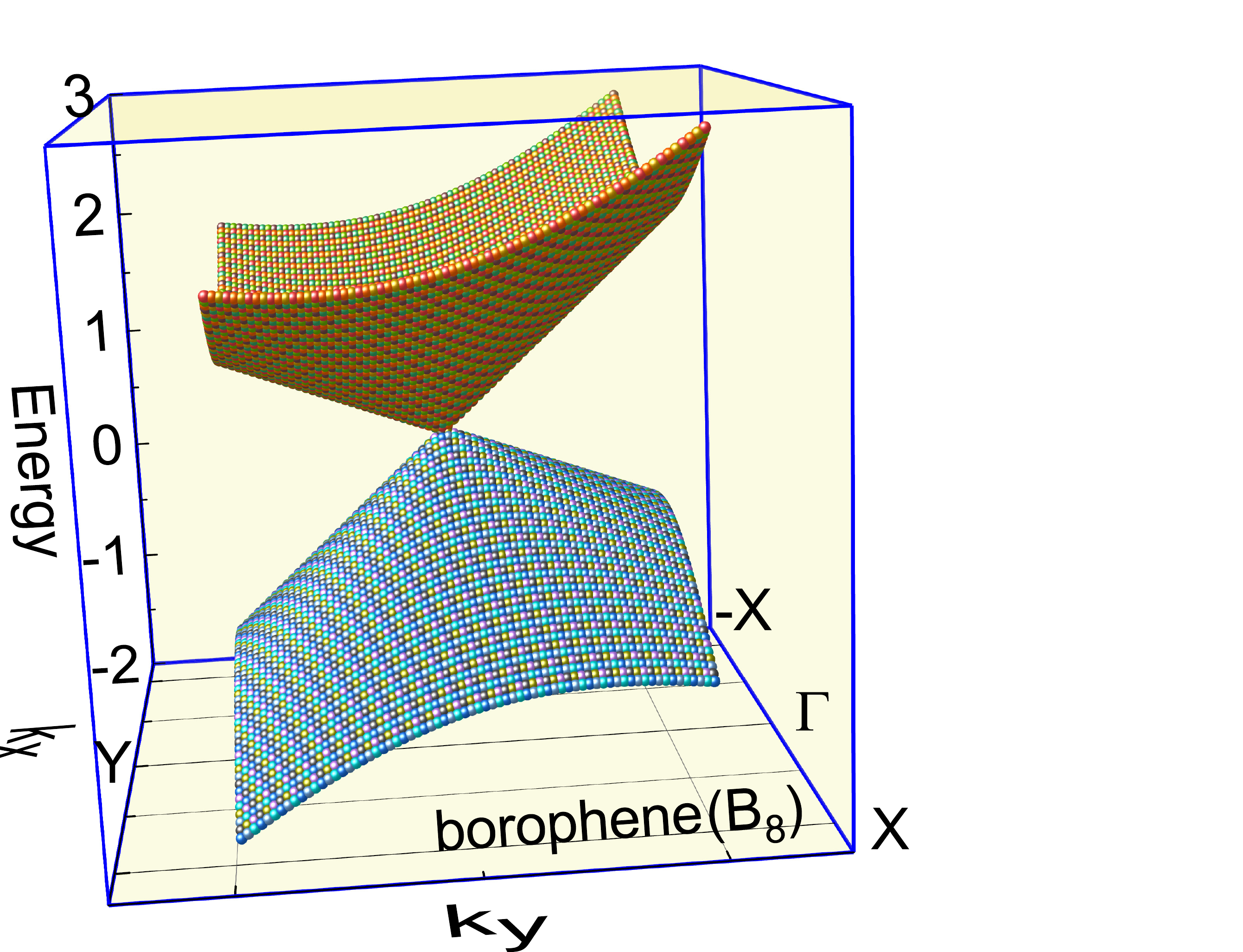}}
  \\
  \subfloat[]{   \includegraphics[width=0.35\textwidth]{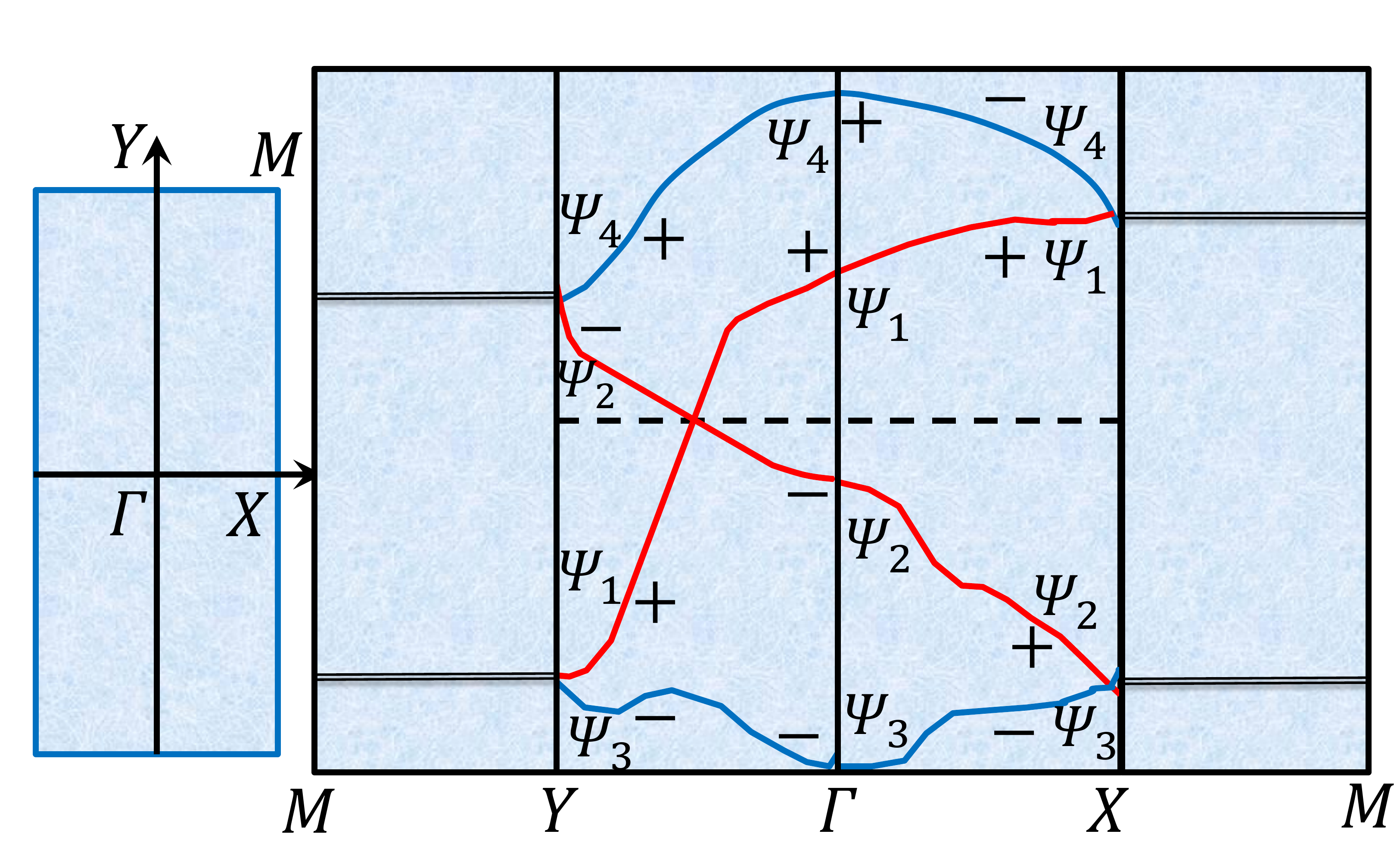} }
  \subfloat[] {     \includegraphics[width=0.34\textwidth]{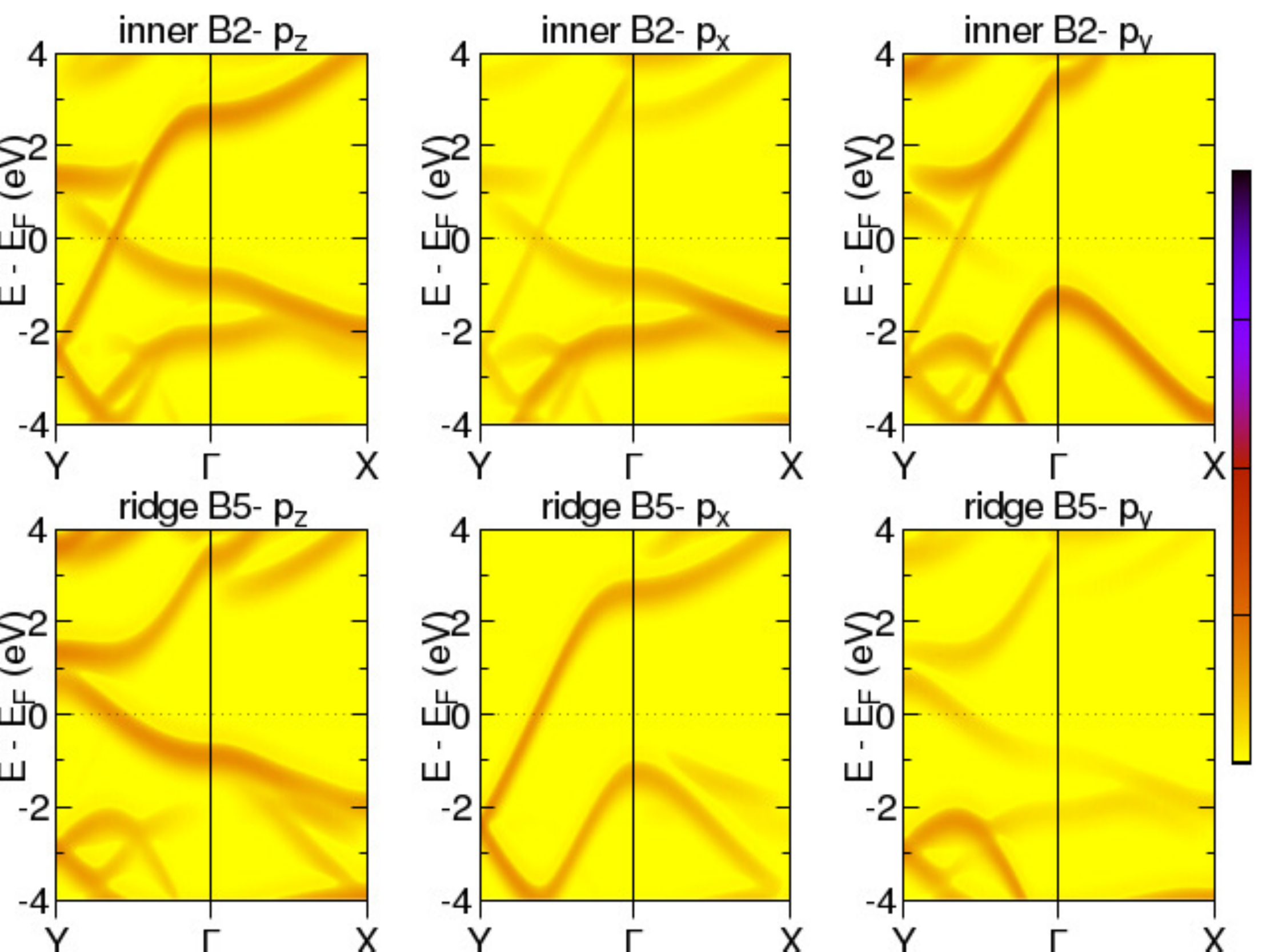}}
  \caption{(a) Top and side view of crystal structure of $8Pmmn$ borophene. The gray and faint teal
circles denote inner and ridge B atoms, respectively. (b) DFT-PBE band structure of pristine $8Pmmn$ borophene
and (c) its three dimensional reconstruction.
(d) First Brillouin zone and  decomposition of elementary band representation
the $8Pmmn$ group. \hlt{The $\pm$ are the eigenvalues of $\tilde{C}_{2x}$, and
$\tilde{C}_{2y}$ operations at high-symmetry points.}
(e) The orbital-projected band structures for two atoms of $8Pmmn$ borophene based on DFT-PBE.}
\label{Fig1.fig}
\end{figure}

In this work we decipher the microscopic mechanism that determines the tilt parameter $\bs\zeta$.
For this purpose we focus on the SG number 59 that describes the $8Pmmn$ structure~\footnote{The tilted Dirac
cone has been experimentally observed in the so called $\chi_3$ structure of borophene~\cite{Feng2017}}
of borophene~\cite{Farajollahpour2019,Bradley,Krowne_2021}. But the logic employed here is generic and is applicable to other SGs,
and can even be used to discover new examples of GQM. 
We show that in this structure, the low-energy and high-energy
degrees of freedom are nicely separated into two sublattices denoted by gray and teal circles in Fig.~\ref{Fig1.fig}(a), respectively.
When the high energy sites available in this particular SG are integrated out, the underlying honeycomb lattice will be promoted to
a "honeycomb graph". Molecular orbitals play significant role in attaching new "graph edges" to the simple underlying honeycomb lattice. 
This graph has the ability to lead to a controlable tilt and hence spacetime  structure in the long distance. 
On such a graph, even if instead of fermions one places a circuit, again a tilted Dirac cone can be obtained~\cite{Emad2021}.
Therefore GQMs themselves can be emulated by circuits with a larger degree of tunability.

\section{Results and discussion}
\subsection{Protection of the Dirac node} The relevant orbitals in the boron (as well as C) atom are $2p$ orbitals.
The possibility of formation of $sp^2$ and $sp^3$ hybridization establishes the honeycomb lattice~\cite{KatsnelsonBook2012}, and
structures such as $8Pmmn$ that involve buckled honeycomb networks as natural lattices for these atoms. As can be seen in Fig.~\ref{Fig1.fig}(a),
there is a backbone (buckled) honeycomb sub-lattice denoted with gray circles that are
called inner (I) sites. The rest of the lattice sites are called ridge (R) sites and are denoted by teal circles.
First principle calculations indicate that the resulting Dirac cone is tilted~\cite{Zhou2014,Lopez2016} which is shown as
red (low-energy) band in Fig.~\ref{Fig1.fig}(b) and the three dimensional reconstruction of the band structure is
shown in panel (c). The first thing that the $8Pmmn$ SG implies about the tilted Dirac cone is that on the $MX$ and $MY$ lines in the
border of the Brillouin zone (BZ) of Fig.~\ref{Fig1.fig}(d), two bands "stick together" (as Kittel puts it~\cite{KittelQTS})
where the two-dimensional irreducible representations are protected by non-symmorphic elements~\cite{KittelQTS}.
Then the compatibility relations gives the qualitative band picture of cat's cradle shape~\cite{Fan2018} shown in panel (d).
For pedagogical details of the derivation of this figure with group theory methods see the supplementary information (SI).

Of course the protection of the tilted Dirac cone by the underlying SG is interesting, but it is not essential for the 
main purpose of this paper, namely the "tilt" of the Dirac cone. The symmetry considerations do not
{\em explain} why and how the tilt parameter $\bs\zeta$ is formed.
In order to "manipulate"
the tilt of the Dirac cone at will, one needs a microscopically detailed understanding of the root cause of formation of tilt in the Dirac cone.
To achieve this, we need to identify the low-energy degrees of freedom that give rise to the
tilted Dirac cone dispersion. For this purpose in Fig.~\ref{Fig1.fig}(e) we have plotted an orbital projected
representation of the band structure that resolve the contribution of atoms 2 (I) and 5 (R).
As can be seen, the dominant contributions to the tilted Dirac dispersion comes from the $p_z$ orbitals of the I atoms (top left plot in panel (e)).
The R atoms also contribute via their $p_z$ and $p_x$ orbitals (bottom left and center). The remaining two boxes on the top row indicate
little contributions from the $p_x$ and $p_y$ orbitals of the I atoms that arises from the buckling of the honeycomb sublattice.

\begin{figure}[]
  \centering
  \includegraphics[width=0.5\linewidth]{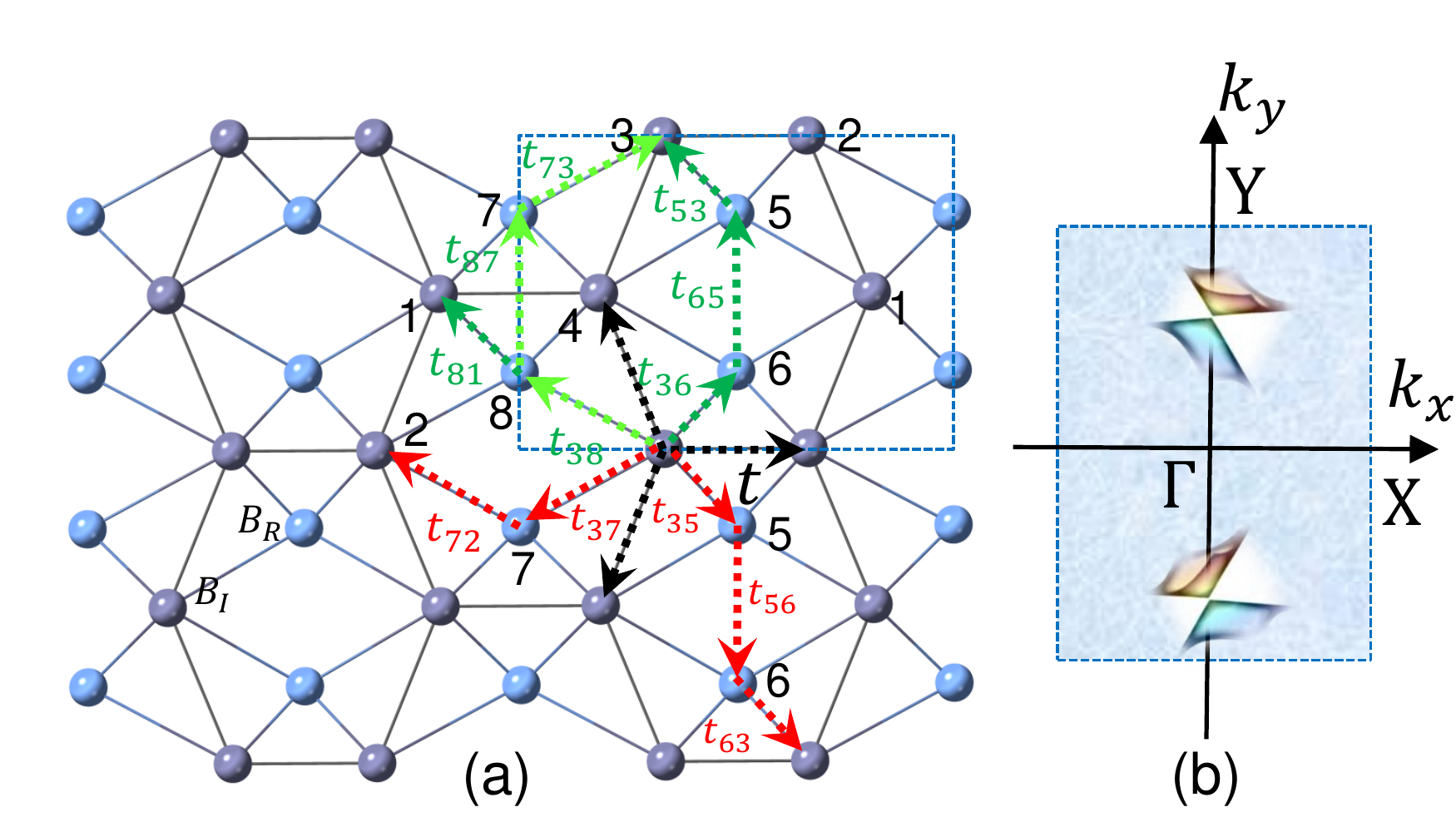} \\
  \includegraphics[width=0.5\linewidth]{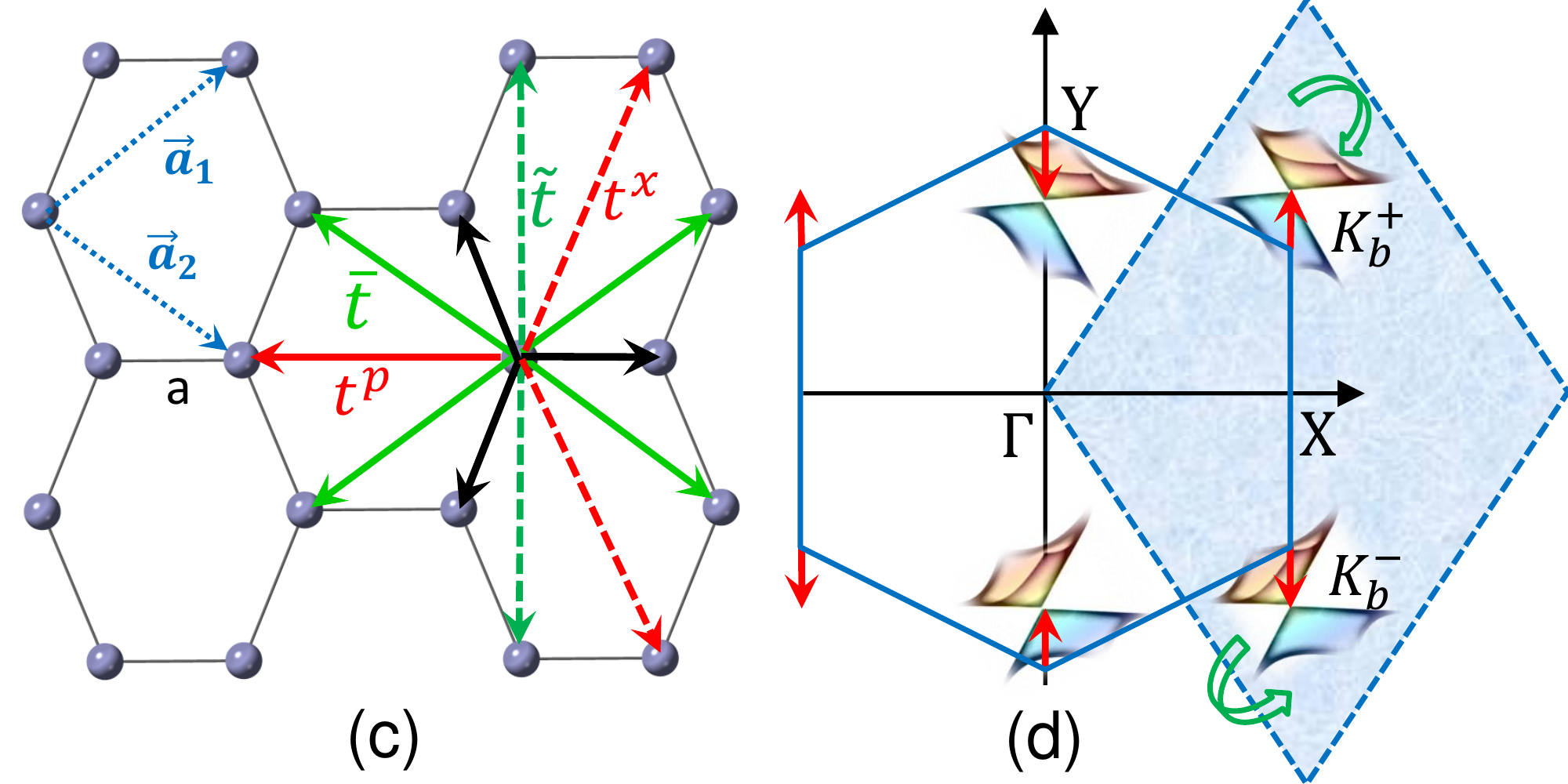}
  \caption{(a) The lattice structure, unit cell, and hopping processes in the $8Pmmn$ lattice and (b) its BZ with two tilted Dirac cones.
(c) The "effective" honeycomb graph of I atoms (low-energy degrees of freedom) obtained by elimination of R sites from the $8Pmmn$ lattice.
(d) two representations (rhombus and hexagonal) of the BZ of the effective
honeycomb structure. The effective second/third neighbor hoppings (green/red) in (c) arise from the corresponding hopping path in (a) via the
process of renormalization (see the text). As indicated in (d), the red (third neighbor) hopping shifts the Dirac cone, while the green (second neighbor) hoppings tilt the cone. }
\label{Fig2.fig}
\end{figure}

\begin{table}[]
\centering
\caption{(top) \emph{ab initio} hopping matrix elements (in $eV$) for borophene and C-doped borophene obtained
from Wannier functions. Conventions for labeling of the hopping matrix elements are given in Fig.~\ref{Fig2.fig}(a).
(bottom) Renormalized parameters of Fig.~\ref{Fig2.fig}(c). $\zeta_y$ is the tilt parameter and $k_D/k_Y$ quantifies the location
of Dirac node.}
\begin{tabular}{c|ccccccccc}
\hline
(atomic)           &$t_{36}$&$t_{65}$&$t_{53}$&$t_{81}$&$t_{38}$&$t_{37}$&$t_{72}$&$t_{78}$\\
\hline
pure borophene $B_8$           &  2.09  & -2.66  & 2.09 & 2.09  & -1.87  & -1.87   & -1.87&-2.54 \\
B$_{6}$C$_{2}$-I-[C2$\&$C3]&  1.93  & -2.52  & 1.96 & 1.92  & -1.52  & -1.52   & -1.55&-2.34  \\
B$_{6}$C$_{2}$-R-[C5$\&$C6]&  2.14  & -2.33  & 2.12 & 2.15  & -2.23  & -2.21   & -2.20&-2.43  \\
\hline
(renormalized)            &  $t$ &$t^{p}$ &$t^{x}$&$\tilde{t}$&$\bar{t}$& $\zeta_y$ & $k_D/k_{Y}$& $\zeta_y^{DFT}$ \\
\hline
pure borophene $B_8$ &-2.21 & -2.36  & -1.07  & -1.99 & -2.51& 0.46 & 0.48 & 0.49\\
B$_{6}$C$_{2}$-[C2$\&$C3]   &-2.37 & -1.75  & -0.95  & -1.62 & -2.05& 0.36 & 0.66 & 0.47\\
B$_{6}$C$_{2}$-R-[C5$\&$C6]   &-2.05  & -2.49  & -1.09  & -2.24 & -2.85& 0.59 & 0.32& 0.66\\
\hline
\end{tabular}
\label{hoppings.tab}
\end{table}

In panel (a) of Fig.~\ref{Fig2.fig} we have labeled atomic hopoings paths with green and red dashed lines that
connect I sites by virtual hoppings through R sites. There is only first neighbor direct hopping between the I sites
denoted by black arrows. The appropriate atomic orbitals involved in forming the above microscopic $t_{ij}$
hoppings can be extracted from the DFT calculation. Working in a gauge that all $t_{ij}$'s are real, the
hermiticity implies $t_{ij}=t_{ji}$. Panel (b) depicts the BZ of original $8Pmmn$ lattice.
The insight from the projected bands in the left and center columns in the second row of Fig.~\ref{Fig1.fig}(e) is that both
$p_x$ and $p_z$ orbitals of the R atoms are of comparable importance and must be incorporated into the atomic scale computation of the
$t_{65}$ (dark green) and $t_{87}$ (light green). Similarly the first column of Fig.~\ref{Fig1.fig}(e) suggests that the $p_z$ orbitals of
R and I atoms dominate in $t_{36}=t_{35}$ hopping process. Tab.~\ref{hoppings.tab} shows the calculated
values of these hopping parameters for pristine borophene (B$_8$) and C-doped borophene B$_6$C$_2$ using Wannier function \cite{Mostofi,Freimuth}.
The purpose of substituting C for B is to study its effect in the tilt of the Dirac cone.
The substituted carbon dimers are placed in R and I positions, respectively. Using these parameters one can re-construct
the {\em ab initio} bands with four $p_z$ orbitals of four inner sites, and eight $p_x$ and $p_z$ orbitals of
ridge sites. The nice coincidence of the original DFT bands with Wannier-interpolated bands in Fig.~S9 (see SI) shows that the obtained $t_{ij}$ values are reliable and the used atomic orbitals are adequate.

Although such a atomic picture might be satisfactory if one wishes to focus on the low-energy features of the
tilted Dirac cone around the Fermi surface, but still working with a $12$-band Hamiltonian is neither
convenient, nor the essential long-range physics depends on so many short-distance details.
To achieve an effective two-band model, we need to decimate the $8Pmmn$ lattice into an effective honeycomb
graph shown in Fig.~\ref{Fig2.fig}(c) where two possible ways of representing its BZ are depicted in panel (d).
On such a coarse-grained lattice, the virtual atomic hopping paths will be replaced by {\em effective}
hoppings of the same color in panel (c). The connection between these effective hoppings and atomic hoppings $t_{ij}$
is a nice example of renormalization that will be discussed now.

\subsection{Renormalization via molecular orbitals:}
Anderson in his book maintains that renormalization is one of the pillars of condensed matter physics~\cite{Anderson2018}.
In this section we will show that the same concept is encoded into a local quantum chemistry of the $8Pmmn$ borophene in a remarkable way.
Since we are not interested in higher energy features that take place away from the tilted Dirac node,
we build an effective picture based on Fig.~\ref{Fig2.fig}(c) where the "effective" hoppings must be
evaluated from the atomic scale data in the upper part of Tab.~\ref{hoppings.tab}.
The low-energy degrees of freedom are dominated by the $p_z$ orbitals of the I sites.
So we decimate the $8Pmmn$ lattice by elimination of the R sites. In doing so, the effective
coarse-grained system becomes the honeycomb graph in Fig.~\ref{Fig2.fig}(c).
The nearest neighbor hoppings denoted by black arrows take place within the low-energy subspace and
are responsible for the formation of a parent Dirac cone from the $p_z$
orbitals of the I sites. Slight anisotropy in the black arrows is known to shift the location of the Dirac cone~\cite{Vozmediano2010}
within the rhombus BZ of panel (d) that we ignore here.
Green/red hopping processes are associated with the second/third neighbors of the honeycomb backbone
that originate from the corresponding atomic process $t_{ij}$ of panel (a) via the renormalization.

\begin{figure} []
  \centering
  \includegraphics[width=0.70\textwidth]{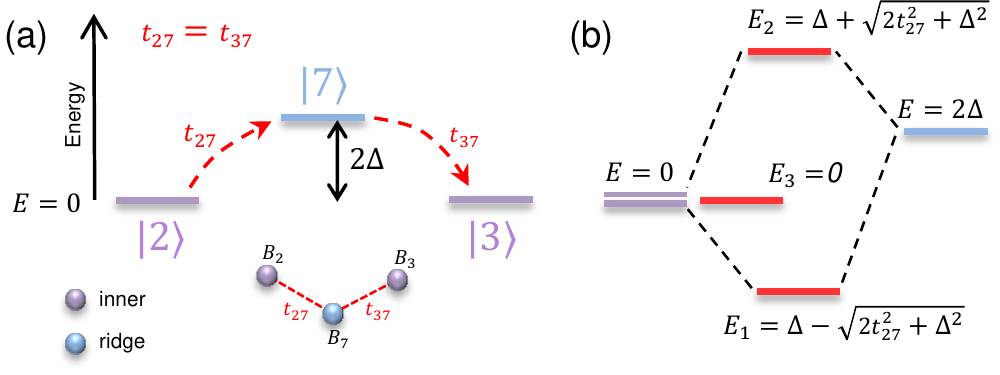}
  \caption{(a) Virtual hopping via the ridge site $7$ generates an effective hopping $t^p$ between $2$ and its {\em third neighbor} site $3$ -- i.e. site $3$
  in the adjacent unit cell in Fig.~\ref{Fig2.fig} --
  (b) The odd parity combination of $2$ and $3$ remains decoupled, but the even parity combination hybridizes with ridge site
  to gain energy. }
\label{Fig3.fig}
\end{figure}

Let us see how the atomic processes $t_{ij}$ in Fig.~\ref{Fig2.fig}(a) are related to the
effective hopping processes in Fig.~\ref{Fig2.fig}(c).
For example consider the simplest third neighbor R sites $2$ and $3$ of adjacent unit cells in Fig.~\ref{Fig2.fig}(a) that
becomes possible via atomic hoppings $t_{27}$ and $t_{37}$ of Fig.~\ref{Fig2.fig}(a).
The hopping process via the R site $7$ is depicted in Fig.~\ref{Fig3.fig}(a). Assuming an on-site energy
offset $2\Delta$ for the R sites with respect to I site, an electron starting at the site $2$ virtually hops to R site $7$
and then returns to low-energy sector at site $3$. Through this process it gains the following energy
\be
   t^p=\Delta-\sqrt{2t_{27}^2+\Delta^2}<0,
   \label{23effhop.eqn}
\ee
Even in the limit of $\Delta\to 0$ this gives an energy lowering $-\sqrt 2|t_{27}|$ that can be regarded as effective hopping between
the third neighbor sites $2$ and $3$ of the I sublattice.
To intuitively understand this formula, note that in the absence of
the I site, $t_{27}=t_{37}=0$, and hence both even and odd combinations of third neighbor atomic orbitals $|2\rangle\pm |3\rangle$
remain inert. But once the inner site is present, the atomic hopping $t_{27}$ causes a coupling of $|7\rangle$ with the even-parity
state $|2\rangle+|3\rangle$, leaving the odd parity state $|2\rangle-|3\rangle$ decoupled at zero energy. The coupling with the
even parity combination provides a channel to lower the energy given by the effective hopping in Eq.~\eqref{23effhop.eqn}.
Furthermore, in the limit of $\Delta\gg|t_{27}|$ the above formula reduces to the perturbatively appealing form $-2t_{27}^2/\Delta$.
For pedagogical details of the above computation, please refer to section II of SI.

\begin{figure} []
  \centering
  \includegraphics[width=0.70\textwidth]{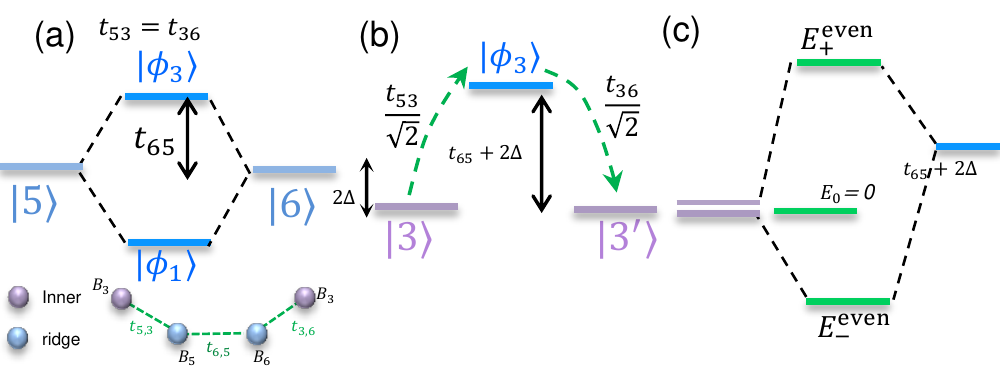}
  \caption{(a) Formation of bonding and anti-bonding orbitals between the ridge atoms $5$ and $6$. (b) Effective hopping $\tilde t$ between second neighbor
  inner sites $3$ and $3'$ can be formed by virtual hopping via the anti-bonding orbital $|\phi_3\rangle$ of two ridge atoms. }
\label{Fig4.fig}
\end{figure}

As a second example, consider the I site $3$ in Fig.~\ref{Fig2.fig}(a) and the same site in the unit cell above it (that we denote by $3'$).
These sites will be second neighbor on the coarse grained lattice and there are two hopping pathways $3\to6\to5\to 3'$ and $3\to8\to7\to3'$
connecting them that are denoted by light and dark green dashed lines in Fig.~\ref{Fig2.fig}(a). The process of the calculation of the
effective hopping amplitude $3\to3'$ is similar for the above two paths. So we focus on the first one. In principle one must consider the
energy gained by the lowest molecular orbitals formed by the above chain of sites. This has been done in section III of SI. But the end result
allows for a nice interpretation depicted in Fig.~\ref{Fig4.fig}: First, due to hopping $t_{65}$, the R sites $5$ and $6$ form bonding
and anti-bonding molecular orbitals denoted by $|\phi_1\rangle$ and $|\phi_3\rangle$ in Fig.~\ref{Fig4.fig}(a). Then as depicted in panel (b),
an electron gains energy by virtually hopping via the anti-bonding orbital $|\phi_3\rangle$ whose energy is now offset by $t_{65}+2\Delta$. The hoppings
connecting site $3$ and $3'$ to $|\phi_3\rangle$ are $t_{65}/\sqrt 2$ where the $\sqrt 2$ factor comes from the normalization 
$|\phi_3\rangle=(|\phi_5\rangle-\phi_6\rangle)/\sqrt 2$.
In the final step as shown in Fig.~\ref{Fig4.fig}(c), the even-parity combination of $|3\rangle$ and $|3'\rangle$ is mixed with $|\phi_3\rangle$
to give an energy gain $E_-^{\rm even}=(t_{56}/2+\Delta)-\sqrt{(t_{56}/2+\Delta)^2+t^2_{63}}$. A similar contribution arises from the second path.
Adding the two contributions we obtain the total renormalized hopping $\tilde t$ of Fig.~\ref{Fig2.fig}(c) as
\bea
  \tilde t=&&\frac{t_{56}+t_{87}+4\Delta}{2} \label{33effhop.eqn}\\
  &&-\sqrt{(\frac{t_{56}}{2}+\Delta)^2+t_{63}^2}-\sqrt{(\frac{t_{87}}{2}+\Delta)^2+t_{73}^2}\nn
\eea
In the above formula, the hopping parameters $t_{65}$  and $t_{87}$ are dominantly contributed by the $p_x$ and $p_z$ orbitals
as the intensity of the $p_x,p_z$ orbitals in the second row of Fig.~\ref{Fig1.fig}(e) is dominant, while $p_y$ is faint.
Similarly the two other renormalized hopping parameters $t^x$ and $\bar t$ can be computed. For details please refer to SI
where we have shown that the elimination of higher energy states actually corresponds to the above simple molecular orbital analysis.

This is how the renormalized parameters of the coarse grained honeycomb graph model in the bottom part of Tab.~\ref{hoppings.tab}
are computed from the {\em ab initio} data of the top part of the table.
Note that if instead of $8Pmmn$ structure, we had a simple honeycomb lattice (such as in graphene),
such a large values of second or third neighbor hopping given in Tab~\ref{hoppings.tab} would be unthinkable
as hopping between the {\em atomic} orbitals exponentially decays with distance. Therefore the virtual hopping
via the ridge sites attaches a great importance to them as providers of channels for energy gain and ultimate
formation of renormalized hoppings on a coarse grained lattice of inner sites.
Furthermore, this is an example of how the molecular orbital play a significant role in the formation of longer range
hoppings on the decimated $8Pmmn$ lattice.

\subsection{Effective coarse-grained model}
Now that we have identified the $p_z$ orbitals residing at the inner sites as the
low-energy degrees of freedom, and have computed various renormalized hoppings between second and third neighbors, we are
ready to write down a physically clear low-energy effective model that will straightforwardly
demonstrate how the renormalized parameters $t^p, t^x,\tilde t$ and $\bar t$ can provide information about the position and the
amount of the tilt.

\begin{figure*}[]
\begin{center}
\subfloat[]{ \includegraphics[scale=0.15]{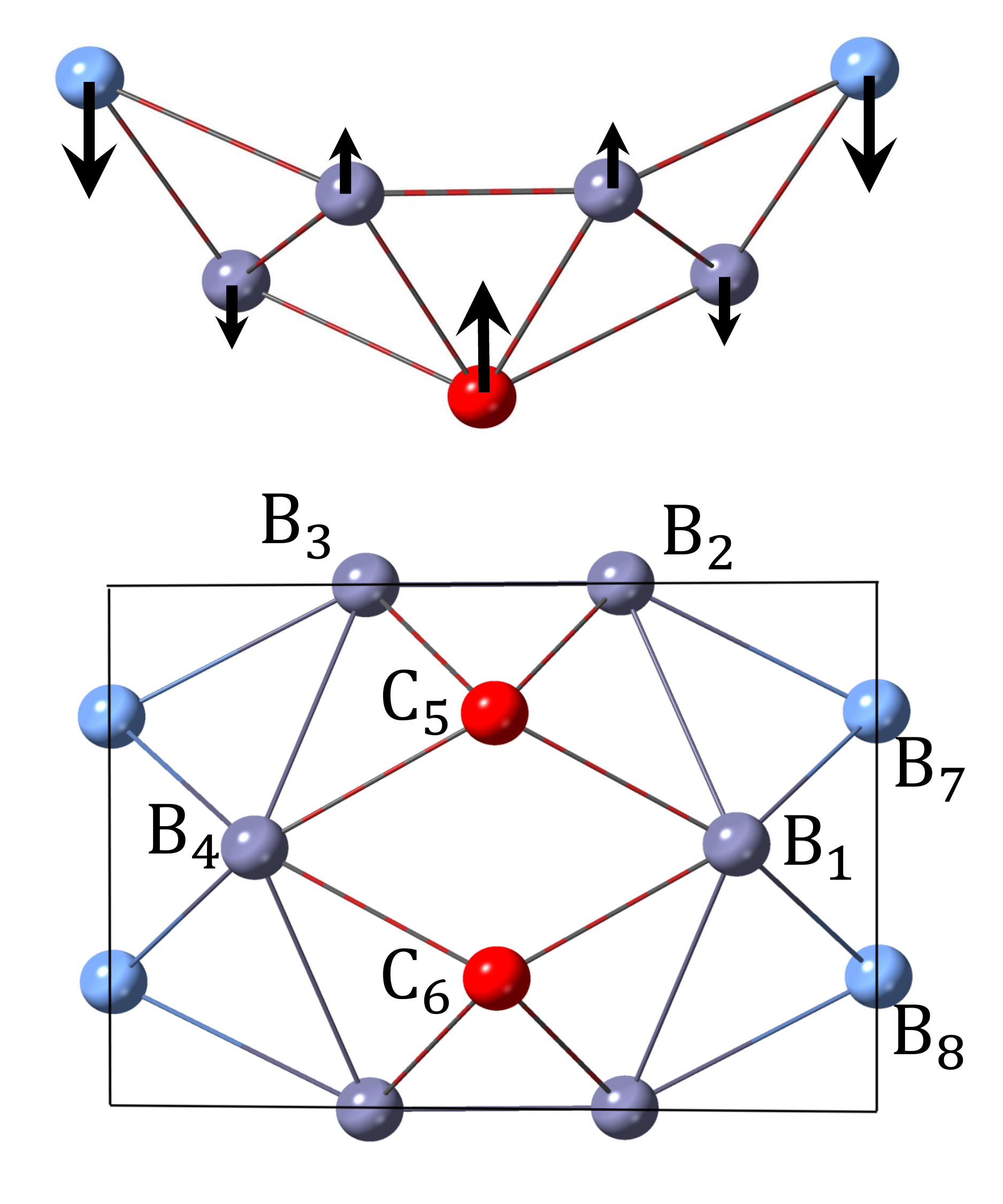}}
  \subfloat[]{ \includegraphics[scale=0.20]{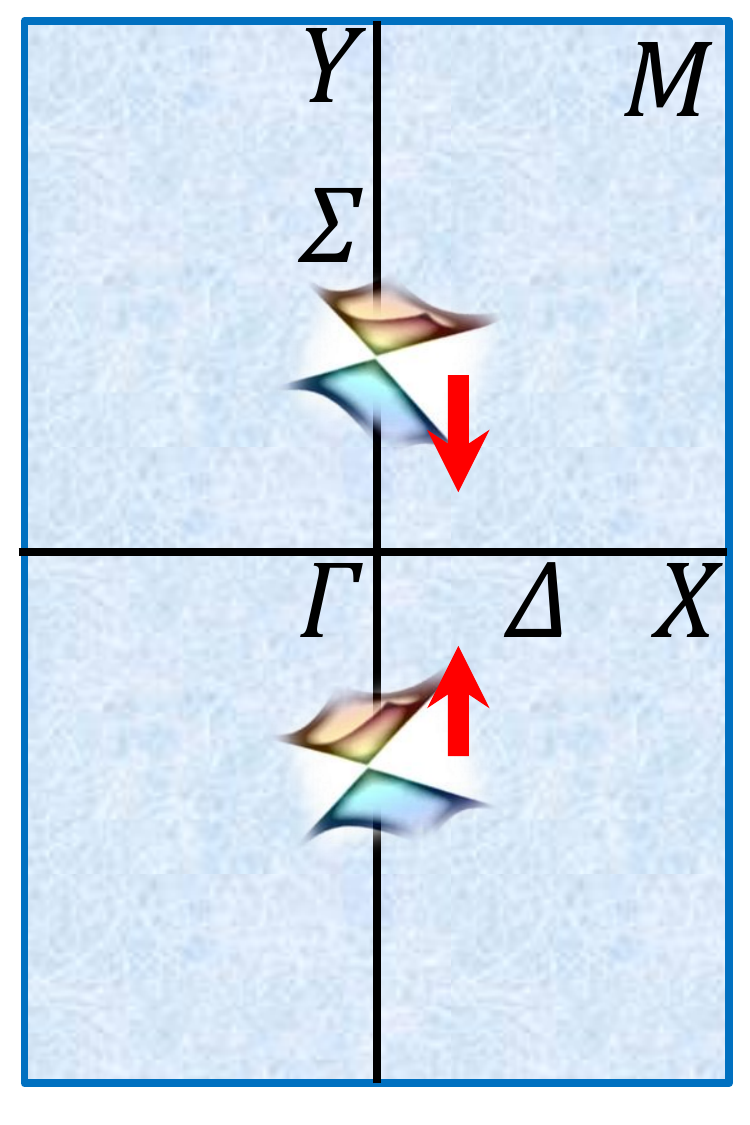}}
  \subfloat[]{ \includegraphics[scale=0.19]{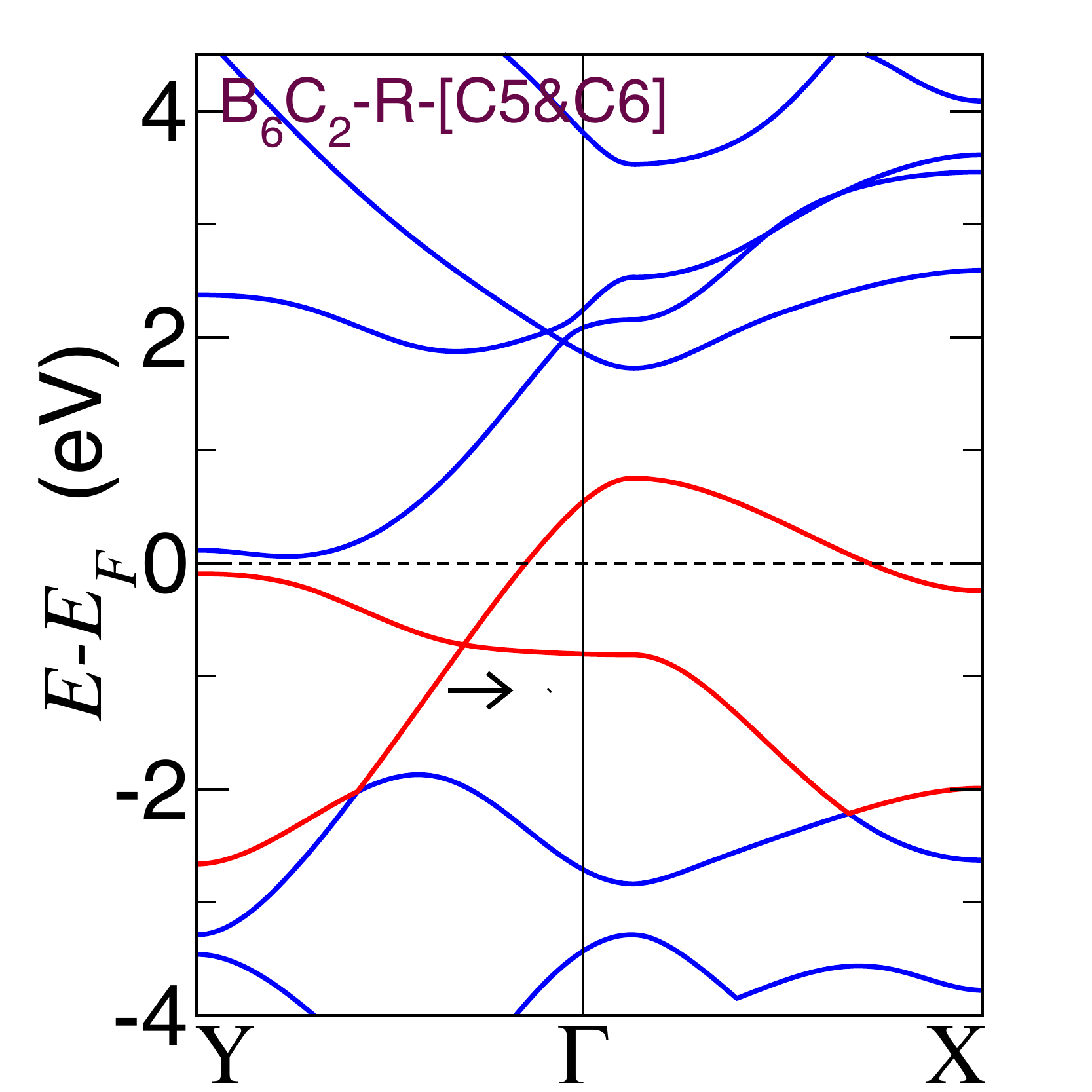} }
  \subfloat[] { \includegraphics[scale=0.25]{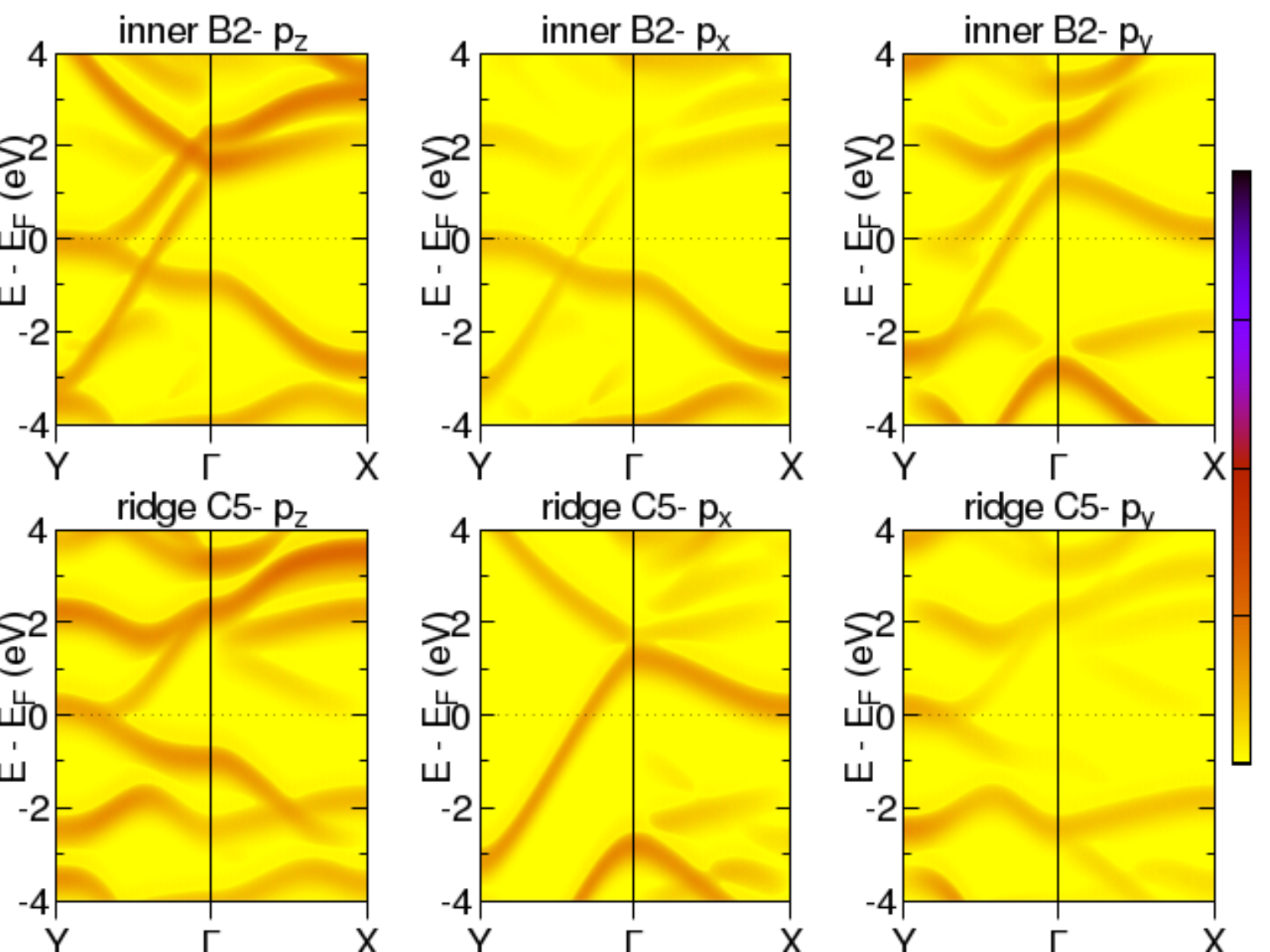}}
  \subfloat[] { \includegraphics[scale=0.18]{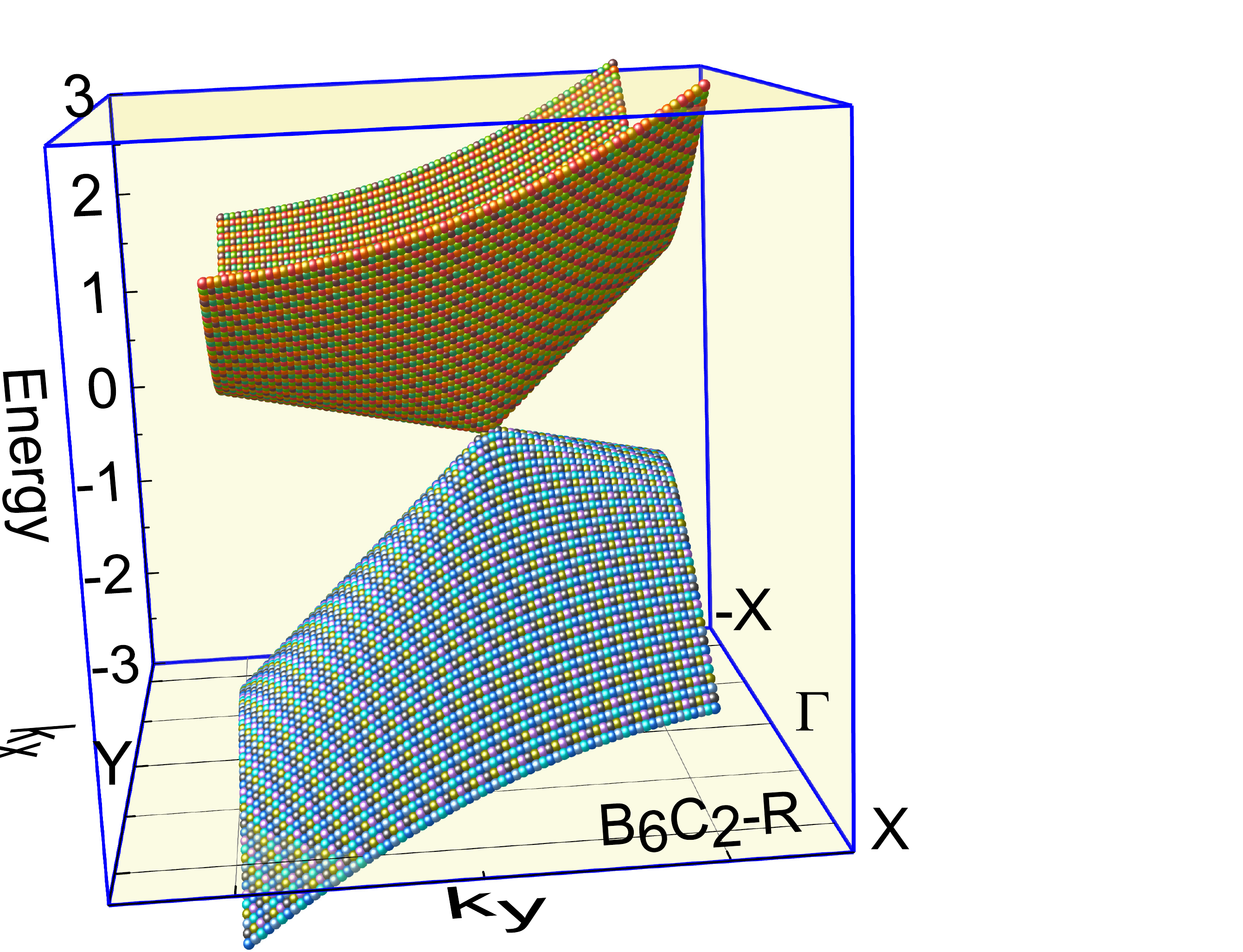}}
\\
\subfloat[]{ \includegraphics[scale=0.15]{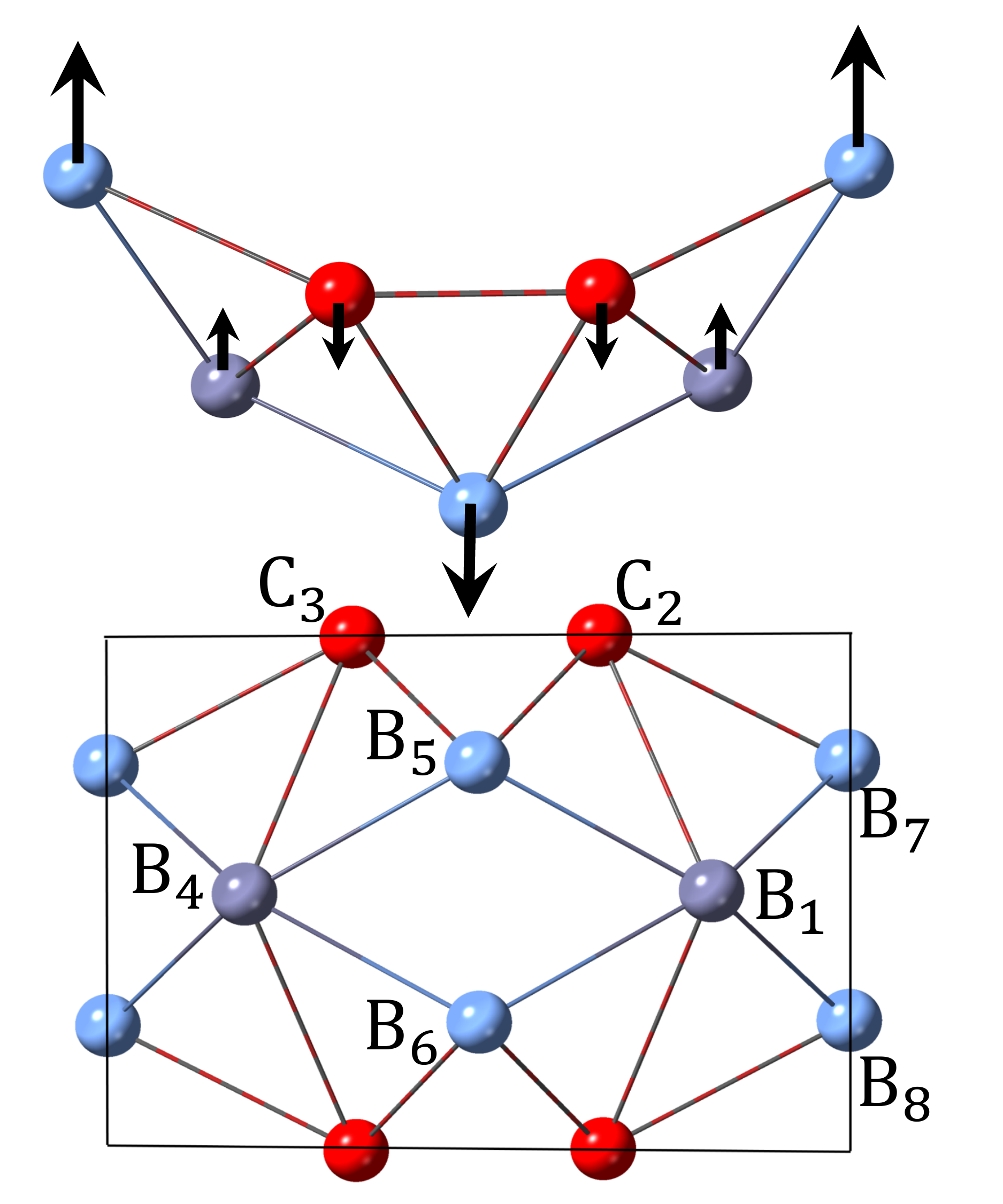}}
  \subfloat[]{ \includegraphics[scale=0.20]{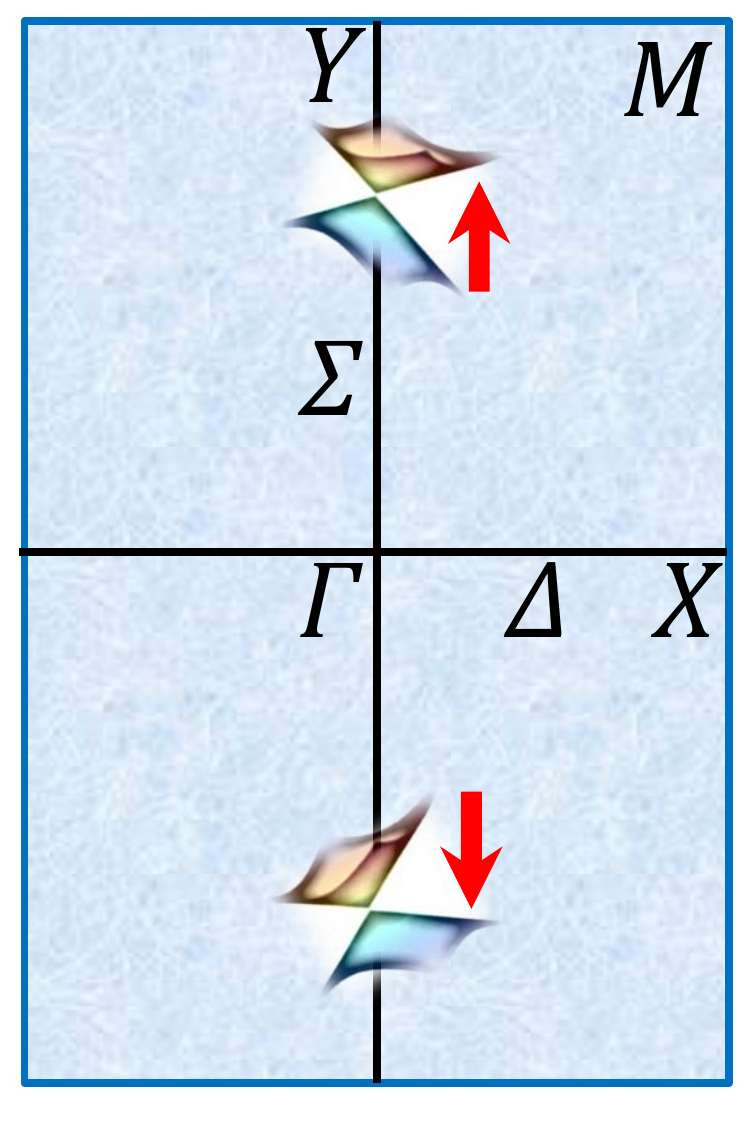}}
  \subfloat[]{ \includegraphics[scale=0.19]{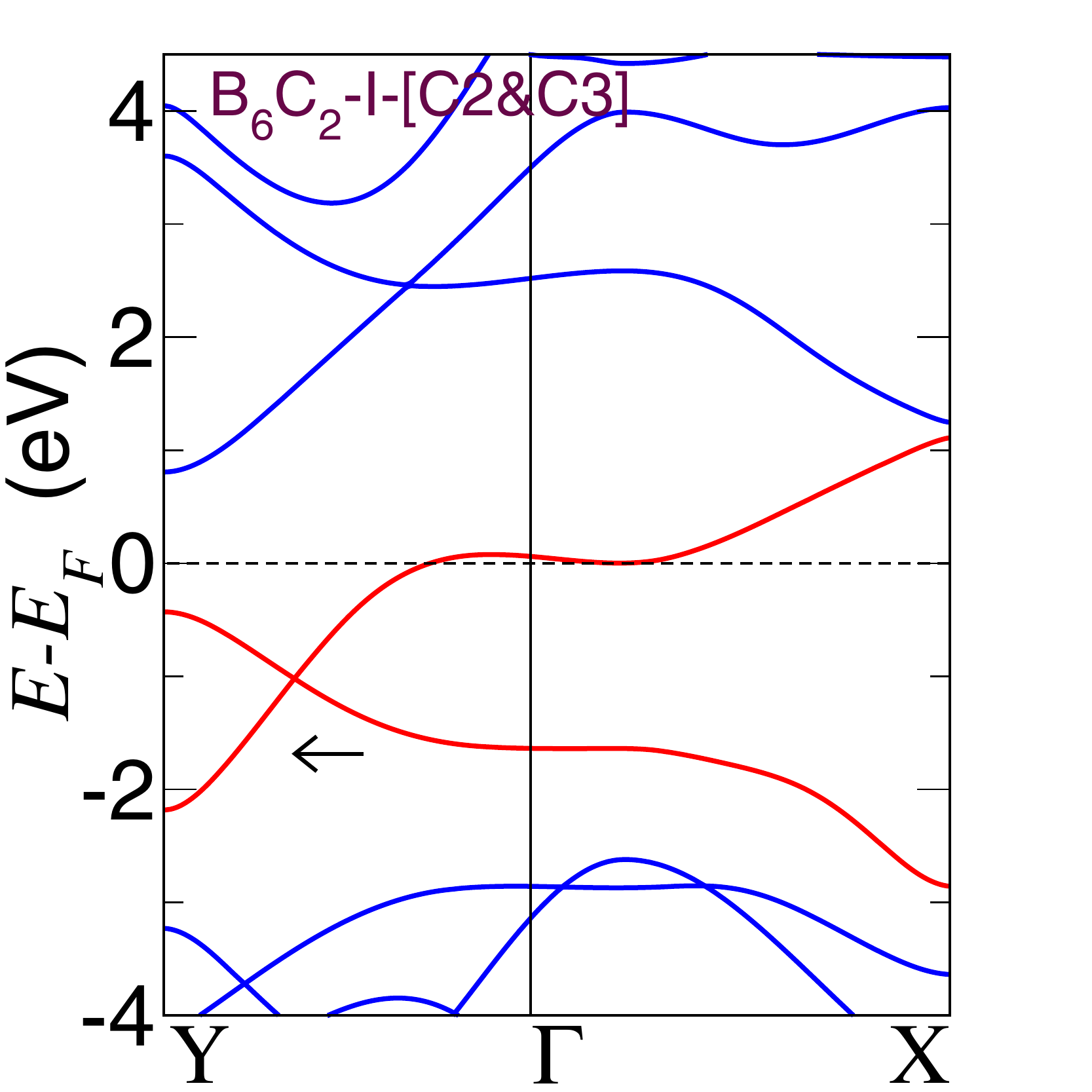}}
  \subfloat[] { \includegraphics[scale=0.25]{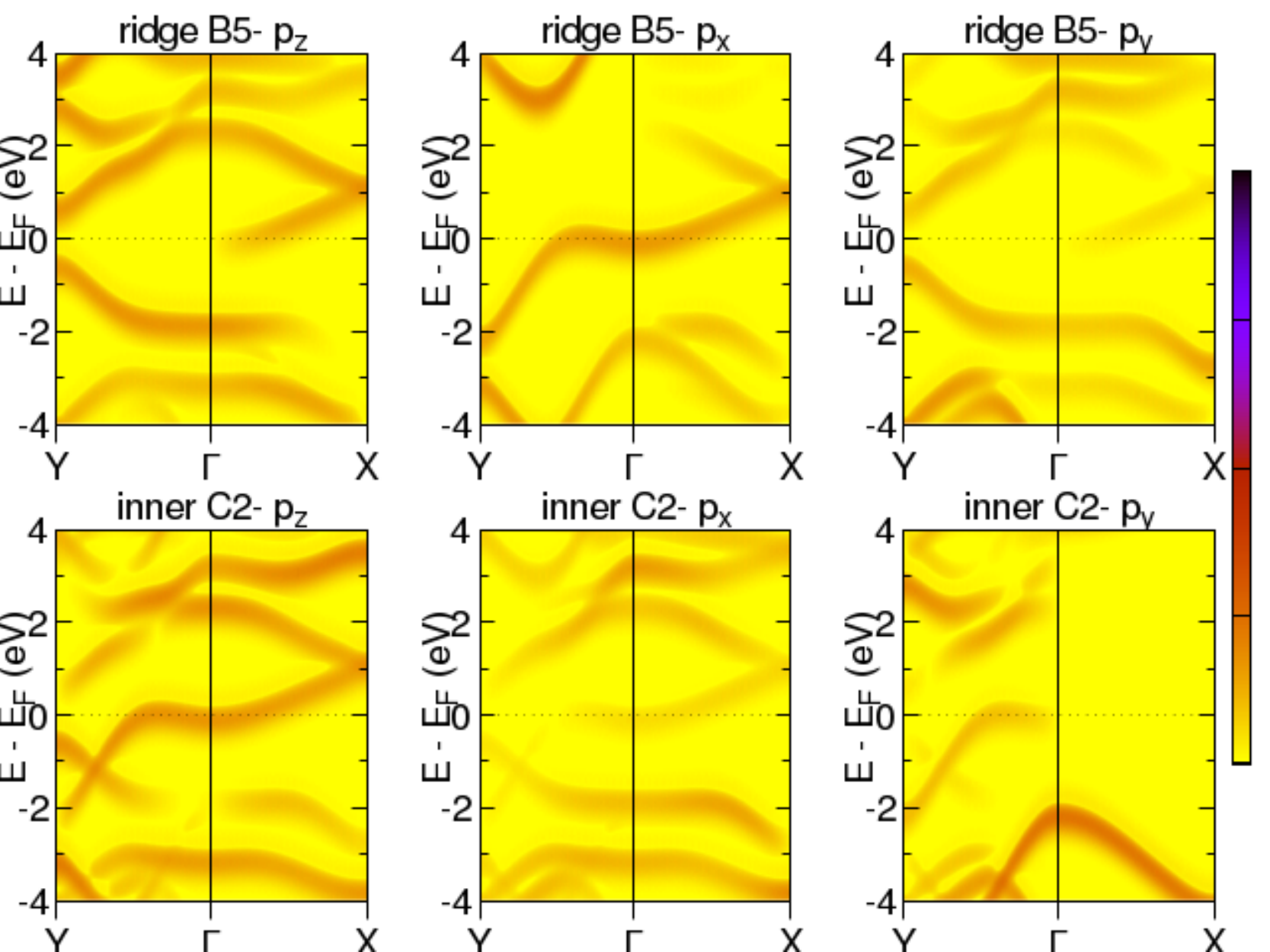}}
  \subfloat[] { \includegraphics[scale=0.18]{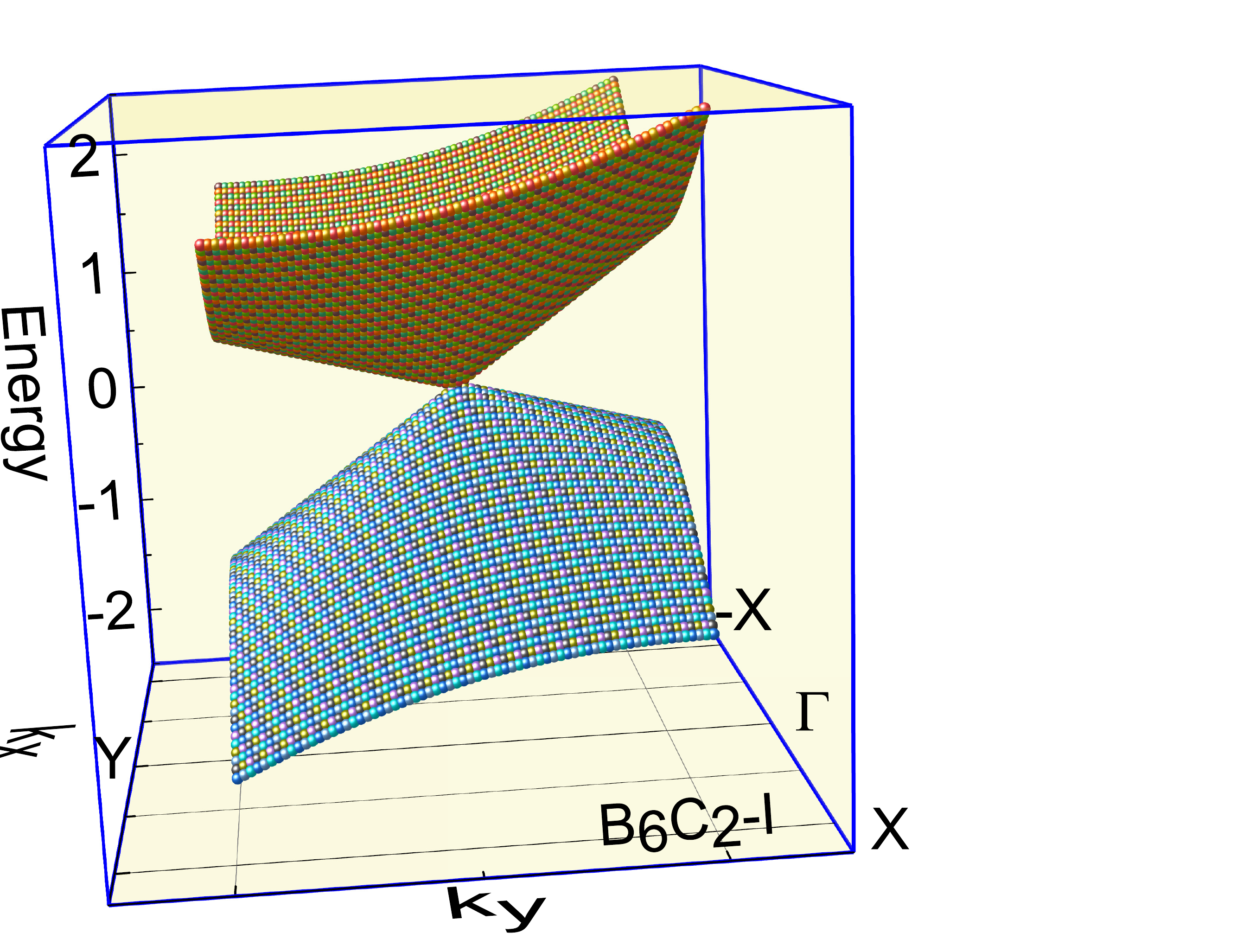}}
\end{center}
\vspace*{-0.6cm}
\caption{(Colors online) (a) Top view of crystal structure of B$_{6}$C$_{2}$-R-[C5$\&$C6] and (b) its Brillouin zone. The red circles denote carbon atoms. (c) DFT-PBE band structure of B$_{6}$C$_{2}$-R-[C5$\&$C6], (d) its orbital-projected band structures for two
atoms and (e) its three dimensional reconstruction. (f)-(j) the same as (a)-(e) for B$_{6}$C$_{2}$-I-[C2$\&$C3].
The horizontal arrows in panels (c) and (h) indicate the direction of the displacement of the Dirac node with respect to the parent B$_8$.
The vertical arrows in (a) and (f) show the direction of movement of the lattice sites upon carbon substitution.
}
\label{Fig5.fig}
\end{figure*}

The first neighbor hoppings denoted by black arrows in Fig.~\ref{Fig2.fig}(c)
and are present even when there are no R sites. In the two-dimensional Hilbert space of A and B sublattice
of such a honeycomb lattice, these hoppings contribute the usual off-diagonal term
   $F_0(\bs k)=-\sum_{\alpha=1,2,3}e^{i\bs k.\bs \delta_\alpha}$ to the Bloch Hamiltonian,
where the sum over $\alpha$ runs over three first neighbors in Fig.~\ref{Fig2.fig}(c).
This is responsible for the formation of a pair of upright Dirac cones similar to graphene. Slight anisotropy of the honeycomb lattice of R sites will amount to
a shift in the location of the Dirac cone~\cite{Vozmediano2010} which is irrelevant for our purposes. For concrete calculations, we assume
a regular effective honeycomb lattice of bond length $a$ for which the Dirac nodes in the rhombus BZ are at
$K_b^\pm=\frac{2\pi}{3a}(\hat i\pm\frac{1}{\sqrt 3}\hat j)$~\cite{KatsnelsonBook2012}.

Next let us consider the third neighbor hoppings denoted by red arrows in Fig.~\ref{Fig2.fig}(c) labeled
by $t^p$ ($p$ for pseudo, as they give rise to pseudogauge fields that shift the location of the Dirac node)
and $t^x$ ($x$ to emphasize the role of $p_x$ orbitals of R sites). They contribute another off-diagonal term
denoted by $=F_{xp}(\bs k)=t^p e^{2iak_x}+2t^x e^{-iak_x}\cos(\sqrt 3 a k_y)$ to the Bloch Hamiltonian.
Expanding this form factor around the Dirac nodes $K_b^\pm$ above, gives rise to (i) a shift $\Delta k_D=2(t^p-t^x)/(\mp 3at\pm a t^x)$ of the Dirac node,
where $t$ is the nearest neighbor hopping (assumed to be $1$) and (ii) anisotropy in the Fermi velocity
given by
   $v_{Fx}\to v_{Fx}\left(1+\frac{4t^p+2t^x}{3t} \right)$ and
   $v_{Fy}\to v_{Fy}\left(1-\frac{6t^x}{3t} \right)$.
Therefore the first and third neighbor hoppings
establish the location of the (still upright) Dirac cone and determine its Fermi velocities.

Now let us focus on the second neighbors (green arrows) in Fig.~\ref{Fig2.fig}(c) that are the root cause of the
tilt formation~\cite{Goerbig2008}. Since these hoppings are driven via virtual hopping through different arrangements of molecular orbitals,
there are two types of them denoted by solid ($\bar t$) and dashed ($\tilde t$) green lines. 
These hoppings being second neighbor, connect two atoms on the same sublattice, and therefore
contribute to the diagonal terms, namely $AA$ and $BB$ components of the effective $2\times 2$ Hamiltonian matrix.
Among the matrices $\sigma_\mu$ with $\mu=0\ldots 3$, only $\sigma_0$ and $\sigma_z$ can contribute
diagonal terms. So now one has to decide whether these diagonal terms come with the same sign ($\sigma_0$ term $\rightarrow$ tilt)
or opposite signs ($\sigma_z$ term $\rightarrow$ gap). There are two ways to see that this term must be proportional to the $\sigma_0$:
(i) Analysis of the irreducible representations and compatibility relations of the original $8Pmmn$ structure in Fig.~\ref{Fig1.fig}(d)
shows that the crossing of the red bands is protected by the glide elements of the $8Pmmn$ lattice (see SI for details). Since the effective
theory has to obey this protection against gap opening, the $\sigma_z$ term is ruled out. (ii) Consider the renormalized lattice itself and focus on the
solid green line in Fig.~\ref{Fig2.fig}(c). If the hopping between $1$ and $3$ in Fig.~\ref{Fig1.fig}(a) contributes to AA term, the hopping between $2$ and $4$
contribute to BB term. Both these contributions arise from the $p_z$ orbitals of these atoms via intermediate hopping through $p_z$ orbital of atom $5$.
Apparently for $p_z$ orbitals the "northwest" ($1\to 5\to 3$) and "northeast" ($4\to 5\to 2$) hoppings are identical. Similar arguments holds for the
dashed green line in Fig.~\ref{Fig2.fig}(c). In this case AA (BB) term is generated by $3\to 6\to 5\to 3$ ($2\to 6\to 5\to 2$) path. Again the $p_x$ orbitals
of the $5,6$ atoms symmetrically connect $3\to 3'$ and $2\to 2'$ (remember $3'$ is the same as $3$ but in adjacent unit cell. Similarly for $2'$.), thereby giving identical AA and BB terms in the effective Hamiltonian.

Therefore the effective Hamiltonian becomes
\bea
&&H_{\rm eff}(\bk)=\begin{pmatrix}
    f_{\rm tilt}(\bs k)  & f(\bs k)\\
    f^*(\bs k)           & f_{\rm tilt}(\bs k)
\end{pmatrix}, ~ f(\bs k)=F_{xp}(\bs k)+F_0(\bs k),\\
    &&f_{\rm tilt}(\bs k)=2\tilde t\cos(\sqrt 3 ak_y)+4\bar t\cos(3ak_x/2)\cos(\sqrt 3 ak_y/2).\nn
\eea
Taylor expanding the diagonal $f_{\rm tilt}$ term around the Dirac node formed by off-diagonal terms gives,
\be
   \zeta_x=0,~~\zeta_y=\pm 2\frac{\tilde t-\bar t}{t},
   \label{tiltformula.eqn}
\ee
where $\pm$ corresponds to the valley around which the expansion is performed.

The above equation indicates that the tilt arises from the difference of the second neighbor hoppings $\tilde t$ and $\bar t$
that in turn are generated via the $p_x$ and $p_z$ orbitals of the R atoms.
An immediate suggestion of the above model is to (partially) replace the R-site boron atoms by carbon atoms to see
whether the tilt is changed or not. In Fig.~\ref{Fig5.fig} we have replaced two of the boron atoms with C atoms. For this
purpose there are two choices: (i) To place a carbon dimer on the R sites as in the panel (a) or (ii) to place the carbon
atoms in the I sites as in panel (f). The results are summarized in Tab.~\ref{hoppings.tab}. For case (i), the location of the
Dirac node is shifted towards the $\Gamma$ pint and its tilt increases.
Placing carbon atoms in the R sites shifts the $5$ and $6$ sites to higher energies, thereby generating larger $\tilde t$ and $\bar t$ (see Tab.~\ref{hoppings.tab})
that ultimately increases the tilt from the $\zeta_y=0.46$ of the pristine borophene to $\zeta_y=0.59$ in B$_6$C$_2$-R-[C5$\&$C6]. Our picture provides also a way to decrease the
tilt parameter. In this case, the hybridization of the $p_z$ orbitals of sites $2,3$ with the R sites ($5,6$) reduces as in Tab.~\ref{hoppings.tab} and
hence the resulting $\bar t$ and likewise $\tilde t$ are scaled down, thereby reducing the tilt to $0.36$ in B$_6$C$_2$-I-[C2$\&$C3]. The above values of the tilt $\zeta_y$
are calculated based on Eq.~\eqref{tiltformula.eqn} and as detailed in SI are in good agreement with the corresponding values directly
extracted from DFT bands.

As shown in Fig.~S6 of the SI, for doping of two carbon atoms into the structural unit of B$_8$, the R-site configurations (c) and (k) have the lower
energy than all other configurations, with staggered configuration (k) having slightly lower energy than (c). For a perfect (translationally invariant) doping of 
two C into the $8Pmmn$ borophene structure, the changes in the tilt are discrete. However, the concentration of C atoms can be continuously varied.
Statistical averaging for a concentration $x$ that continuously varies between $0$ and $2/8$ is expected to generate a continuously varying tilt. 
The dominant configurations will be the R-site dimers and staggered R-site configurations of panel (c) and (k) of Fig.~S6. The lack of perfect
backscattering of Dirac electrons is expected to protect the Dirac cone against the Anderson localization.

\section{Summary and outlook}
Based on our {\em ab initio} calculations,
we have identified the $p_z$ orbitals of the I sites as real space sublattice on which the low-energy degrees of freedom in $8Pmmn$ lattice reside.
The effective hoppings between these sites are obtained via renormalization that encodes the virtual hopping via R-sites.
This gives a physically clear picture of the formation of the tilt in two-dimensional Dirac cone of $8Pmmn$ borophene.
In our picture, tilt arises from a competition between two renormalized hoppings $\tilde t$ and $\bar t$ between the second neighbor I-sites.
The former involves virtual hopping via two R sites (and $2p_x$ orbitals of the R sites), while the later involves one R site.
This builds in a natural difference between $\bar t$ and $\tilde t$. That is why the pristine borophene has a substantial tilting.
Within this picture it is natural to expect that
replacing the ridge (inner) atoms by C increase (decreases) the tilt.
The rest of the effective hoppings determine the location of the (protected) Dirac node and the anisotropy in the Fermi velocity.

The logic of our work can be applied to other SGs to discover more GQM within a class of 2D materials that afford to provide a 
"parent" upright Dirac cone~\cite{YasinPmm2}:
Molecular orbitals in peculiar SGs can mediate longer range hoppings on a backbone lattice of Dirac fermions by promoting it
to appropriate graph. The resulting graph in turn deforms the Minkowski spacetime of the Dirac fermions into a metric involving
spatio-temporal elements. The ability to tune the tilt parameter is tantamount to controllability of the ensuing solid-state spacetime structure. 
Dirac materials subject to "gravitational" (i.e. geometric) disorder~\cite{Foster2020Lensing,Foster2021Geodesic} 
find their salient materialization in the present context when the carbon is randomly substituted for boron.
Assuming a doping fraction $x$ of carbon atoms that can be continuously tuned, the most favorable configurations for the 
C atoms are R-site dimers and staggered R-site (Fig. S6 (c) and (k), respectively). Random substitutions of this form is expected to 
lead to continuous variation of tilt as a function of $x$. The absence of backscattering for Dirac fermions protect it against 
Anderson localization~\cite{Amini2009}. 

The relation between certain graphs and their continuum limit as space geometries is well known~\cite{Boettcher2020,Baek2009,Kollar2019}.
Our current results suggests a solid-state platform to promote a simple lattice that supports Dirac fermions into a rich graph 
where vierbeins can be attached to the Dirac fermions~\cite{Yepez2011,Hughes2013}. This enriches the physics of Dirac materials and promotes them to GQMs
where a variety of spacetime geometries can be fabricated that might not even have any analogue in the cosmos. 
On the 2D materials side, there will be a plenty of room to explore the effects of "geometric" forces or even 
synthesise non-Abelian gauge fields~\cite{TohidPRR} that are likely to lead to better control in electronic/optical devices,
as the effects of spacetime curvature~\cite{ExiriNature,ExiriCurvature} can be much stronger GQMs.
Furthermore, the geometry of our emergent spacetime in GQMs roots in the Coulomb forces
that are much stronger than the gravitational forces. 
Furthermore, unlike the gravity, the Coulomb forces can be both attractive and repulsive. 
Therefore the geometric structure of GQMs seems to be much richer than the Einsten's gravity. 
In three dimensional tilted Dirac/Weyl fermions the resulting geometric theory can be even more interesting:
The fermions in such materials are chiral and hence a chiral geometric theory can be relevant to GQM. 
The developments in GQMs will allow us to emulate aspects of spacetime geometry that is not 
easily accessible in the cosmos. 
GQMs have a potential to equip Dirac fermions with types of vielbeins~\cite{Yepez2011} that are beyond the spatial 
veilbeins of the strain/dislocation paradigm~\cite{Hughes2013}.

\bibliography{mybib}

\newpage
\setcounter{equation}{0}
\setcounter{figure}{0}
\renewcommand{\theequation}{S\arabic{equation}}
\renewcommand{\thefigure}{S\arabic{figure}}

\begin{center}
{\bf Supplemental materials to the manuscript: Tunning the tilt of a Dirac cone by atomic manipulations: application to 8Pmmn borophene}
\end{center}

Since we expect dual readership both from condensed matter and gravity research,
in this supplementary information (SI), we provide pedagogical derivations of all the relations and figures referred in the text.
To benefit the graduate students, this SI is meant to make the paper self-contained, and understandable on its own trying to minimize 
consultation with other references.

\section{Symmetry analysis}
\label{section1}
In this section, we illustrate the application of group theory methods in classification and labeling of the band structures
by way of example of $8Pmmn$ borophene. This section is based on chapter 10 of C. Kittel's {\em Quantum Theory of Solids}~\cite{Kittel}. For a
quick and concise introduction to the theory of groups and their representations, we recommend chapter 3 of G. Mahan's {\em Applied Mathematics}~\cite{Mahan}.

\begin{figure}[H]
  \centering
  \subfloat[]{   \includegraphics[width=0.29\textwidth]{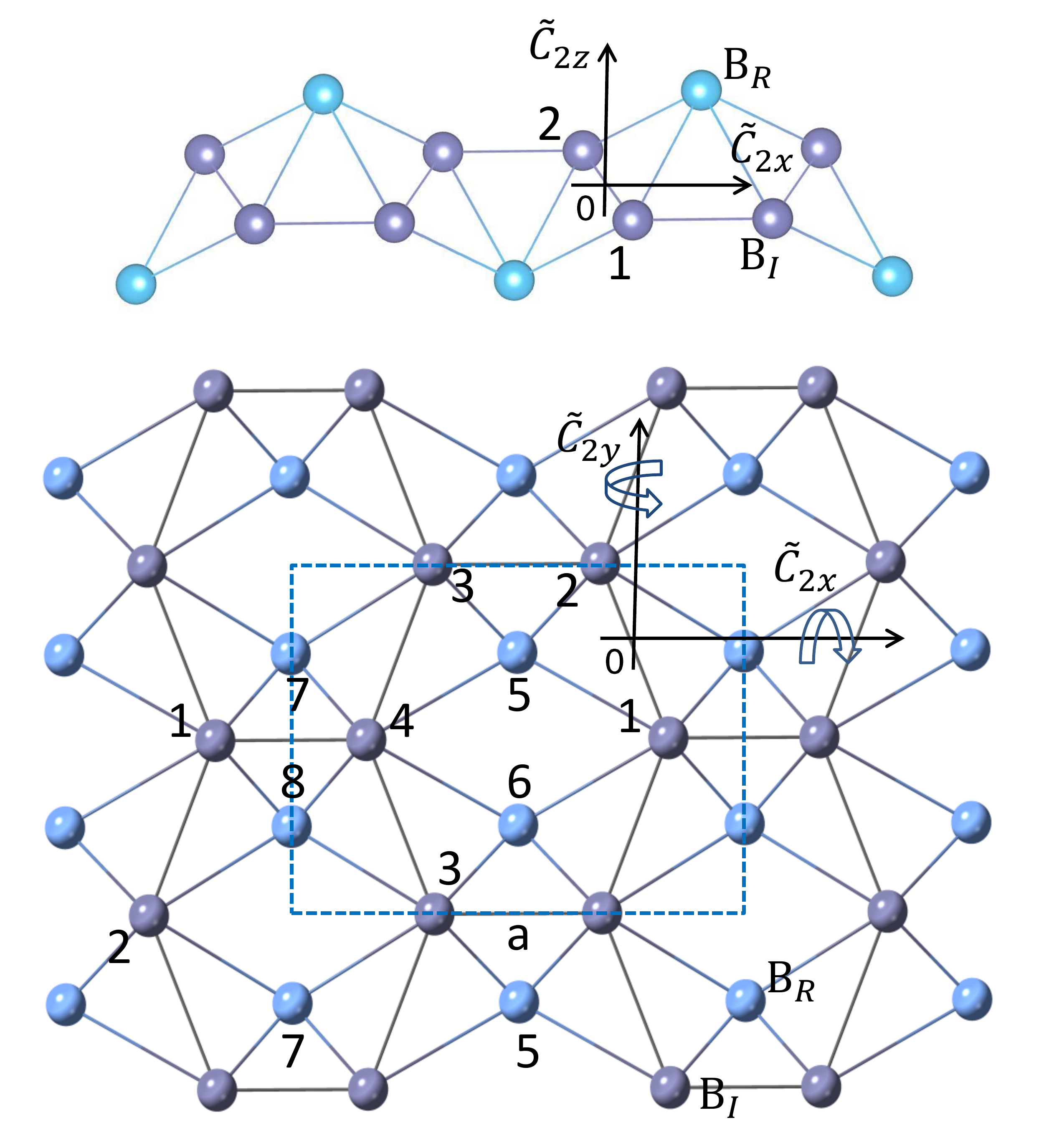} }
  \subfloat[]{   \includegraphics[width=0.37\textwidth]{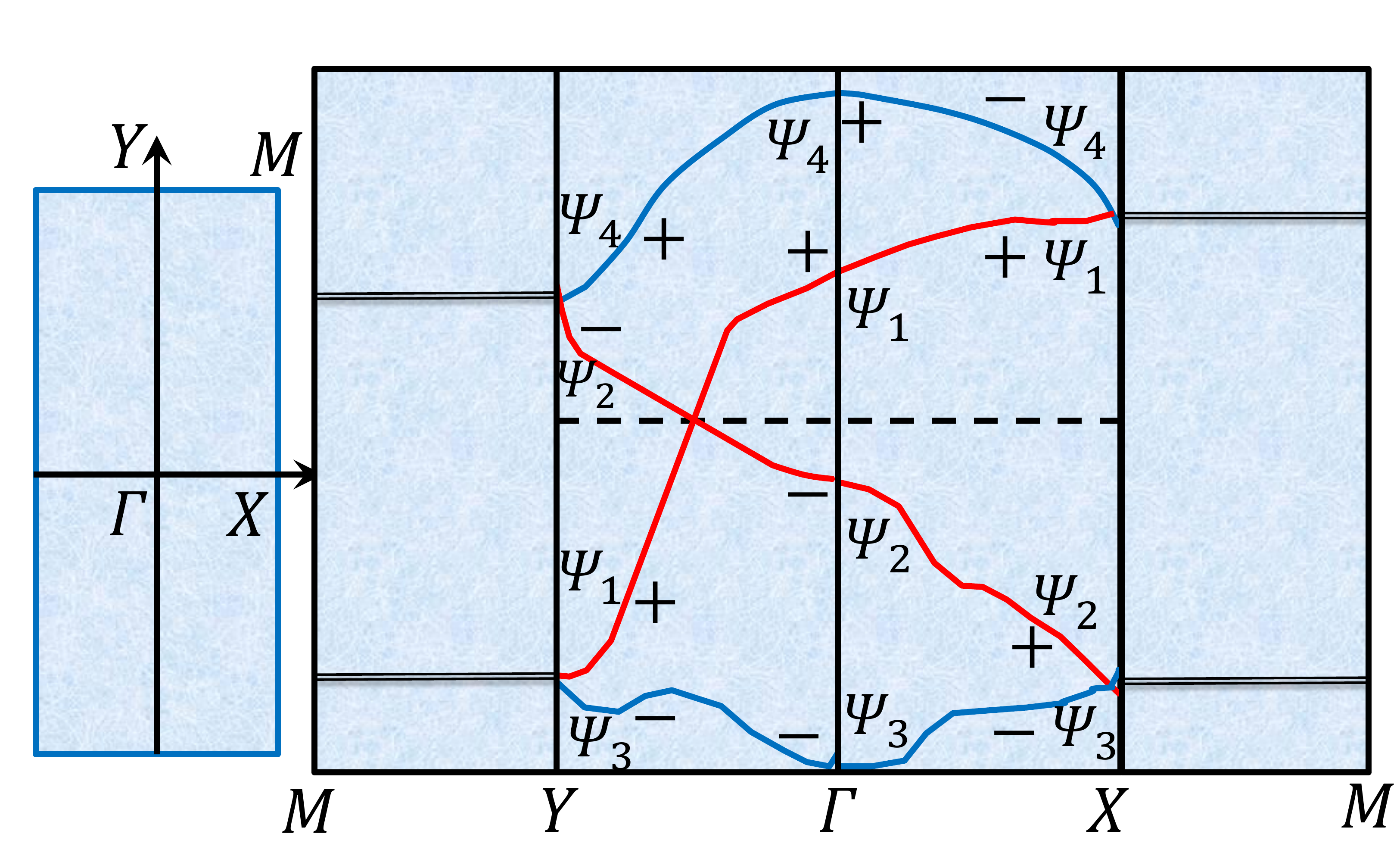} }
  \subfloat[]{   \includegraphics[width=0.25\textwidth]{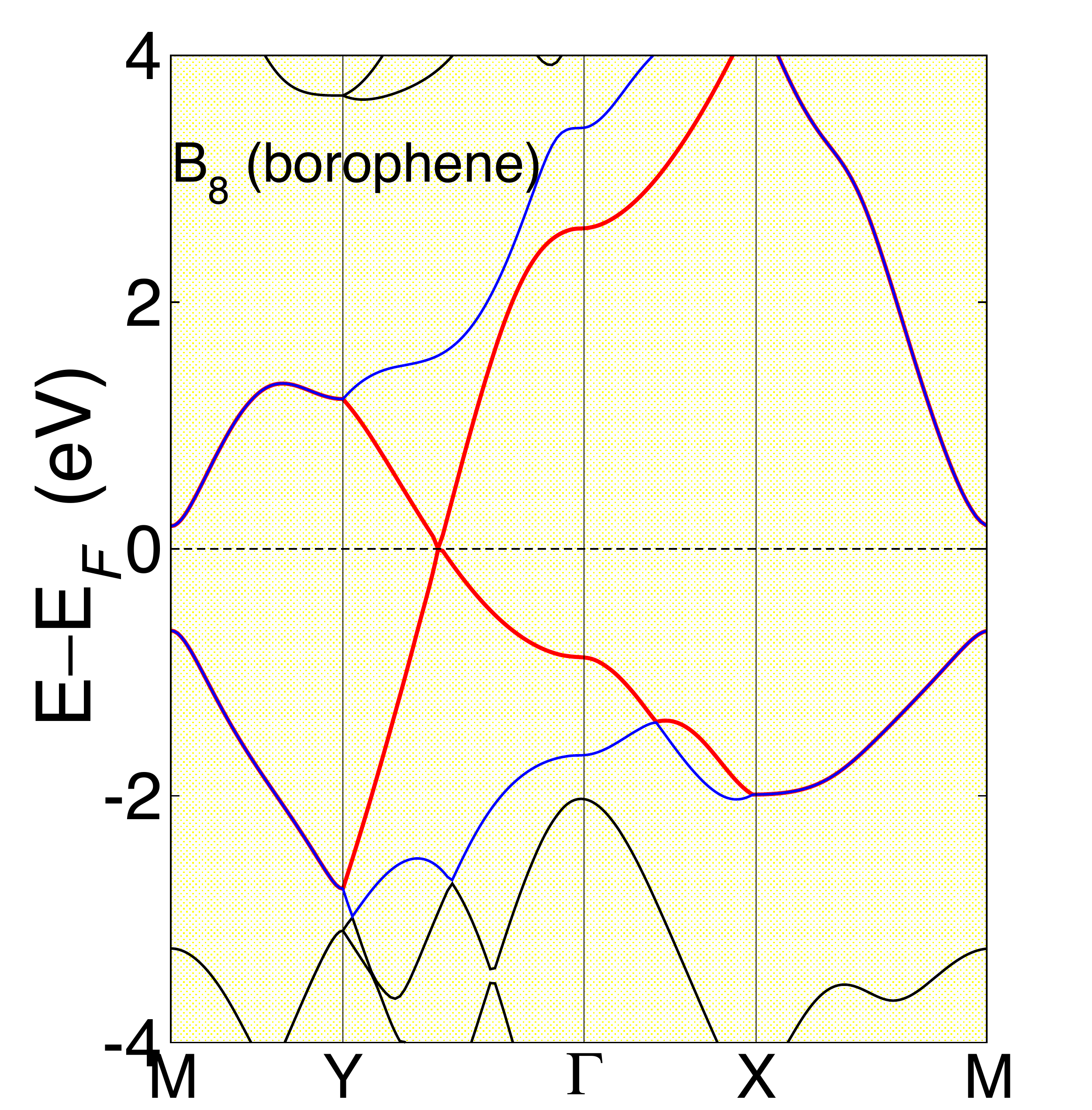} }
  \caption{(a) The crystal structure of the pure $8Pmmn$ borophene B$_{8}$. Dark and
light blue circles exhibit inner boron atoms B$_{I}$ and ridge boron atoms B$_{R}$ respectively. $\tilde{C}_{2x}=\{C_{2x}|\frac{a}{2}00\}$ and $\tilde{C}_{2y}=\{C_{2y}|0\frac{b}{2}0\}$ shown above, are two generators of $Pmmn$ space-group. (b) Rectangular first Brillouin zone of $8Pmmn$ borophene and band representation of
the $8Pmmn$ space group labeled according to notations of Eqs.~\eqref{psi2.eqn} and \eqref{psi3.eqn}. \hlt{The $\pm$ are the eigenvalues of $\tilde{C}_{2x}$, 
and $\tilde{C}_{2y}$ operations at high-symmetry points X and Y, respectively.}
(c) DFT-PBE band structure of pristine $8Pmmn$ borophene.}
\label{fig-sub-1}
\end{figure}

Here, the two-dimensional rectangular lattice of pristine borophene B$_{8}$ has symmetry group of $Pmmn$.
 The crystal structure of the pure $8Pmmn$ borophene and the character table for the D$_{2h}$ ($Pmmn$) point group are presented in Fig.~\ref{fig-sub-1}(a)
 and Table~\ref{table:1} respectively~\cite{Dresselhaus,Bradley-1}.
 The generators of $Pmmn$ space-group are given by two screw operations $\tilde{C}_{2x}=\{C_{2x}|\frac{a}{2}00\}$ and $\tilde{C}_{2y}=\{C_{2y}|0\frac{b}{2}0\}$ together with an inversion operation $\tilde{P}=\{P|000\}$, where $\tilde{C}_{2x}$ ($\tilde{C}_{2y}$) is a nonsymmorphic operator meaning the twofold rotation $C_{2x}$ ($C_{2y}$) around the $x$($y$) axis is followed by a translation $\frac{a}{2}$($\frac{b}{2}$) as indicated in Fig.~\ref{fig-sub-1}(a). 
 \hlt{Note that the inversion center is located at
the crossing point of two screw axes (between bond line of atom 1 and 2) as depicted in Fig.~\ref{fig-sub-1}(a) and the screw is performed around this inversion center.}
 These three independent operations can produce all other symmetry operations of the $Pmmn$ space group as follow:

\begin{equation}
\tilde{M}_{x}=P\tilde{C}_{2x}, \tilde{M}_{y}=P\tilde{C}_{2y}, \\
 \tilde{C}_{2z}=\tilde{C}_{2x}\tilde{C}_{2y}, \tilde{M}_{z}=P\tilde{C}_{2z}\\
\end{equation}

Without taking the spin of the electrons into account, $D_{2h}$($8Pmmn$) group has eight 1D irreducible representations given in character Table~\ref{table:1}. There are two other 2D representations belonging to the double group (i.e. accounting for both orbitals and spin) which is considered in presence of spin-orbit coupling and is not indicated in this table.

First of all, following Kittel~\cite{Kittel} we establish that the non-symmorphic elements $\tilde C_{2y}$ and $\tilde C_{2x}$ that are combination of two-fold rotations
around the axis indicated in Fig.~\ref{fig-sub-1}(a) followed by half-cell translation, guarantee the "sticking together" of the bands along the Brillouine zone (BZ) boundary
as indicated in Fig.~\ref{fig-sub-1}(b). 
Then we show that band inversion happens by connecting two eigenvalues $\pm$ of these operators, as a result of which
a Dirac point between $Y$ and $\Gamma$ is protected by the screw axis operator $\tilde{C}_{2y}$, in agreement with our 
{\em ab initio} band structure, as shown in Fig.~\ref{fig-sub-1}(c).

Suppose that $\Psi(x,y)$ is a solution of the wave equation at $k$-point along boundary of the BZ (i.e. $MX$ or $MY$ line). The screw rotation $\tilde{C}_{2x}$, mirror $\tilde{M}_{x}$, and inversion $\tilde{P}$ operations transform the wave function according to $ \tilde{C}_{2x}\Psi(x,y)=\Psi(x+a/2,-y)$,  $\tilde{P}\Psi(x,y)=\Psi(-x,-y)$, and $\tilde{M}_{x}\Psi(x,y)=\Psi(-x+a/2,y)$ respectively which result in $\tilde{M}_{x}\Psi(x,y)=\tilde{C}_{2x}\tilde{P}\Psi(x,y)$, or the multiplication rule $\tilde C_{2x}=\tilde M_x\tilde P$. 
Other multiplication rules similarly follows. 

\hlt{Now to prove that along the $M X$ line defined by $k_x=\pi/a$, there is a double degeneracy,
we show that $\tilde{P}$, $\tilde M_{x}$, and $i\tilde C_{2x}$ satisfy the algebra of Pauli matrices. 
The definition of $\tilde{C}_{2x}$ implies ($a$ is the length of the unit cell in $x$ direction)
\begin{eqnarray}
\tilde{C}_{2x}^{2}\Psi(x,y)=\tilde{C}_{2x}\Psi(x+a/2,-y)=\Psi(x+a,y)
=e^{ik_{x}a}\Psi(x,y)=e^{i\pi}\Psi(x,y)
=-\Psi(x,y)
\end{eqnarray}
where, in the the third equality have used the Bloch theorem and in the fourth equality we have used that on $M X$ line we have $k_xa=\pi$.}

\hlt{Subsequently, from the definition of $\tilde P$ and $\tilde M_{x}$,  We have
$\tilde{P}^{2}\Psi(x,y)=\Psi(x,y)$ and $\tilde{M}_{x}^{2}\Psi(x,y)=\Psi(x,y)$.
Then by multiplying these operators two by two, we find their anticommunication relations as follows:
\begin{eqnarray}
&&\tilde{P}\tilde{C}_{2x}\Psi(x,y)=\tilde{P}\Psi(x+a/2,-y)=\Psi(-x-a/2,y), \\
&&\tilde{C}_{2x}\tilde{P}\Psi(x,y)=\tilde{C}_{2x}\Psi(-x,-y)=\Psi(-x+a/2,y) = e^{ik_xa}\Psi(-x-a/2,y)
=-\Psi(-x-a/2,y)
\end{eqnarray}
which indicate that along the $M X$ line defined by $k_x=\pi/a$, it gives anticommunication relation $\{\tilde{P},\tilde{C}_{2x}\}=0$.}

\hlt{For $\tilde{P}$ and $\tilde{M}_{x}$ operators, the same procedure yields  
\begin{eqnarray}
   &&\tilde{P}\tilde{M}_{x}\Psi(x,y)=\tilde{P}\Psi(-x+a/2,y)=\Psi(x-a/2,-y), \\
   &&\tilde{M}_{x}\tilde{P}\Psi(x,y)=\tilde{M}_{x}\Psi(-x,-y)=\Psi(x+a/2,-y)=e^{ik_xa} \Psi(x-a/2,-y)=-\Psi(x-a/2,-y)
\end{eqnarray}
where again along the $M X$ line, we have $\{\tilde{P},\tilde{M}_{x}\}=0$.}

\hlt{Finally, considering $\tilde{C}_{2x}$ and $\tilde{M}_{x}$ operators results in
\begin{eqnarray}
  &&\tilde{M}_{x}\tilde{C}_{2x}\Psi(x,y)=\tilde{M}_{x}\Psi(x+a/2,-y)=\Psi(-x,y),\\
  &&\tilde{C}_{2x}\tilde{M}_{x}\Psi(x,y)=\tilde{C}_{2x}\Psi(-x+a/2,y)=\Psi(-x+a,-y)=e^{ik_xa}\Psi(-x,-y)=-\Psi(-x,-y)
\end{eqnarray}
which implies $\{\tilde{C}_{2x},\tilde{M}_{x}\}=0$ along the $M X$ line defined by $k_x=\pi/a$.}

\hlt{Naming $\gamma_{1}=\tilde{P}$, $\gamma_{2}=\tilde{M}_{x}$, and
$\gamma_{0}=\tilde{C}_{2x}$, we find that $\tilde{C}_{2x}$, $\tilde{P}$,  and $\tilde{M}_{x}$ obey the Clifford algebra defined by
anticommutation relations: $\{\sigma_{\mu},\sigma_{\nu}\}=2I\eta_{\mu\nu}$ where $I$ is the identity matrix and $\eta^{\mu\nu}={\rm diag}(-1,1,1)$ is the Minkowski metric. 
As is well known the representations of the above algebra are even dimensional, the lowest dimensional representation of which will be
two-dimensional where $\gamma^\mu$ matrices will be represented by Pauli matrces as e.g. $\gamma^0=i\sigma^z,\gamma^1=\sigma^x,\gamma^2=\sigma^y$.
Therefore along the $MX$, the bands are at least doubly degenerate. Similar arguments applies to three operators $\tilde M_y,\tilde C_{2y}$ and $\tilde P$
along the $MY$ direction. 
}

Now let us see how the degeneracies can split by moving along the $X\Gamma Y$ path. For this we need to
label the eigenstates at the $\Gamma$ point in terms of the eigenvalues of $\tilde C_{2x(y)}$. 
Since the $p_z$ states of inner sites in backbone
honeycomb sub-lattice are responsible for formation of Dirac cone, 
the most generic molecular orbitals or Bloch states at the vicinity of $\Gamma$ point composing the bottom of conduction band $\ket{\Psi_1}$ and $\ket{\Psi_4}$
and those at the top of valence band $\ket{\Psi_2}$ and $\ket{\Psi_3}$ can be constructed from the binding (anti-binding) combinations $|p_z^1\rangle\pm |p_z^2\rangle$
and $|p_z^3\rangle\pm |p_z^4\rangle$ of the $12$ and $34$ bonds, respectively as,

\begin{eqnarray}
&&\ket{\Psi_{1}}=\frac{1}{2}[\ket{p_{z}^{1}}-\ket{p_{z}^{2}}-\ket{p_{z}^{3}}+\ket{p_{z}^{4}}], ~~
\ket{\Psi_{4}}=\frac{1}{2}[\ket{p_{z}^{1}}-\ket{p_{z}^{2}}+\ket{p_{z}^{3}}-\ket{p_{z}^{4}}]
\label{psi2.eqn}\\
&&\ket{\Psi_{2}}=\frac{1}{2}[\ket{p_{z}^{1}}+\ket{p_{z}^{2}}-\ket{p_{z}^{3}}-\ket{p_{z}^{4}}],~~
\ket{\Psi_{3}}=\frac{1}{2}[\ket{p_{z}^{1}}+\ket{p_{z}^{2}}+\ket{p_{z}^{3}}+\ket{p_{z}^{4}}]
    \label{psi3.eqn}
\end{eqnarray}
Note that the above ordering that places $|\Psi_1\rangle$ and $|\Psi_4\rangle$ at the bottom of conduction band can be obtained from
a simple tight binding (molecular) picture involving atoms $1,2,3,4$ in Fig.~\ref{fig-sub-1}(a).
Likewise, it is quite reasonable to place $|\Psi_3\rangle$ (the most "even" combination of atomic orbitals with no minus sign) at the
lowest energy, followed by $|\Psi_2\rangle$ at slightly higher energy. 
Now we apply the action of the point group elements on this molecular orbitals to construct the elementary band representation. 
As depicted in Fig.~\ref{fig-sub-1}(a),  $\tilde{C}_{2x}$ replaces the site numbers as $1\leftrightarrow 3$ and $2\leftrightarrow 4$. 
Moreover, the $|p_z\rangle$ orbitals having $\propto z$ character, change sign under a $\pi$ rotation around $x$ (as well as $\pi$ rotation around $y$). 
Therefore the effect of $\tilde C_{2x}$ on the above wave functions becomes

\begin{equation}
\tilde{C}_{2x} \ket{\Psi_{1}}= +\ket{\Psi_{1}}, \tilde{C}_{2x} \ket{\Psi_{2}}= +\ket{\Psi_{2}}, \tilde{C}_{2x} \ket{\Psi_{3}}= -\ket{\Psi_{3}}, \tilde{C}_{2x} \ket{\Psi_{4}}= -\ket{\Psi_{4}},
  \label{psix.eqn}
\end{equation}
Similarly for the action of  $\tilde{C}_{2y}$ we have
$1\leftrightarrow 2$ and $3\leftrightarrow 4$, followed by $|p_z\rangle \to -|p_z\rangle$ (again since $p_z$ orbital has $\propto z$ character). 
Therefore its effect on the above four states becomes
\begin{equation}
\tilde{C}_{2y} \ket{\Psi_{1}}=  \ket{\Psi_{1}}, \tilde{C}_{2y} \ket{\Psi_{2}}= -\ket{\Psi_{2}}, \tilde{C}_{2y} \ket{\Psi_{3}}= -\ket{\Psi_{3}}, \tilde{C}_{2y} \ket{\Psi_{4}}=  \ket{\Psi_{4}} .
  \label{psiy.eqn}
\end{equation}
The effect of $\tilde{P}$ operator on the above wave functions are the same as $\tilde{C}_{2y}$. 
The eigenvalues of operators $\tilde{C}_{2x}$, $\tilde{C}_{2y}$, and $\tilde{P}$ for bottom of conduction and top of valence
band states (+1,+1,+1) and (+1,-1,-1) respectively in agreement with Ref.~\cite{Fan},

\begin{table}[H]%
\centering
\caption{The character table for
group of $D_{2h}$($8Pmmn$).} \label{table:1}
\begin{ruledtabular}
\begin{tabular}{ccccccccc}
 Rep & $E$ & $\tilde{C}_{2z}$& $\tilde{C}_{2y}$& $\tilde{C}_{2x}$ & $\tilde{P}$& $\tilde{M}_{z}$& $\tilde{M}_{y}$ & $\tilde{M}_{x}$\\
\hline
$\Gamma_{1}^{+}$ & 1 & 1 & 1 & 1 & 1 & 1 & 1 & 1\\
$\Gamma_{1}^{-}$ & 1 & 1 & 1 & 1 & -1 & -1 & -1 & -1\\
$\Gamma_{2}^{+}$ & 1 & 1 & -1 & -1 & 1 & 1 & -1 & -1\\
$\Gamma_{2}^{-}$ & 1 & 1 & -1 & -1 & -1 & -1 & 1 & 1\\
$\Gamma_{3}^{+}$ & 1 & -1 & 1 & -1 & 1 & -1 & 1 & -1\\
$\Gamma_{3}^{-}$ & 1 & -1 & 1 & -1 & -1 & 1&-1& 1 \\
$\Gamma_{4}^{+}$ & 1 & -1& -1& 1 & 1&-1&-1& 1 \\
$\Gamma_{4}^{-}$ & 1 & -1& -1& 1 &-1& 1& 1&-1\\
\end{tabular}
\end{ruledtabular}
\end{table}


Now we need all the possible paths where the crystal symmetries connect bands with corresponding eigenvalues of the above three operators in a consistent way.
Compatibility relations~\cite{Kittel} express that the irrepresentation along the high-symmetry lines are completely determined by the irrepresentation that appear at high-symmetry points (little groups of high-symmetry points/lines). We apply it to the line $X\Gamma$ and $\Gamma Y$ according to the Eqs.~\eqref{psix.eqn} and \eqref{psiy.eqn}. 
Taking one possible path as an example, as shown in Fig.~\ref{fig-sub-1}(b) for $\ket{\Psi_{1}}$, one of the conduction bands from $X$ point (the eigenfunction of $\tilde{C}_{2x}$ with eigenvalue of +1) is connected to the $\Gamma$ point (the eigenfunction of $\tilde{P}$ with eigenvalue of +1), and then inverted with the valence band from $Y$ point (the eigenfunction of $\tilde{C}_{2y}$ with eigenvalue of +1) that gives (tilted) cone crossing along the $\Gamma Y$ path when we consider the same procedure for $\ket{\Psi_{2}}$.

For the formation of cone crossing a "sticking together" (as Kittel puts it) of conduction states and valence states is necessary. Such sticking together
is protected by glid/screw elements~\cite{Kittel}. In our case it is the screw-symmetry $\tilde{C}_{2y}$.
Other possibilities are to have the crossing along the $\Gamma X$ path, or to have a crossing between the blue and red bands
of Fig.~\ref{fig-sub-1}(b). These possibilities for band inversion between states $\Psi_\alpha\rangle$ with $\alpha=1,2,3,4$ is ruled out by the
specific ordering given in Eqs.~\eqref{psi2.eqn} and~\eqref{psi3.eqn}.

\section{8Pmmn motivated Tight-binding model for tilted Dirac fermions}
\label{tbmode.sec}
As pointed out in the main text, effective degrees of freedom are $p_z$ orbitals of inner sites the projection of which on the
$xy$ plane will be a honeycomb lattice. In this effective model, the buckling of honeycomb lattice sites are ignored. This can be
absorbed into effective hybridization with $p_x$ orbitals of the ridge sites. Furthermore, the projection of the inner sites of
$8Pmmn$ lattice into $xy$ plane will not be a regular honeycomb lattice. Ignoring this will amount to a shift in the location of
the Dirac node. This shift can also be absorbed into various effective hoppings. Therefore for modeling purposes, it is enough
to start with an "effective regular honeycomb lattice" on which the relevant degrees of freedom (the $p_z$ orbitals of the inner sites) reside.
In this section we pedagogically derive all the details related to the formation, movement and titling of the Dirac cone based on
the effective honeycomb lattice. As is shown in the main text, the picture arising from an effective model can accurately
reproduce the DFT tilt parameters. Furthermore, having a tight binding two-band model allows to study various effects such as
disorder/interactions/symmetry breaking.

\subsection{Theory without ridge atoms}
\begin{figure}[H]
  \centering
  \includegraphics[width=0.72\textwidth]{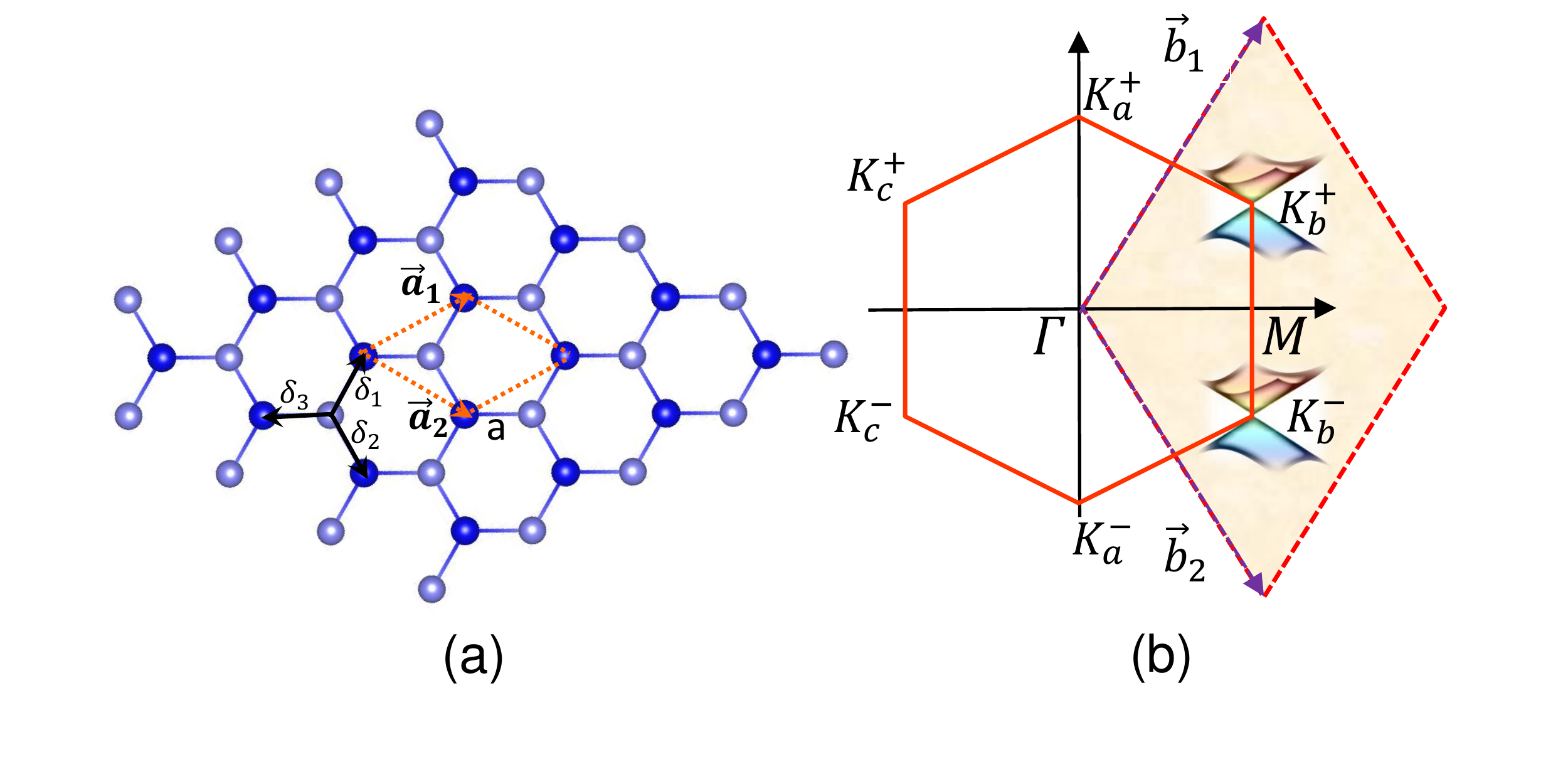}
  \caption{(a) The honeycomb lattice with two electrons per unit cell. The lattice constant is $a$ and three vectors connecting to nearest neighbors are denoted by $\delta_i,i=1,2,3$ (b) Two representations of the first Brillouin zone. The hexagonal Brillouin zone is the Wigner-Seitz cell in the reciprocal lattice and manifests the symmetries better. The rhombic Brillouin zone better indicates two independent Dirac cones at wave vectors at $K_b^\pm$.}
\label{fig-sub-2}
\end{figure}

Without ridge atoms (faint teal sites in Fig.~1(a) and Fig.~2(a) of the main text), life is simple and one
basically needs a theory of graphene which will be pedagogically and quickly reviewed below to build upon.
We therefore start with Fig.~\ref{fig-sub-2} that depicts a regular honeycomb lattice and its associated
first Brillouin zone.

The three vectors connecting every site from a given sublattice to three nearest neighbors from the opposite
sublattice in Fig.~\ref{fig-sub-2}(a) are
\begin{equation*}
\begin{aligned}
&\delta_{1}=\frac{a}{2}(1,\sqrt{3}), \\
&\delta_{2}=\frac{a}{2}(1,-\sqrt{3}), \\
&\delta_{3}=-a(1,0).
 \end{aligned}
\end{equation*}
The primitive lattice basis vectors  are
\begin{equation*}
\begin{aligned}
&\vec{a}_1=\frac{3}{2}a\hat{i}+\frac{\sqrt{3}}{2}a\hat{j}, \\
&\vec{a}_2=\frac{3}{2}a\hat{i}-\frac{\sqrt{3}}{2}a\hat{j}.
  \end{aligned}
\end{equation*}
Using $\vec a_i.\vec b_j=2\pi \delta_{ij}$ gives the following basis for the reciprocal space,
\begin{equation*}
\begin{aligned}
&\vec{b}_1=2\pi(\frac{1}{3a}\hat{i}+\frac{1}{\sqrt{3}a}\hat{j}), \\
&\vec{b}_2=2\pi(\frac{1}{3a}\hat{i}-\frac{1}{\sqrt{3}a}\hat{j}),
  \end{aligned}
\end{equation*}
where $a$ is the distance between two neighbouring atoms.
In the hexagonal Brillouin zone it appears the there are six corners $K$. But in in fact two of them are independent and fall into two categories
related to each other by a reciprocal lattice vector constructed from the above vectors. For example
the corners of Brillouin zone shown in Fig.~\ref{fig-sub-1}(b) labeled by $K_{a}^{\pm}$ and $K_{b}^{\pm}$
are given by,
\begin{equation*}
\begin{aligned}
&K_{a}^{+}=\frac{4\pi}{3\sqrt{3}a}\hat{j}, K_{a}^{-}=-\frac{4\pi}{3\sqrt{3}a}\hat{j} \\
&K_{b}^{+}=\frac{2\pi}{3a}(\hat{i}+\frac{1}{\sqrt{3}}\hat{j}), K_{b}^{-}=\frac{2\pi}{3a}(\hat{i}-\frac{1}{\sqrt{3}}\hat{j})
  \end{aligned}
\end{equation*}
As can be seen adding $\vec b_2$ to $K_a^+$ gives $K_b^-$ and hence they are equivalent.
Similarly $K_a^-$ and $K_b^+$ are equivalent. So in the calculations, one can pick a convenient one.
The effective theories around equivalent $K$ points are the same within a gauge transformation that will be
discussed below.
The $K_b^\pm$ points indicated in the rhombic Brillouin zone of Fig.~\ref{fig-sub-2}(b) manifestly emphasizes
that there are two independent corners that can not be connected to each other by a reciprocal lattice vectors.

Denoting by $a,b$ the field operators for the annihilation of electrons in $p_z$ orbitals of the $A$ and $B$ sublattices of
the honeycomb lattice, the parent Hamiltonian is basically the theory of graphene and is given by,
\begin{equation*}
\begin{aligned}
H_0=-t\sum_{i,j} a_{i}^{\dag} b_{j} + \mbox{h.c.} \\
  \end{aligned}
\end{equation*}
Fourier transformation of the above Hamiltonian becomes,
\begin{equation*}
H=-t\sum_{k,\alpha=1,2,3} a_{k}^{\dag} b_{k} e^{ik.\delta_\alpha} + \mbox{h.c.}
\end{equation*}
that begs for a matrix representation as
\begin{equation*}
H=\sum_{k}\begin{pmatrix} a_{k}^{\dag} & b_{k}^{\dag} \end{pmatrix}
\begin{pmatrix} 0 & F_0(\textbf{\emph{k}}) \\ F_0^{*}(\textbf{\emph{k}}) & 0  \end{pmatrix}.
\begin{pmatrix} a_{k} \\ b_{k} \end{pmatrix}.
\end{equation*}
This is how the spinor structure emerges.
The form factor connects the opposite sublattices, and hence is off-diagonal in the above sublattice space and
is given by,
\begin{equation*}
\begin{aligned}
F_0(\textbf{\emph{k}})= \sum_{\delta} te^{i\textbf{\emph{k}}.\delta}
 \end{aligned}
\end{equation*}
Plugging in the Cartesian representations of
$\delta_{1}$, $\delta_{2}$, and $\delta_{3}$ given above and carrying out the
summation over the three neighbors, the famous formula of graphene literature can be obtained,
\begin{equation*}
\begin{aligned}
 F_0(\textbf{\emph{k}})= te^{-ik_{x}a}+2te^{i\frac{1}{2}k_{x}a}\cos(\frac{\sqrt{3}}{2}k_{y}a)  \\
 \end{aligned}
\end{equation*}
The above form factor vanishes at the corners of the hexagonal Brillouin zone. Expanding it around two
independent corners, e.g. $K_a^\pm$ gives,
\begin{equation*}
\begin{aligned}
&F(K_{a}^{\pm}+\textbf{\emph{k}})=F(K_{a}^{+})+ \frac{\partial F}{\partial k_{x}}|_{\textbf{\emph{k}}=K_{a}^{\pm}}k_{x}+\frac{\partial F}{\partial k_{y}}|_{\textbf{\emph{k}}=K_{a}^{\pm}}k_{y},\\
&=-\frac{3}{2}ati(k_x\mp ik_y).\\
 \end{aligned}
\end{equation*}
The overall factor of $-i=e^{-i\pi/2}$ can be gauge transformed by rotating the overall phase of the wave function in one sublattice.
For example gauge transforming the amplitude of wave function on sublattice $B$ as $\phi_B\to i\phi_B$, eliminates the $-i$ factor above
and we have
\begin{equation*}
   F_0(\textbf{\emph{k}})\to \frac{3at}{2}(k_x\mp i k_y)
\end{equation*}
Upon the above gauge transformation the behavior of the Hamiltonian around the crossing points is given by,
\begin{equation*}
\begin{aligned}
  &H_0(\textbf{\emph{k}})=\frac{3at}{2}\begin{pmatrix} 0 & k_{x}\mp ik_{y} \\ k_{x}\pm ik_{y} & 0  \end{pmatrix}
=\frac{3at}{2}(k_{x}\sigma_{x}\pm k_{y}\sigma_{y})\\
  \end{aligned}
\end{equation*}
Defining $3ta/2=\hbar v_F$, this formula gives two representations of the Dirac theory $H_{\rm Dirac}(\textbf{\emph{k}})=\alpha_x k_x+\alpha_y k_y+\beta m$
with $\alpha_x=\sigma_x, \alpha_y=\pm \sigma_y$ and $m=0,\beta=\sigma_z$, hence massless Dirac electrons. These two representations are connected to
each other by time-reversal operator $\vec k\to -\vec k$  followed by ${\cal T}=i\sigma_y{\cal K}$ where ${\cal K}$ is complex conjugation.

\begin{figure}[H]
  \centering
  \includegraphics[width=0.44\textwidth]{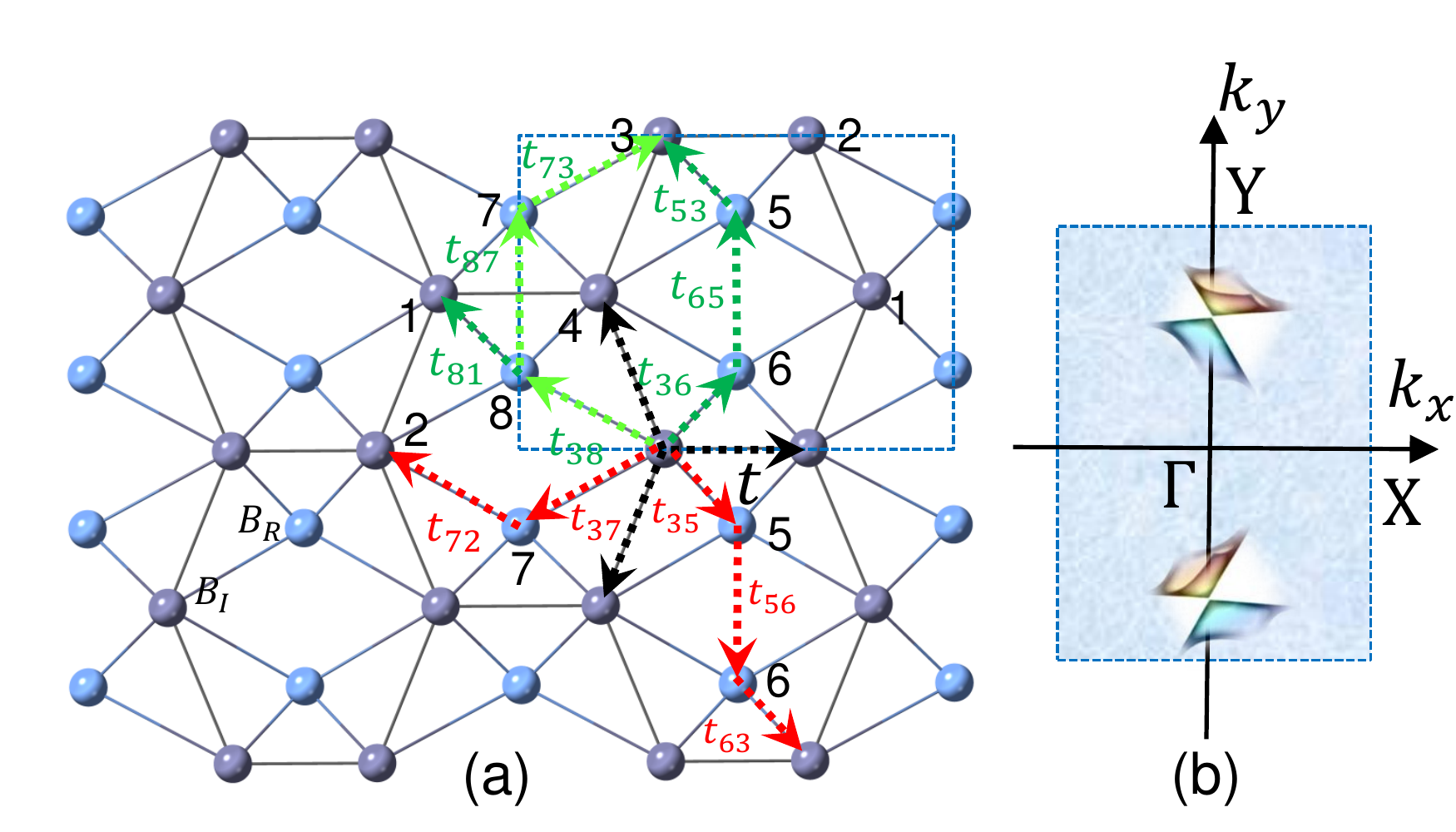}
  \includegraphics[width=0.45\textwidth]{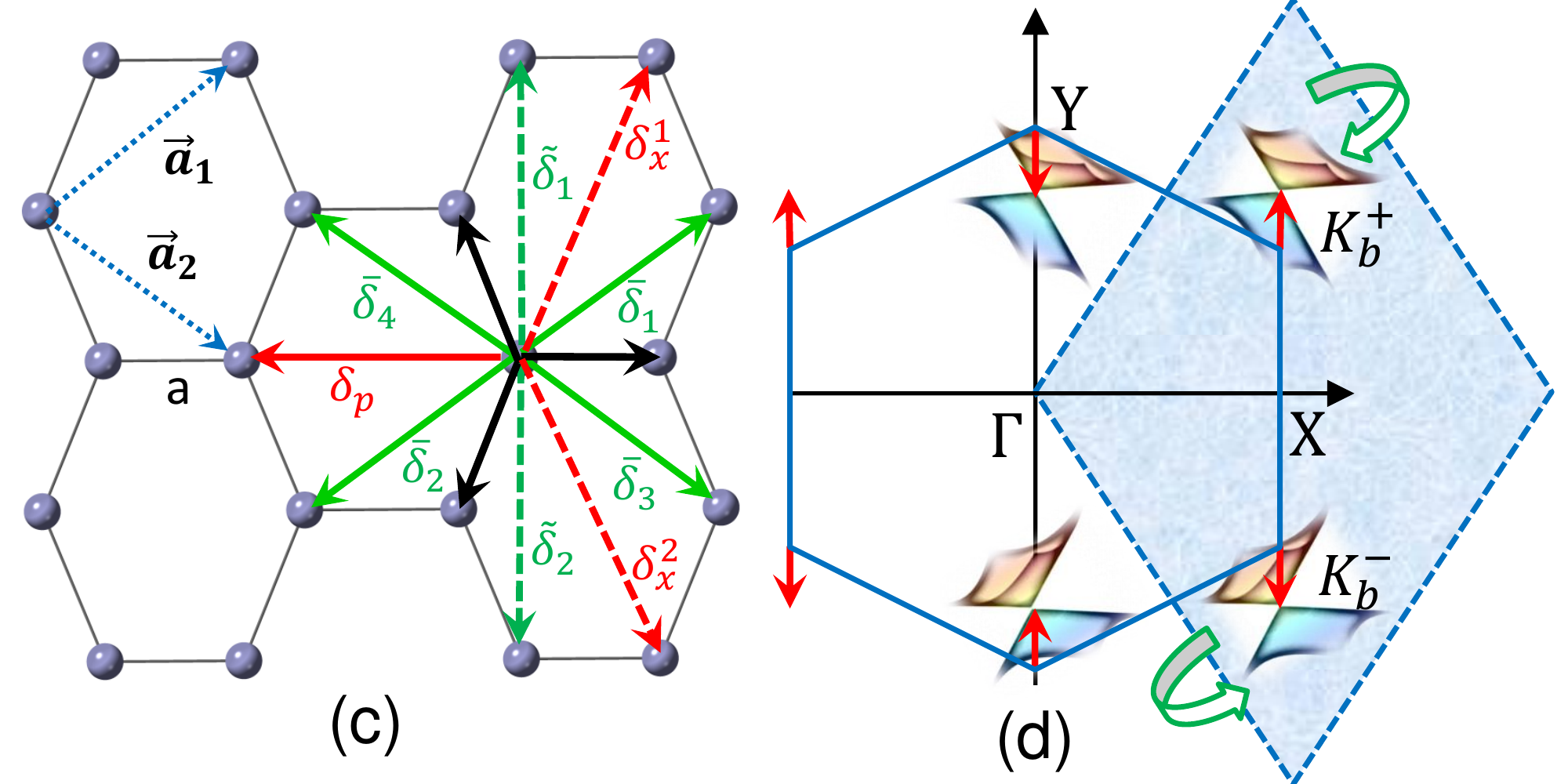}
  \caption{(a) The crystal structure of $8Pmmn$ borophene which consists of two different types of B atoms (B$_{R}$ and B$_{I}$).
  The coarse grained lattice of inner sites form a honeycomb structure. Black arrows indicate direct hopping between the atomic $p_z$ orbitals.
  There are additionally green path (second neighbor) and red path (third neighbor) hopping on the effective honeycomb lattice that becomes
  possible via the molecular orbitals involving $p_x$ and $p_z$ orbitals of ridge site atoms.
  (b) The first Brillouin zone of original borophene (B$_{8}$) lattice.
  (c) The effective honeycomb-like crystal structure of $8Pmmn$ borophene after decimation.
 The role of higher energy ridge site atoms is replaced by effective hoppings denoted by a single path of corresponding color, namely
 green for second neighbors and red for the same neighbors. The dashed and solid lines correspond to two different types of virtual processes.
 (d) The first Brillouin zone of effective hexagonal system. The two Dirac cones in (d) are born from two Dirac cones at (b), albeit shifted and
 tilted by effective green and red path hoppings of (c).}
  \label{fig-sub-3}
\end{figure}

\subsection{Role of ridge atoms}
The $8Pmmn$ borophene's lattice consists of two different types of B atoms, which we will call
inner B (B$_I$) and ridge B (B$_R$) atoms in Fig.~\ref{fig-sub-3}(a). The B$_{I}$ atoms form a hexagonal graphene-like lattice
and B$_{R}$ atoms look like one-dimensional chains passing through the effective honeycomb lattice as in Fig.~\ref{fig-sub-3}(c).
As pointed out, in absence of B$_R$ atoms,  the structure looks like a distorted graphene lattice. This lattice naturally includes the nearest neighbor (black path hoppings in panel (c)) that gives rise to two Dirac cones described above. Deviations of the honeycomb lattice from planar structure allows for mixing with $p_x$ orbitals of the ridge
sites, while the deviations of the projected honeycomb lattice from perfect $C_6$ symmetry of graphene accounts for shift in the location of the Dirac cones. Therefore
the schematically depicted Dirac cones in panel (d) above are shifted and tilted as described below to form the two Dirac cones in panel (b) of the original $8Pmmn$ lattice.

If there were not ridge sites, in a pure honeycomb lattice, hoppings between 2nd and 3rd neighbors via atomic orbitals would be negligibly small.
Therefore there would be no green and red hoppings as in panel (c). However, ridge atoms provide dotted hopping paths of panel (a) that facilitate
a hopping between 2nd and 3rd neighbors of the parent honeycomb-like lattice via virtual hopping through the ridge atoms. The connection between the microscopic paths in (a) and effective hoppings in (c) will be discussed in next section. For the purpose of present sub-section, we only need to focus
on the effective green and red hoppings in panel (c) above. The corresponding hoppings as depicted in Fig.~2(c) of the main text are $\bar t$ and $\tilde t$
for the green path (2nd neighbor). Two different symbols correspond to two different virtual processes depicted in panel (a) above. Similarly for the red paths
there are two hoppings $t^p$ and $t^x$ as depicted in Fig.~2(c) of the main text. As we will see shortly, $t^p$ (and $t^x)$ acts like a {\em pseudo-gauge} field that shifts the
location of the Dirac cone. The superscript $x$ in $t^x$ is meant to emphasize a special role played by the $p_x$ orbitals of the ridge atoms that will be discussed in the next section. The $t^p$ and $t^x$ being 3rd neighbor hoppings connect a site from A sublattice to a one in the B sublattice. Therefore the corresponding
form factor $F_{xp}(\textbf{\emph{k}})$ will contribute off-diagonally to the effective Hamiltonian. Furthermore, the hoppings $\bar t$ and $\tilde t$ being 2nd neighbor hoppings
connect atoms on the same site, and therefore the corresponding form factors $\bar F(\textbf{\emph{k}})$ and $\tilde F(\textbf{\emph{k}})$
contribute diagonally to the effective Hamiltonian in the sublattice space. So we end up with
\begin{equation*}
\begin{aligned}
&H=\sum_{k}\begin{pmatrix} a_{k}^{\dag} & b_{k}^{\dag} \end{pmatrix}
\begin{pmatrix} \tilde{F}(\textbf{\emph{k}})+\bar{F}(\textbf{\emph{k}}) & F_{0}(\textbf{\emph{k}})+F_{xp}(\textbf{\emph{k}}) \\ F_{0}^{*}(\textbf{\emph{k}})+F_{xp}^{*}(\textbf{\emph{k}}) & \tilde{F}(\textbf{\emph{k}})+\bar{F}(\textbf{\emph{k}})  \end{pmatrix}
\begin{pmatrix} a_{k} \\ b_{k} \end{pmatrix} \\
  \end{aligned}
\end{equation*}
where $F_0(\textbf{\emph{k}})$ provides a upright Dirac cone to begin with (as discussed in the previous sub-section).
\begin{equation*}
\begin{aligned}
&F_{0}(K_{a}^{\pm}+\textbf{\emph{k}})=-\frac{3ati}{2}(k_x\mp ik_y)
 \end{aligned}
\end{equation*}
Using the third nearest-neighbor vectors $\delta_{p}=-\frac{2}{3}(\vec{a_{1}}+\vec{a_{2}})=-2a\hat{i}$, $\delta_{x}^{1}=a\hat{i}+\sqrt{3}a\hat{j}$, $\delta_{x}^{2}=a\hat{i}-\sqrt{3}a\hat{j}$ where "$a$" is an "effective" lattice constant for the parent honeycomb lattice, the $F_{xp}(\textbf{\emph{k}})$ form factor and its Taylor expansion
near the Dirac crossing become,
\begin{equation*}
\begin{aligned}
&F_{xp}(\textbf{\emph{k}})=t^{p}e^{2iak_{x}}+2t^{x}e^{-iak_{x}}\cos(\sqrt{3}ak_{y})\\
&F_{xp}(K_{a}^{\pm}+\textbf{\emph{k}})=t^{p}-t^{x}\pm3at^{x}k_{y}+ia(2t^{p}+t^{x})k_{x}
\end{aligned}
\end{equation*}
As can be seen the difference $t^p-t^x$ between the 3rd neighbor hoppings generated via two different microscopic paths effectively shifts the
$k_y$ of the opposite valleys by opposite values. Furthermore, the coefficients of the $k_x$ and $k_y$ above that arise from $t^x$ and $t^p$
contribute to the re-definition and anisotropy of the Fermi velocity.
Remembering to affect the gauge transformation $\phi_B\to -i\phi_B$ of the previous sub-section, the off-diagonal form factor
giving still an upright (but shifted) Dirac cone becomes,
\begin{equation}
f_{\rm Dirac}(K_{a}^{\pm}+\textbf{\emph{k}})\to a(\frac{3t}{2}+2t^{p}+t^{x})k_{x}-i(\mp\frac{3at}{2}\pm 3at^{x})
(k_y+\frac{t^{p}-t^{x}}{\mp\frac{3at}{2}\pm 3at^{x}})
\label{shifted.eqn}
\end{equation}
This form makes it manifest how the third neighbor hoppings shift the Dirac cones and make their velocities anisotropic.
The difference $t^p-t^x$ is responsible for the mutual movement of the Dirac cones alone the $k_y$ direction:
The Dirac point at $K_{a}^{+}$ move downward by $(t^{p}-t^{x})/(3at^{x}-\frac{3at}{2})$  and Dirac point at $K_{a}^{-}$
move upward by the same value. An illustration of this shift is indicated in Fig.~\ref{fig-sub-3}(d) by red arrows.
So, by increasing difference $(t^{p}-t^{x})/(3at^{x}-\frac{3at}{2})$, the Dirac-nodes at $K_{a}^{\pm}$  move toward $\Gamma$. Equivalently, this means that the
Dirac-nodes at $K_{b}^{\pm}$  move away from $\Gamma$ as indicated by red arrows.


{\bf Formation of the tilt:} We are now ready to discuss the green path hoppings in Fig.~\ref{fig-sub-3}(c)
that are second neighbor hoppings on the effective honeycomb structure.
The form factors $\tilde{F}(\textbf{\emph{k}})$ and $\bar{F}(\textbf{\emph{k}})$ correspond to the real-space displacements
\begin{equation*}
\begin{aligned}
&\tilde{\delta}_{1}=\vec{a}_1-\vec{a}_2=\sqrt{3}a\hat{j}, \tilde{\delta}_{1}= (0,\sqrt{3}a) \\
&\tilde{\delta}_{2}=\vec{a}_2-\vec{a}_1=-\sqrt{3}a\hat{j}, \tilde{\delta}_{2}= (0,-\sqrt{3}a)
  \end{aligned}
\end{equation*}
and,
\begin{equation*}
\begin{aligned}
&\bar{\delta}_{1}=\vec{a}_1= \frac{3}{2}a\hat{i}+\frac{\sqrt{3}}{2}a\hat{j}\\
&\bar{\delta}_{2}=-\vec{a}_1 \\
&\bar{\delta}_{3}=\vec{a}_2 = \frac{3}{2}a\hat{i}-\frac{\sqrt{3}}{2}a\hat{j}\\
&\bar{\delta}_{4}=-\vec{a}_2,
 \end{aligned}
\end{equation*}
respectively. The $\tilde{F}(\textbf{\emph{k}})$, $\bar{F}(\textbf{\emph{k}})$, in the diagonal part of Hamiltonian are thus
\begin{equation*}
\begin{aligned}
&\tilde{F}(\textbf{\emph{k}})=\tilde{t}(e^{i\sqrt{3}ak_{y}}+ e^{-i\sqrt{3}ak_{y}})=2\tilde{t}\cos(\sqrt{3}ak_{y})\\
&\bar{F}(\textbf{\emph{k}})=2\bar{t}(\cos(\textbf{\emph{k}}.\vec{a_{1}})+\cos(\textbf{\emph{k}}.\vec{a_{2}}))
=4\bar{t}\cos(\frac{3}{2}ak_{x}) \cos(\frac{\sqrt{3}}{2}ak_{y})
 \end{aligned}
\end{equation*}
that add up to give a tilt form factor, $f_{\rm tilt}$
\begin{equation*}
f_{\rm tilt}(\textbf{\emph{k}})= \tilde{F}(\textbf{\emph{k}})+\bar{F}(\textbf{\emph{k}})=2\tilde{t}\cos(\sqrt{3}ak_{y})+4\bar{t}\cos(\frac{3}{2}ak_{x}) \cos(\frac{\sqrt{3}}{2}ak_{y})
\end{equation*}
Expanding around the Dirac point $K_{a}^{\pm}$, the total AA matrix element becomes,
\begin{equation}
f_{\rm tilt}(K_{a}^{\pm}+\textbf{\emph{k}})=-(\tilde{t}+2\bar{t})\pm 3a(\tilde{t}-\bar{t})k_{y}
  \label{equ2}
\end{equation}
Note that the BB matrix element is the same as above. Remember that the gauge transformation $\phi_B\to -i\phi_B$ does not change
the BB matrix element as a factor $(-i)^*(-i)$ gives $1$.
From the above formula we read:
\begin{equation}
   \zeta_x=0,~~\zeta_y=\pm 2\frac{\tilde t-\bar t}{t}.
   \label{tilt.eqn}
\end{equation}

Eqs.~\eqref{shifted.eqn} and~\eqref{tilt.eqn} derived for our model system are great guiding principles.
Eq.~\eqref{shifted.eqn} shows that the 3rd neighbor (red) hoppings are responsible for the location and anisotropy of the upright Dirac cone.
Eq.~\eqref{tilt.eqn} simply means that the tilt (a long distance property) depends on the difference
between the microscopic parameters $\tilde t-\bar t$, namely the difference in the hopping via two different
green paths in Fig.~\ref{fig-sub-3}(c). This model shows how the presence of ridge atoms move and tilt the Dirac cone.
In the following section we are going to show how the effective hoppings in panel (c) of Fig.~\ref{fig-sub-3}(c) arise from microscopic hoppings in panel (a).

\section{Effective hopping terms: renormalization via molecular orbitals}
\label{rg.sec}
If one considers the honeycomb-like structure formed by inner sites only, the hopping between 2nd and 3rd neighbors
are quite negligible. This is because in a tight-binding approach, the overlap between the atomic orbitals
exponentially decays with the distance. But when the ridge sites (denoted by teal color in Fig.~\ref{fig-sub-3}(a)) are added,
they act as virtual sites thorough which a renormalized hopping between the ($p_z$) atomic orbitals of the inner sites
is formed. In this section we provide a simple picture of renormalization based on the molecular orbitals theory
that explains why relatively large hoppings between 2nd and 3rd neighbors are obtained.
The remarkable agreement of the tilt parameters extracted within such a local quantum chemistry picture indicates its validity.
So in this section using molecular orbital theory, we first analytically derive these renormalized hopping parameters of
new effective honeycomb lattice in term of hopping parameters of original $8Pmmn$ lattice. This will show that
the ridge atoms (B$_R$) play a crucial role in mediating strong hoppings between the 2nd and 3rd inner sites atoms.
Then we discuss the connection between the present molecular orbital treatment with a picture of renormalization procedure
by considering Dyson equation for the local quantum clusters of the $8Pmmn$ lattice.

\begin{figure} [H]
  \centering
  \includegraphics[width=0.82\textwidth]{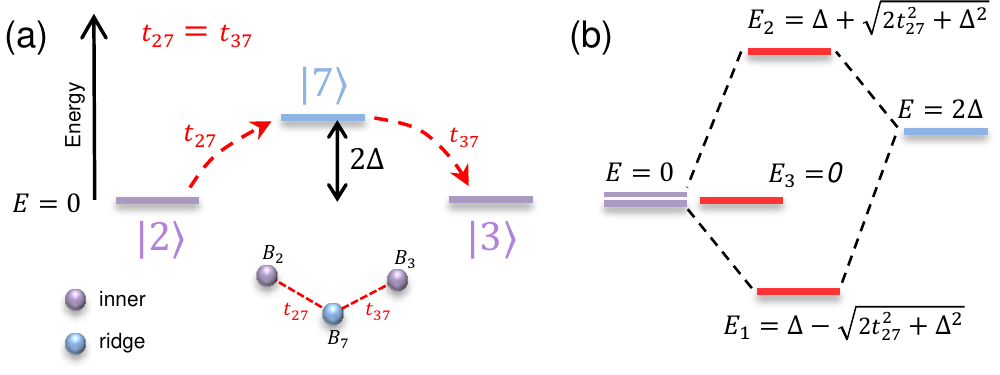}
  \caption{(a) Energy diagram for inner (B$_2$, B$_3$) and ridge (B$_7$) atoms before the formation of molecular states.
  Ridge atoms have an energy offset $2\Delta$ with respect to inner atoms.
  (b) The atomic couplings $t_{27}$ and $t_{37}$ mix the even parity combination of $|2\rangle$ and $|3\rangle$ with the orbital
  $|7\rangle$ of the ridge atoms. The resulting molecular orbitals are denoted by red. The odd-parity combination of $|2\rangle$
  and $|3\rangle$ is decoupled from the rest, and remains at energy $E_3=0$, while a molecular binding orbital constructed from the
  even combination of $|2\rangle$ and $|3\rangle$ and $|7\rangle$ is stabilized by energy $E_1$ with respect to the initial
  level $0$ of the inner atoms. When the ridge site $|7\rangle$ is eliminated, this lowering becomes an effective hopping between the
  atomic orbitals $|2\rangle$ and $|3\rangle$ of the third neighbor sites B$_2$ and B$_3$ of the parent honeycomb lattice. }
\label{fig-sub-4}
\end{figure}

{\bf Formation of effective third neighbor hopping:}
Let us illustrate the basic principle by considering inner sites B$_2$ and B$_3$ in Fig.~\ref{fig-sub-3}(a) and see how the ridge site
B$7$ facilitates a hopping between them. In Fig.~\ref{fig-sub-4}(a) we have depicted an energy diagram where sites B$_2$ and B$_3$ (by symmetry)
are at the same atomic energy levels $|2\rangle$ and $|3\rangle$, respectively. The site B$_7$ is assumed at state $|7\rangle$ energetically lying $2\Delta$
above the inner sites. The atomic hopping $t_{27}$ and $t_{37}$ connecting sites B$_2$ and B$_3$ to site B$_7$ that can be derived from Wannier
states are indicated. These two (real) hoppings are the same by symmetry. The basic principle can be seen in the limit where $t_{27}=t_{37}\ll 2\Delta$.
In this limit a virtual second order in $\Delta$ hopping via higher energy atomic orbital $|7\rangle$ stabilizes the energy of the states
$|2\rangle$ and $|3\rangle$ by $ -2t_{27}^2/\Delta$. The factor $2$ comes from two different ways of performing such a virtual process:
$|2\rangle\to |7\rangle\to |3\rangle$ and $|3\rangle\to |7\rangle\to |2\rangle$.
The analytical extension of this result to arbitrary $2\Delta$ is straightforward.

In Fig.~\ref{fig-sub-4}(b) we have denoted the energy level $E=0$ and $E=2\Delta$ of the inner and atomic sites before the hybridization $t_{27}$
by red and teal colors. The sites B$_2$ and B$_3$ are equivalent and hence symmetry adopted basis based on the even and odd
representation of a two-element group including identity operation and operator that exchanges B$_2$ and B$_3$ is given by,
\begin{equation*}
\begin{aligned}
&|\Phi_{1}\rangle=\frac{|2\rangle + |3\rangle }{\sqrt{2}} \\
&|\Phi_{2}\rangle=|7\rangle \\
&|\Phi_{3}\rangle=\frac{|2\rangle - |3\rangle }{\sqrt{2}} \\
 \end{aligned}
\end{equation*}
The states $|\Phi_1\rangle$ and $\Phi_2\rangle$ are even parity and mix with each other by atomic hopping $t_{27}=t_{37}$, while
the state $|\Phi_3\rangle$ being odd parity decouples from the others. In fact state $|\Phi_{1(3)}\rangle$ is a bonding (anti-bonding) molecular orbital
and is composed of atomic orbitals of inner sites B$_2$ and B$_3$.
The matrix elements and Hamiltonian in this basis are
\begin{equation*}
\begin{aligned}
&\langle\Phi_{2}|H|\Phi_{3}\rangle=\langle 7|H|\left( \frac{|2\rangle - |3\rangle }{\sqrt{2}} \right)=0, ~~~ \langle\Phi_{1}|H|\Phi_{3}\rangle=0, \\
&\langle\Phi_{2}|H|\Phi_{1}\rangle=\langle 7|H|\left( \frac{|2\rangle + |3\rangle }{\sqrt{2}} \right)=\sqrt{2}t_{27}, \\
 \end{aligned}
\end{equation*}
that give,
\begin{equation*}
\begin{aligned}
\begin{bmatrix}
    0 & \sqrt{2}t_{27} & 0   \\
    \sqrt{2}t_{27} & 2\Delta & 0  \\
    0 & 0 & 0 \\
\end{bmatrix}
  \end{aligned}
\end{equation*}
As can be seen when $t_{27}=0$, the bonding/anti-bonding states have zero energy (equal to the energy of atomic orbitals $|2\rangle$ and $|3\rangle$)
and hence bonding/anti-bonding degrees of freedom are inert in this limit. Upon turning on the atomic hopping to ridge site B$_7$, the above structure
of the Hamiltonian, still leaves the anti-bonding orbital $|\Phi_3\rangle$ decoupled and hence it remains at zero energy,
while the bonding (even parity) combination $|\Phi_1\rangle$ is mixed
with $|\Phi_2\rangle=|7\rangle$ giving rise to two split-off states at energies $\Delta\pm\sqrt{2t_{27}^2+\Delta^2}$. The energy of the
$E_1=\Delta-\sqrt{2t_{27}^2+\Delta^2}$ state is always less than energy $E=0$ of the original B$_2$ and B$_3$ sites.
In the low-energy sub-space that -- involving only the inner sites B$_2$ and B$_3$ -- the high-energy site B$_7$ is eliminated (or integrated out in the
renormalization group terminology), this lowering of energy can be interpreted as an effective hopping between B$_2$ and B$_3$, namely,
\begin{equation}
   t^{\rm eff}_{23}=\Delta-\sqrt{2t^2_{27}+\Delta^2}.
   \label{teff23.eqn}
\end{equation}
This is so, because in a sub-space composed of only inner orbitals $|2\rangle$ and $|3\rangle$, placing such a off-diagonal hopping
between the two states at energy zero, correctly reproduces the energy lowering of $E_1$ with respect to $E=0$.
This effective hopping in the main text has been denoted by $t^p$ and is responsible for a pseudo-gauge field that
shifts the location of the Dirac node.
Note that the in the $\Delta\gg t_{27}$ limit, the above espression reduces to the perturbative result $-2t^2_{27}/\Delta$ cited above, while
in the opposite limit $\Delta\ll t_{27}$ it becomes $-\sqrt 2 |t_{27}|$.

\begin{figure} [H]
  \centering
  \includegraphics[width=0.82\textwidth]{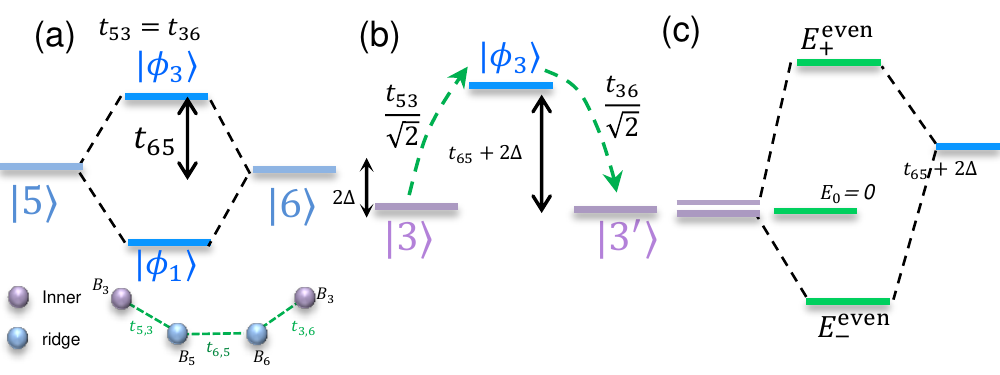}
  \caption{(a) Energy diagram for bonding and anti-bonding orbitals arise from the linear combination of atomic orbitals $|5\rangle$ and $|6\rangle$. (b) Energy levels of two successive inner (B$_3$) and anti-bonding orbitals ($|\Phi_{3}\rangle$) stem from ridge B$_5$ and B$_6$ atoms (c) Energy diagram after the formation of molecular states which connect B$_3$ to one of its successive third nearest neighbors through antibonding orbitals ($|\Phi_{3}\rangle$).}
\label{fig-sub-5}
\end{figure}

{\bf Formation of effective second neighbor hopping:} As a second example of how longer range hoppings on the parent
honeycomb-like lattice of inner atoms are formed, let us consider in Fig.~\ref{fig-sub-1}(a) the effective hopping between site B$_3$ in one unit cell and
the same B$_3$ in the lower unit cell (let's call it $3'$). This process will be achieved by two virtual paths: $3'\to 5\to6\to 3$ and $3'\to8\to7\to 3$. Let us consider the first
virtual path indicated by dotted green path in Fig.~\ref{fig-sub-3}(a).
Denoting by $|3\rangle$ and $|3'\rangle$ the atomic orbitals on the two B$_3$ sites, and by $|5\rangle$ and $|6\rangle$ the atomic orbitals on the
intermediate B$_5$ and B$_6$ sites, respectively, the four dimensional Hilbert space is broken into even sector
\begin{eqnarray}
   |\phi_1\rangle=\frac{|5\rangle+|6\rangle}{\sqrt 2},~~~
   |\phi_2\rangle=\frac{|3\rangle+|3'\rangle}{\sqrt 2},\nn
\end{eqnarray}
and odd sector
\begin{eqnarray}
   |\phi_3\rangle=\frac{|5\rangle-|6\rangle}{\sqrt 2},~~~
   |\phi_4\rangle=\frac{|3\rangle-|3'\rangle}{\sqrt 2}.\nn
\end{eqnarray}
The hopping matrix elements between the atomic orbitals are shown in Fig.~\ref{fig-sub-5} and by symmetry $t_{53}=t_{63}$.
The hopping between B$_5$ and B$_6$ is associated with a $\pi$-bonding between the $p_x$ orbitals of these two sites.
Taking  the energy offset $2\Delta$ for the ridge sites with respect to the inner sites into account, the even block of the Hamiltonian
and its eigenvalues become,
\begin{equation*}
   \begin{bmatrix}
   t_{56}+2\Delta & t_{63}\\
   t_{63}          & 0
   \end{bmatrix},~~~~~
   E^{\rm even}_\pm=\left(\frac{t_{56}}{2}+\Delta\right)\pm\sqrt{\left(\frac{t_{56}}{2}+\Delta\right)^2+t_{63}^2}
\end{equation*}
while for the odd sector we have
\begin{equation*}
   \begin{bmatrix}
   2\Delta & t_{63}\\
   t_{63}          & 0
   \end{bmatrix},~~~~~
   E^{\rm odd}_\pm=\Delta\pm\sqrt{\Delta^2+t_{63}^2}
\end{equation*}
In fact replacement $\Delta\to\Delta+t_{56}/2$ maps the odd sector to even sector. This can be interpreted as the fact that
the even sector take advantage of the $t_{56}$ hopping term between the $p_x$ atomic orbitals of the
ridge atoms and generate larger stabilization. Therefore for the $3'\to 5\to6\to 3$ path,  a contribution to
the effective second neighbor hopping between the two B$_3$ inner sites will become,
\begin{equation}
   t_{33}^{{\rm eff }(3'\to5\to6\to3)}=
   \left(\frac{t_{56}}{2}+\Delta\right)-\sqrt{\left(\frac{t_{56}}{2}+\Delta\right)^2+t_{63}^2}
   \label{t33eff1.eqn}
\end{equation}
A similar contribution arises from the path $3'\to8\to7\to3$ by simply replacing $(5,6)\to(8,7)$ that gives
\begin{equation}
   t_{33}^{{\rm eff }(3'\to8\to7\to3)}=
   \left(\frac{t_{87}}{2}+\Delta\right)-\sqrt{\left(\frac{t_{87}}{2}+\Delta\right)^2+t_{73}^2}
   \label{t33eff2.eqn}
\end{equation}
Adding these two terms gives the green $\tilde t$ in Fig.~2(c) of the main text,
\begin{equation}
  \tilde t=
   \left(\frac{t_{56}}{2}+\Delta\right)-\sqrt{\left(\frac{t_{56}}{2}+\Delta\right)^2+t_{63}^2}+
   \left(\frac{t_{87}}{2}+\Delta\right)-\sqrt{\left(\frac{t_{87}}{2}+\Delta\right)^2+t_{73}^2}
\end{equation}



It is rather surprising to note that the Eq.~\eqref{t33eff1.eqn} for the virtual hopping path $3'\to5\to6\to3$ can be obtained by a simple intuitive picture as follows:
As shown in Fig.~\ref{fig-sub-5}(a), first an antibonding orbital from the ridge sites B$_5$ and B$_6$ is formed that has been denoted by
$\phi_a\rangle$. The upward shift of the antibonding orbital with respect to the initial level $2\Delta$ of the ridge atoms is given by
the off-diagonal element $t_{65}$ in the space of $|5\rangle$ and $|6\rangle$. Then $|\phi_a\rangle=(|5\rangle-|6\rangle)/\sqrt 2$ will have
hopping matrix elements $t_{53}/\sqrt 2$ and $t_{36}/\sqrt 2$ to sites $3'$ and $3$, respectively. Then a virtual hopping with these amplitudes
via the intermediate state $|\phi_a\rangle$ at energy $t_{65}+2\Delta$, precisely gives Eq.~\eqref{t33eff1.eqn}.

{\bf Relation with renormalization:}
Now we are ready to show that the effective hoppings obtained in this way, admit a nice interpretation in terms of renormalization.
As an example, let us elaborate on Eq.~\eqref{teff23.eqn} and show that it can be alternatively obtained from a renormalizatioin picture.
This will establish that the molecular orbital theory that was explained in the previous part is indeed equivalent to a renormalization procedure.

Imagine the situation we discussed for effective third neighbor hopping energy between B$_2$ and B$_3$.
Denoting as before the atomic basis $|2\rangle$, $|3\rangle$, and $|7\rangle$, and breaking the three dimensional Hilbert space into a
two dimensional subspace composed of low-energy states $|2\rangle$, $|3\rangle$ and a one-dimensional high energy subspace described by $|7\rangle$,
the Hamiltonian $H$ can be split as,
\begin{table}[ht]
\centering
\label{my-label}
$ H =\left(
\begin{array}{ll|ccc}
    0 & 0 & t_{27}   \\
    0 & 0 & t_{27}  \\
\hline
    t_{27} & t_{27} & 2\Delta \\
\end{array}
\right)
=
\begin{pmatrix}
   H_{LL}  &  H_{LH}   \\
    H_{HL} & H_{HH}  \\
\end{pmatrix}=
\begin{pmatrix}
   H_{LL}  &  0   \\
    0 & H_{HH}  \\
\end{pmatrix}+
\begin{pmatrix}
   0  &  H_{LH}   \\
    H_{HL} & 0  \\
\end{pmatrix}
$
\end{table}
where the subscripts L (H) stand for low-energy (high-energy) sectors. As described in the textbook of Grosso and Pastori Paravicini \cite{Grosso},
the application of Dyson's equation leads to a renormalized (effective) Hamiltonian in the low-energy subspace of the following form,
\begin{equation*}
   H_{LL}^{\rm eff}(E)=H_{LL}+H_{LH}\frac{1}{E-H_{HH}}H_{HL} \\
\end{equation*}
that depends on energy. Plugging the various sub-matrices in the above equation gives,
\begin{equation*}
H_{LL}^{\rm eff}(E)=
\begin{bmatrix}
0  &  0   \\
0 & 0  \\
\end{bmatrix}+\begin{bmatrix}
t_{27}   \\
t_{27}  \\
\end{bmatrix}\frac{1}{E-2\Delta}\begin{bmatrix}
t_{27}  &  t_{27}   \\
\end{bmatrix}=\frac{1}{E-2\Delta}  \begin{bmatrix}
 t_{27}^{2} & t_{27}^{2}    \\
 t_{27}^{2} & t_{27}^{2}   \\
 \end{bmatrix}
\end{equation*}

Now we have to evaluate the operator, whose matrix elements are energy-dependent as,
\begin{equation*}
\begin{aligned}
   &\frac{1}{E-2\Delta}
   \begin{bmatrix}
   t_{27}^{2} & t_{27}^{2}    \\
   t_{27}^{2} & t_{27}^{2}   \\
   \end{bmatrix}\Psi=E\Psi.
\end{aligned}
\end{equation*}
The eigenvalues are given by,
\begin{equation*}
\begin{aligned}
\det
\begin{bmatrix}
\frac{t_{27}^{2}}{E-2\Delta}-E &\frac{t_{27}^{2}}{E-2\Delta}   \\
     \frac{t_{27}^{2}}{E-2\Delta}   & \frac{t_{27}^{2}}{E-2\Delta}-E   \\
\end{bmatrix} = 0,
  \end{aligned}
\end{equation*}
that immediately becomes,
\begin{equation*}
   E=\Delta\pm\sqrt{2t_{27}^{2}+\Delta^{2}}.
\end{equation*}
This is exactly the result~\eqref{teff23.eqn} of our simple molecular orbital theory.
The same logic applies to all other effective hoppings.
Therefore the process of virtual hopping via the inner sites
that involves molecular orbitals is actually equivalent to formation of renormalized hoppings in the low-energy sector of the theory (residing
on the inner sites) as a result of elimination of higher-energy degrees of freedom that
reside on ridge sites.






\section{{\em Ab initio} calculation: relaxation, electronic structure, and stability}
\begin{figure}[H]
\centering
 \subfloat[]{    \includegraphics[width=0.22\textwidth]{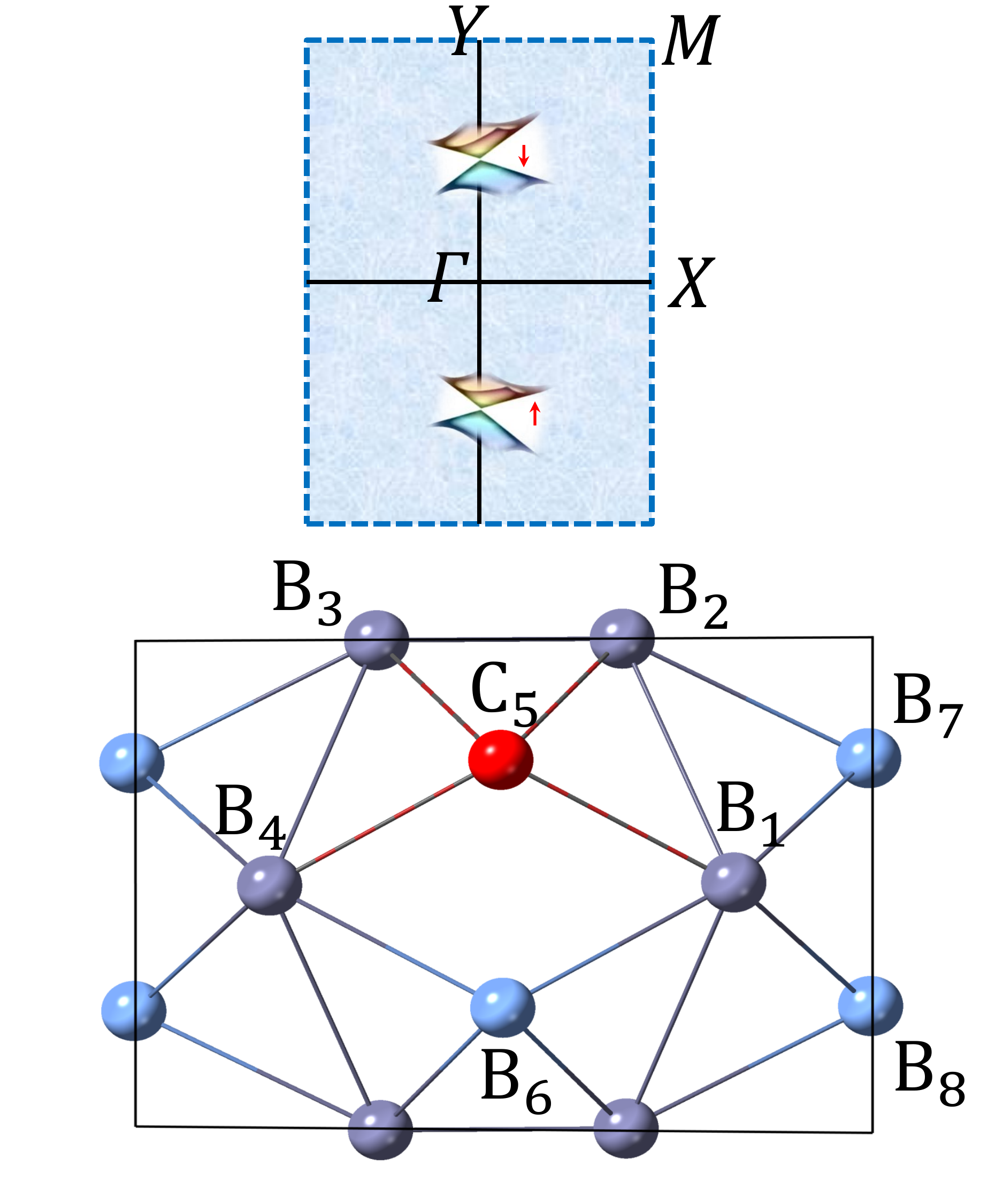}}
    \subfloat[]{ \includegraphics[width=0.22\textwidth]{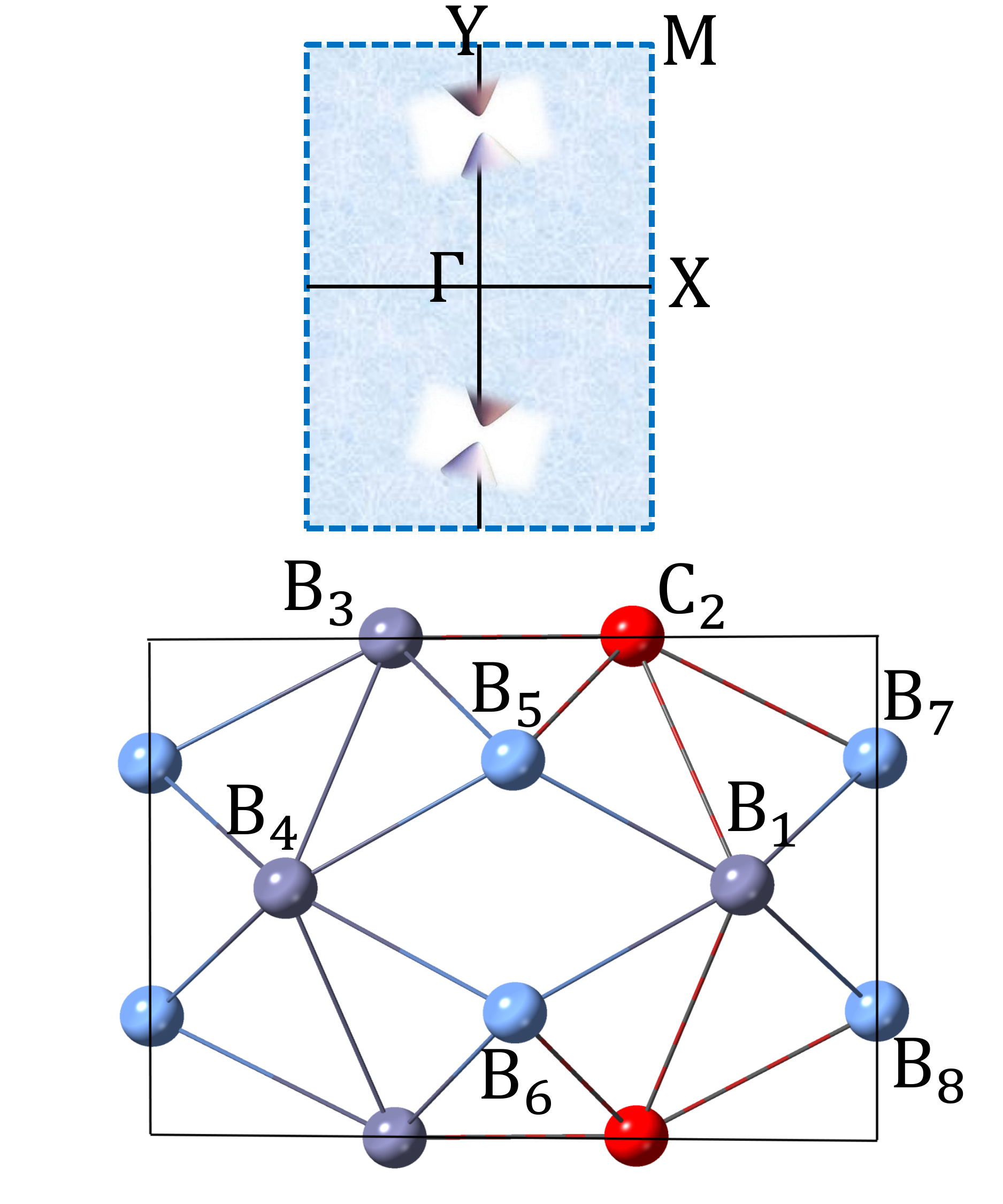}}
  \subfloat[]{   \includegraphics[width=0.22\textwidth]{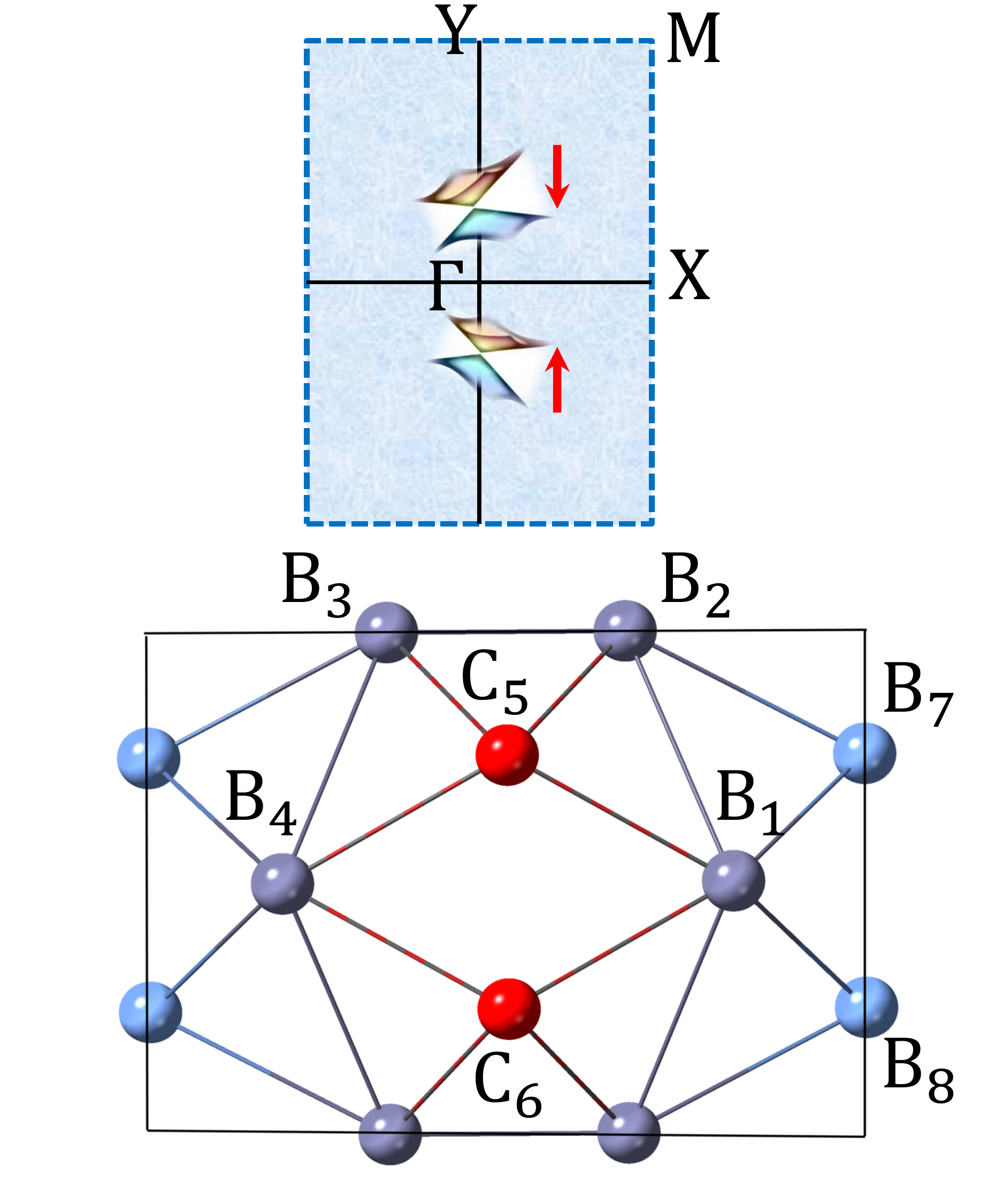} }
  \subfloat[]{   \includegraphics[width=0.22\textwidth]{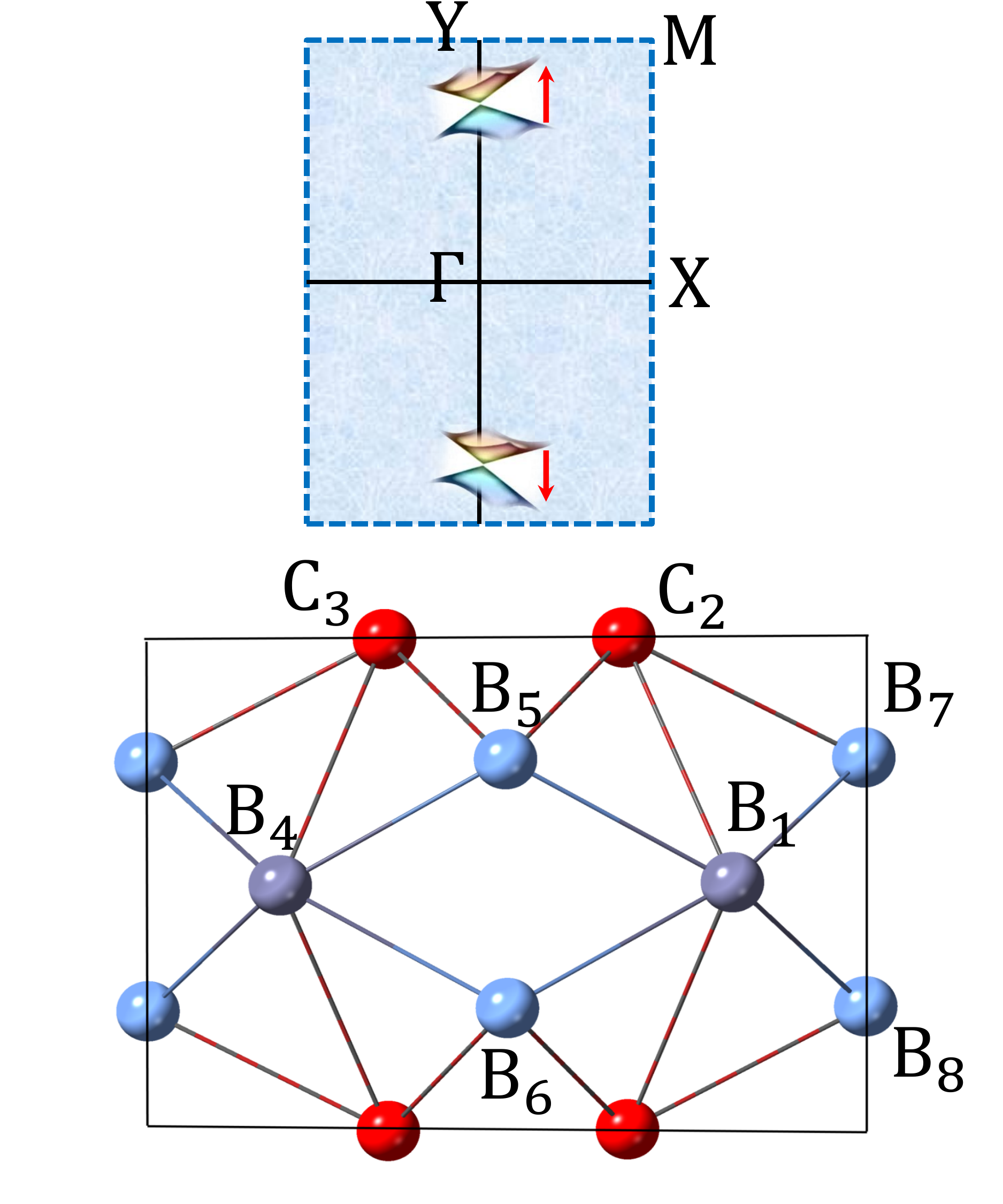} }\\
\subfloat[]{ \includegraphics[width=0.22\textwidth]{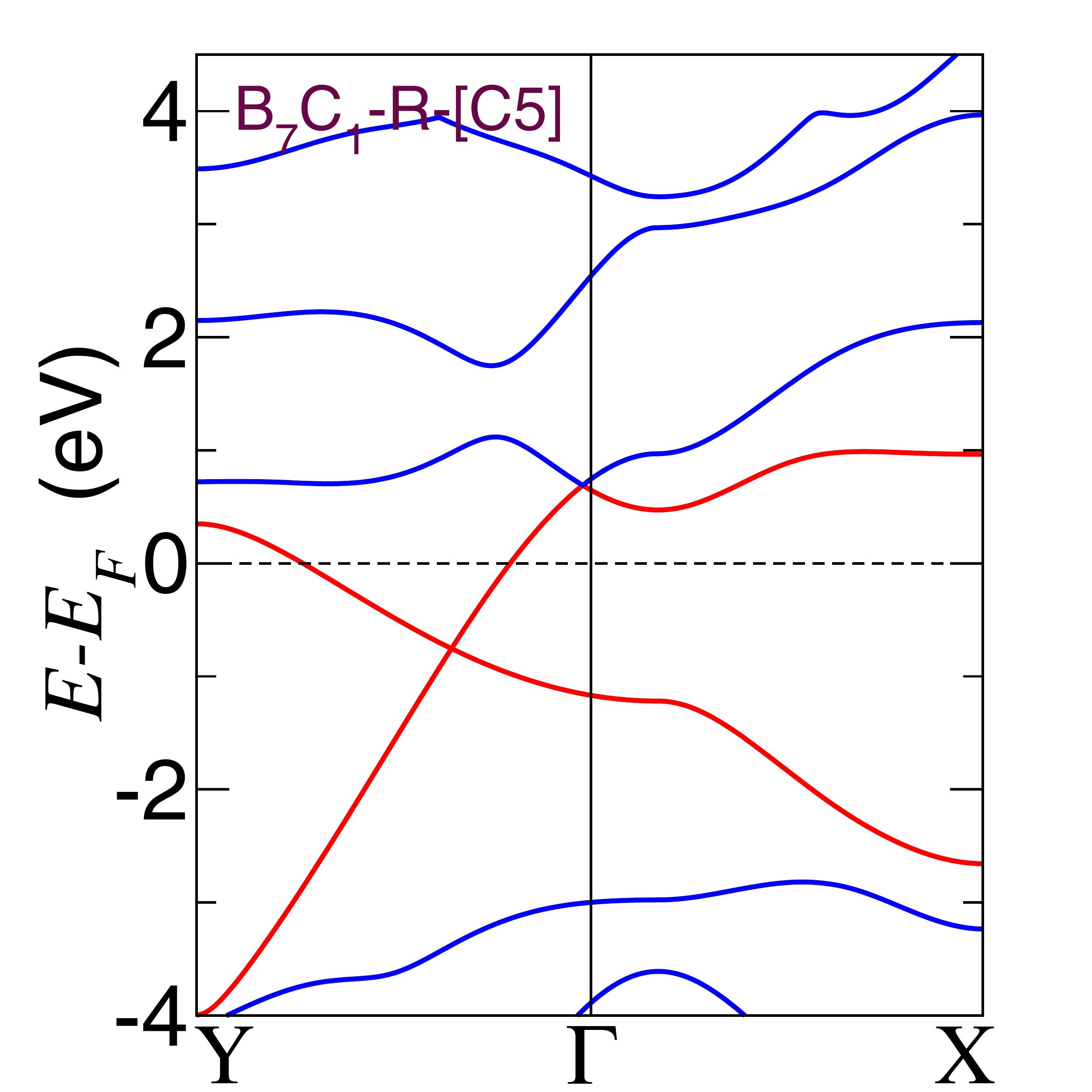}}
  \subfloat[]{    \includegraphics[width=0.22\textwidth]{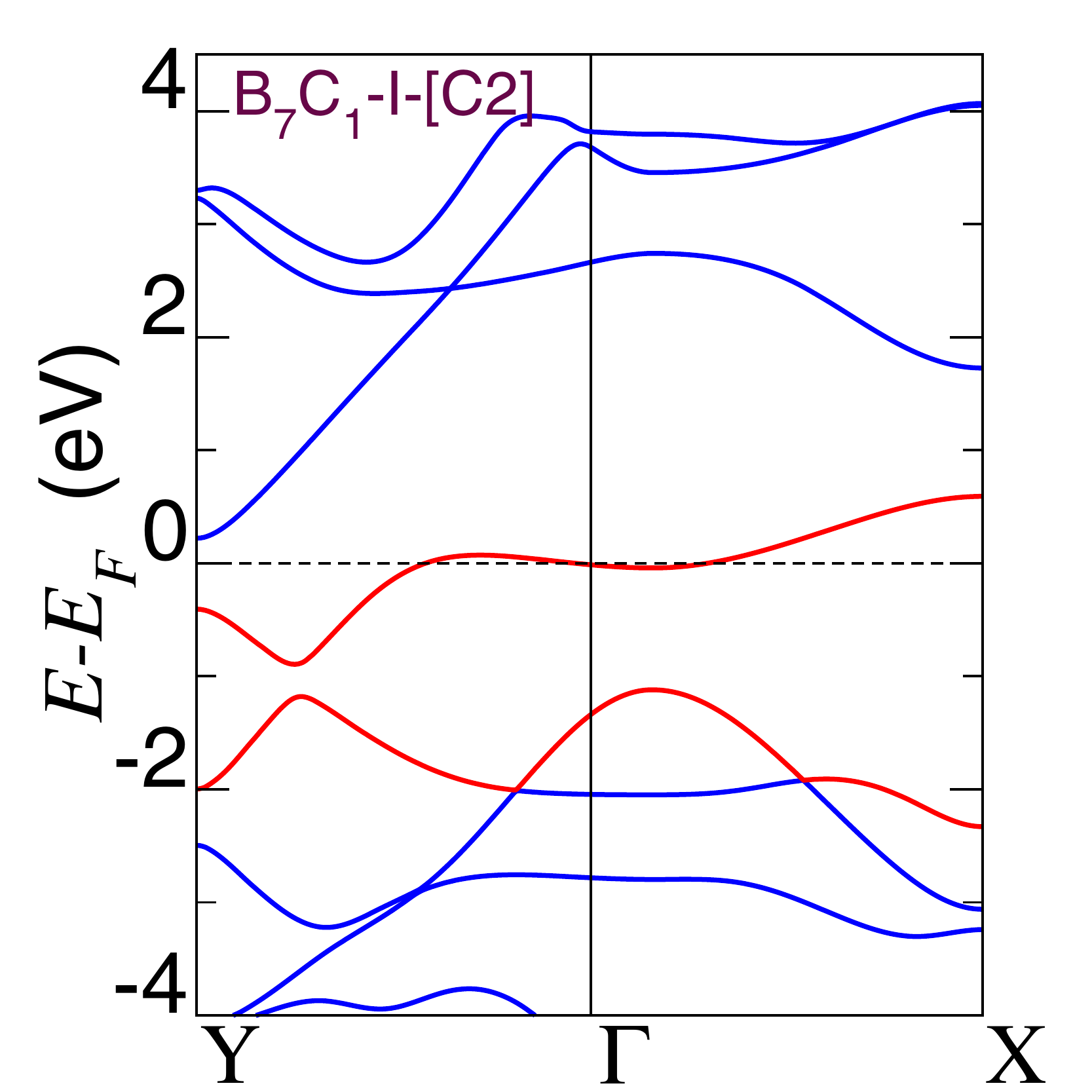}}
  \subfloat[]{   \includegraphics[width=0.22\textwidth]{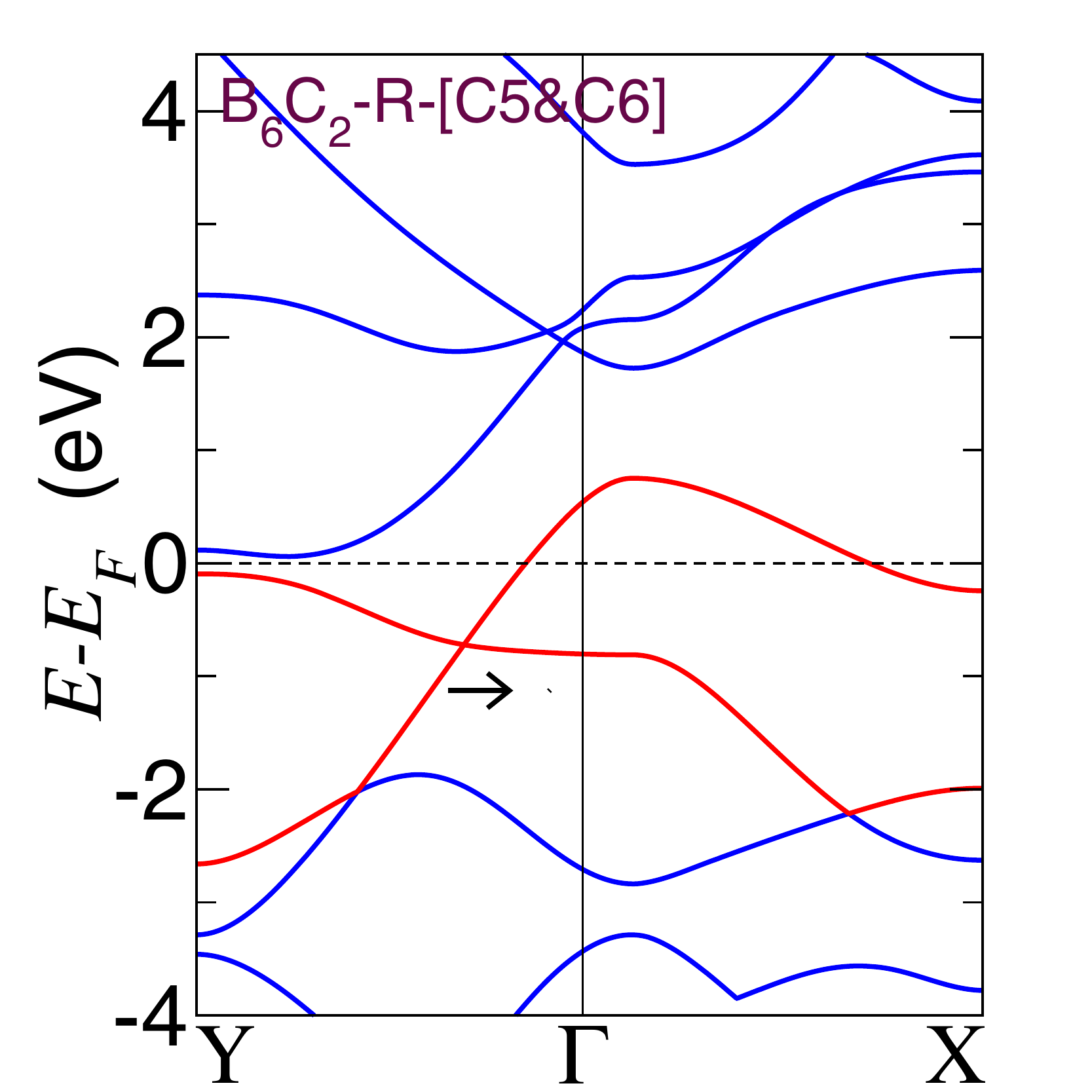} }
  \subfloat[]{   \includegraphics[width=0.22\textwidth]{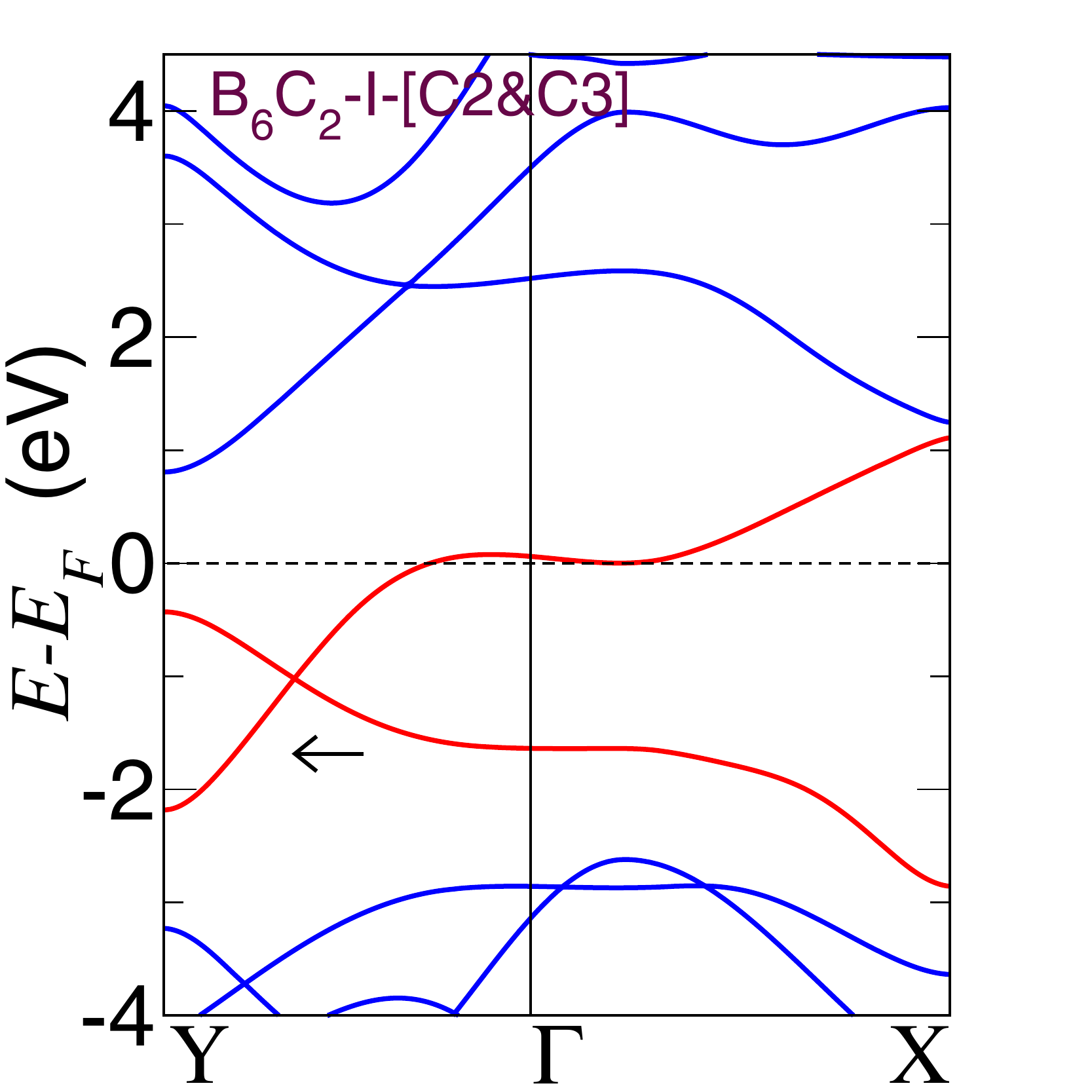} }\\
  \subfloat[]{    \includegraphics[width=0.21\textwidth]{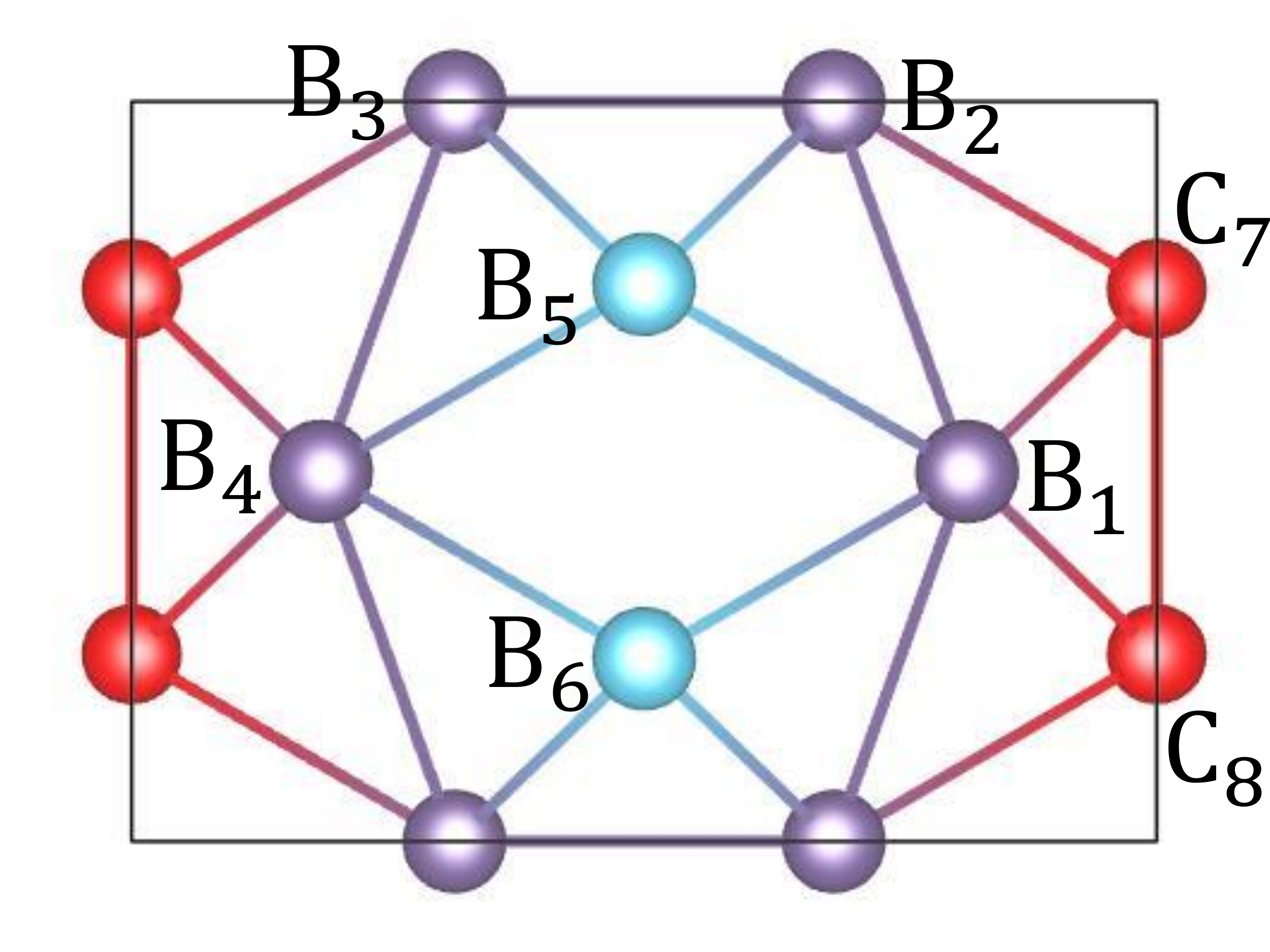}}
    \subfloat[]{ \includegraphics[width=0.21\textwidth]{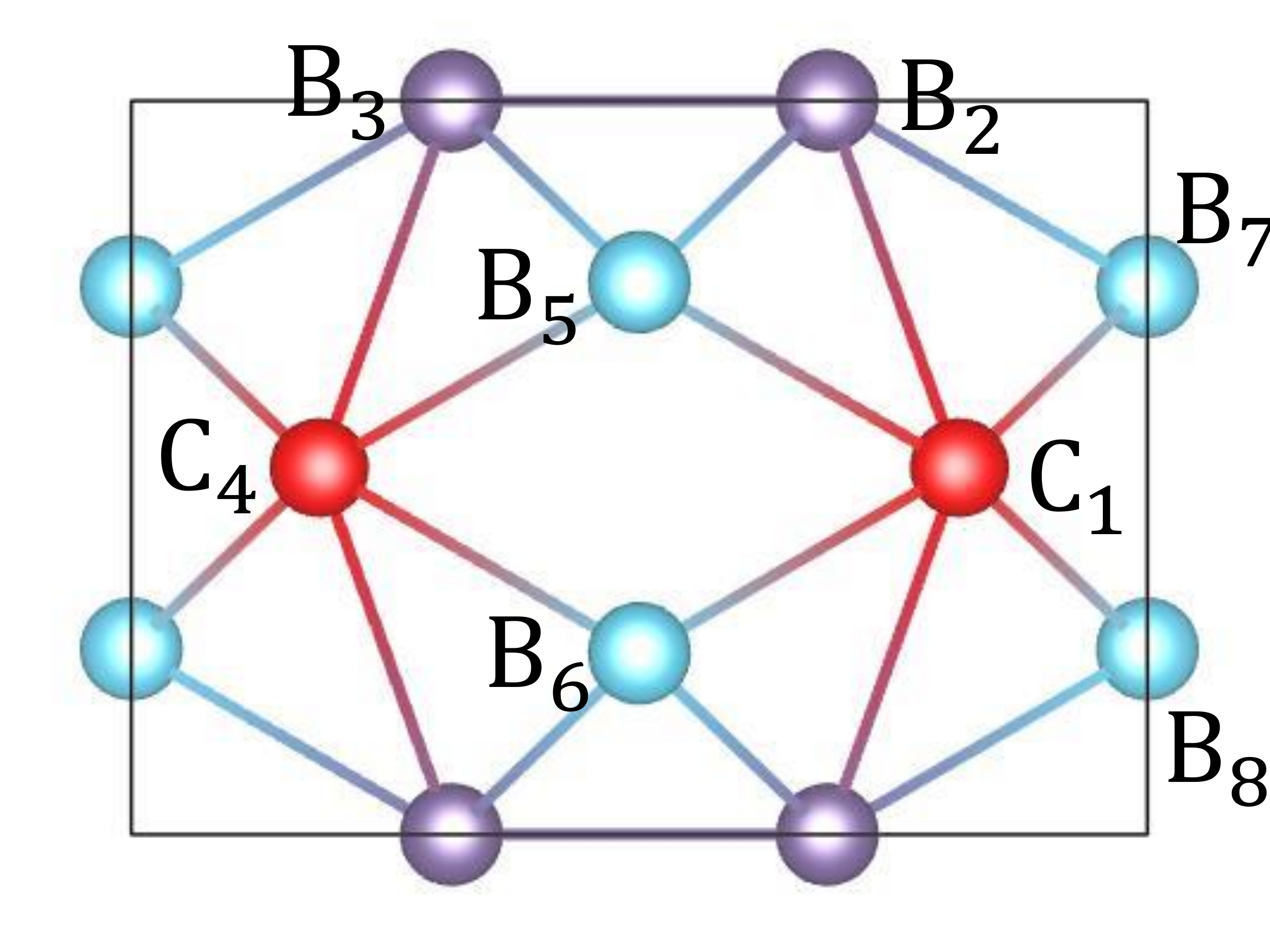}}
  \subfloat[]{   \includegraphics[width=0.21\textwidth]{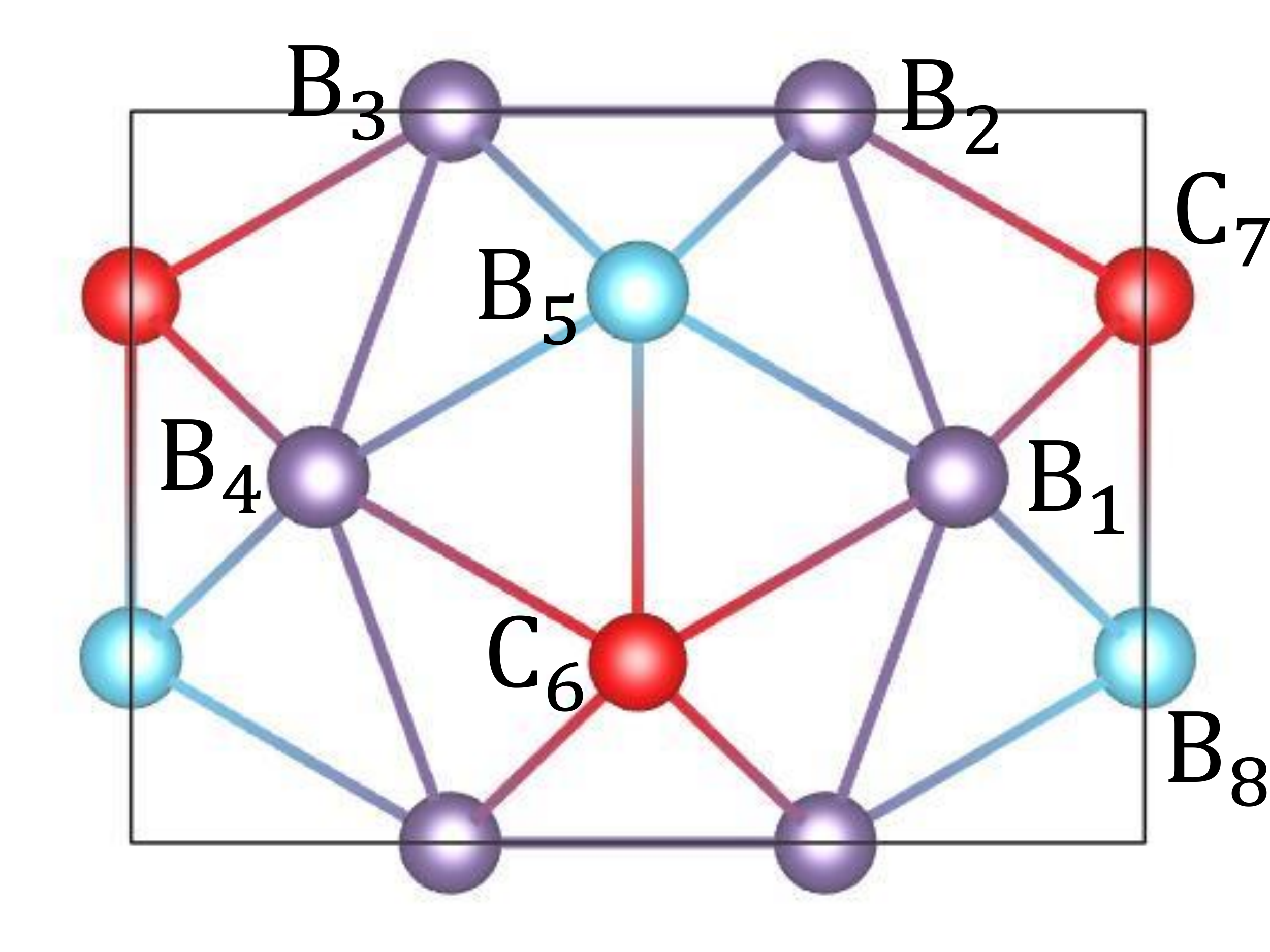} }
  \subfloat[]{   \includegraphics[width=0.21\textwidth]{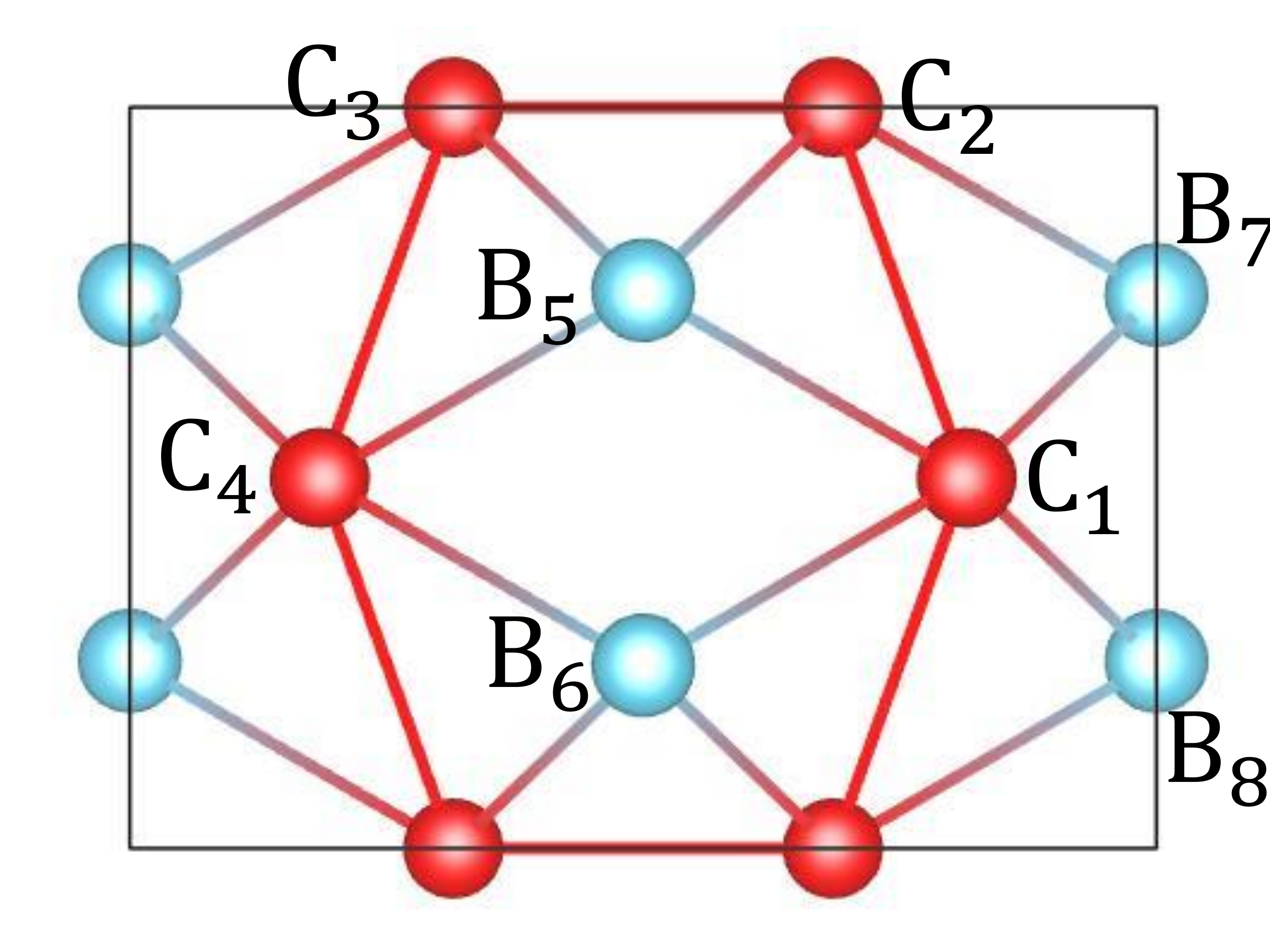} }\\
    \subfloat[]{    \includegraphics[width=0.22\textwidth]{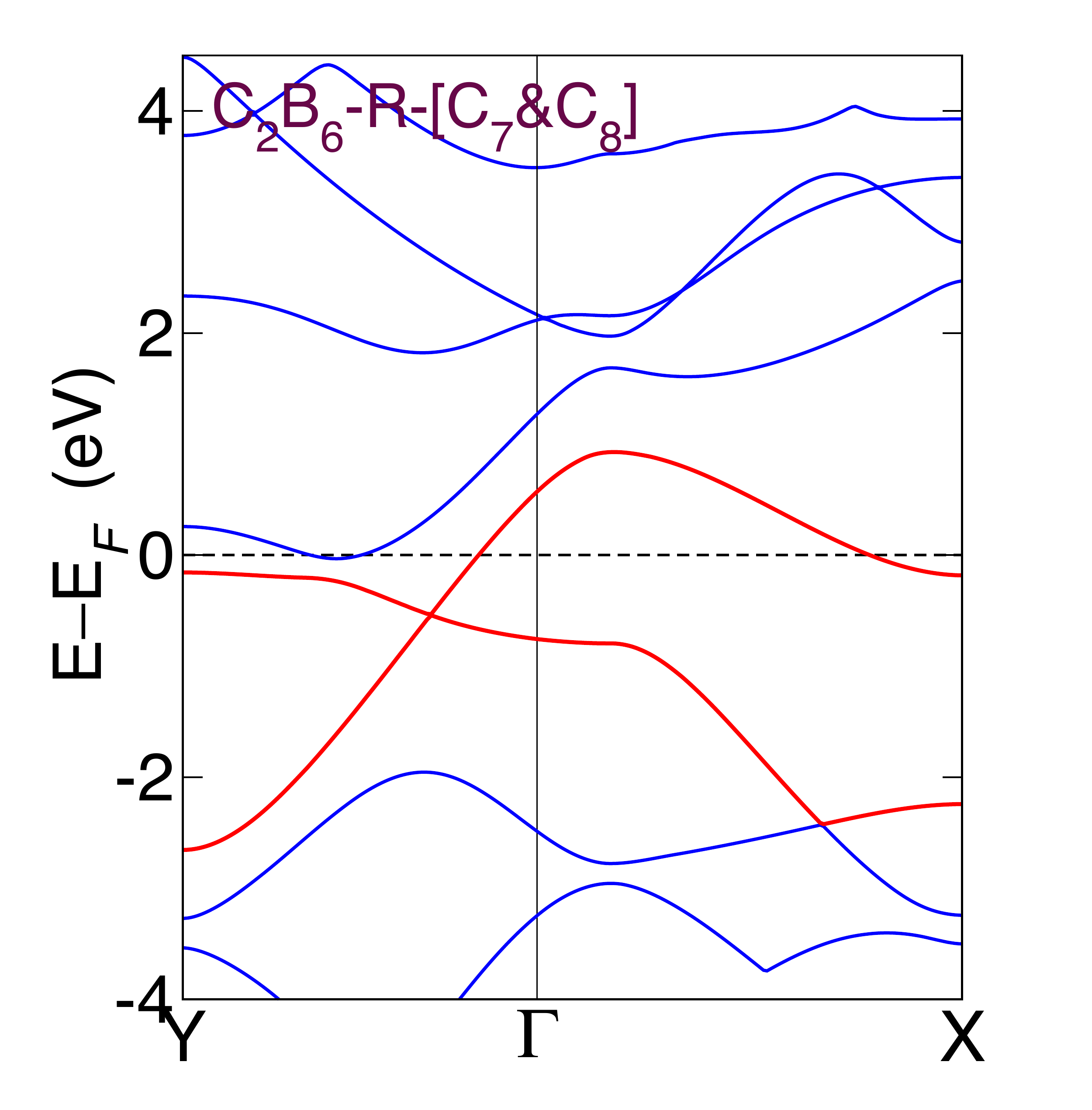}}
    \subfloat[]{ \includegraphics[width=0.22\textwidth]{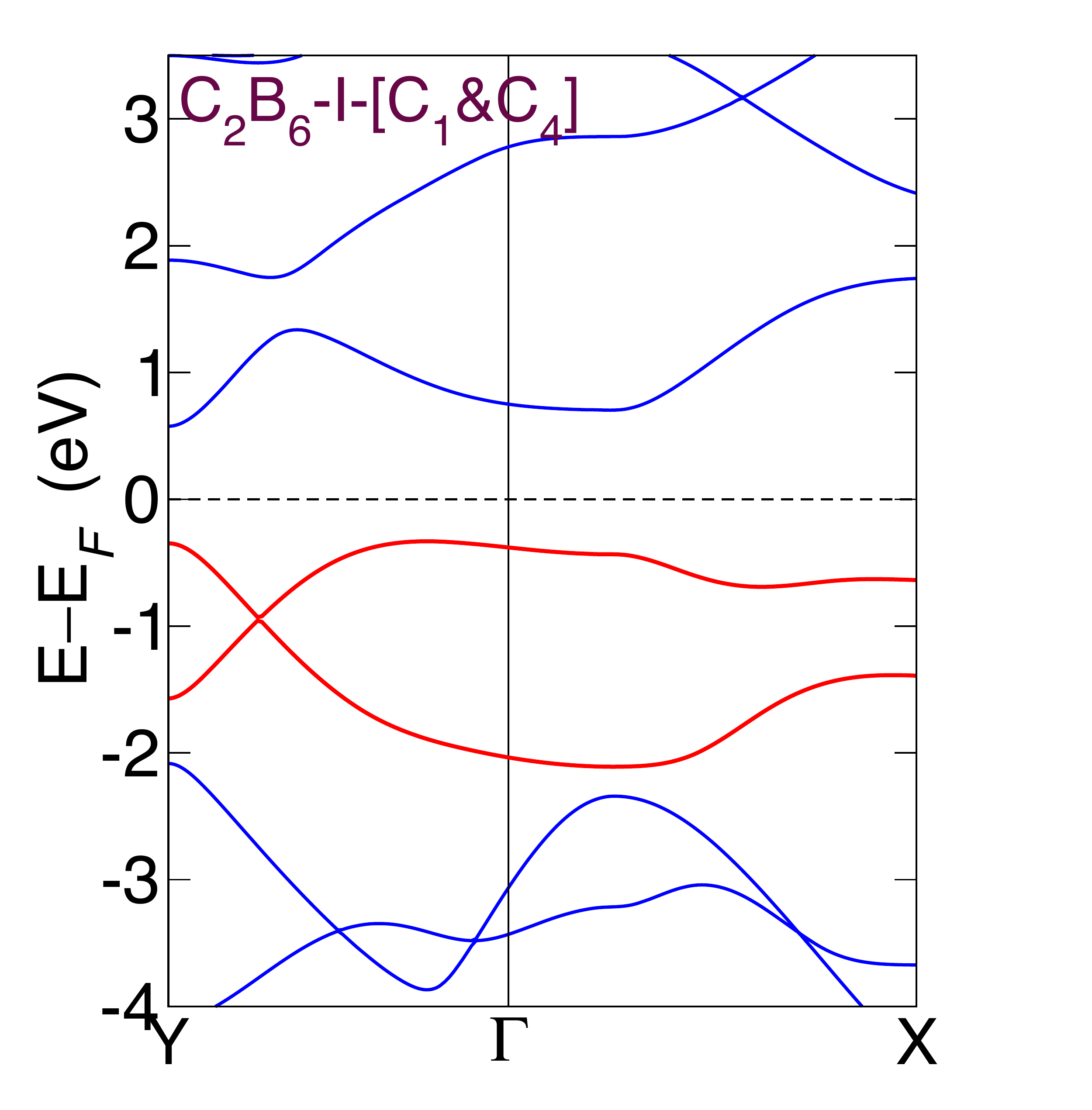}}
  \subfloat[]{   \includegraphics[width=0.22\textwidth]{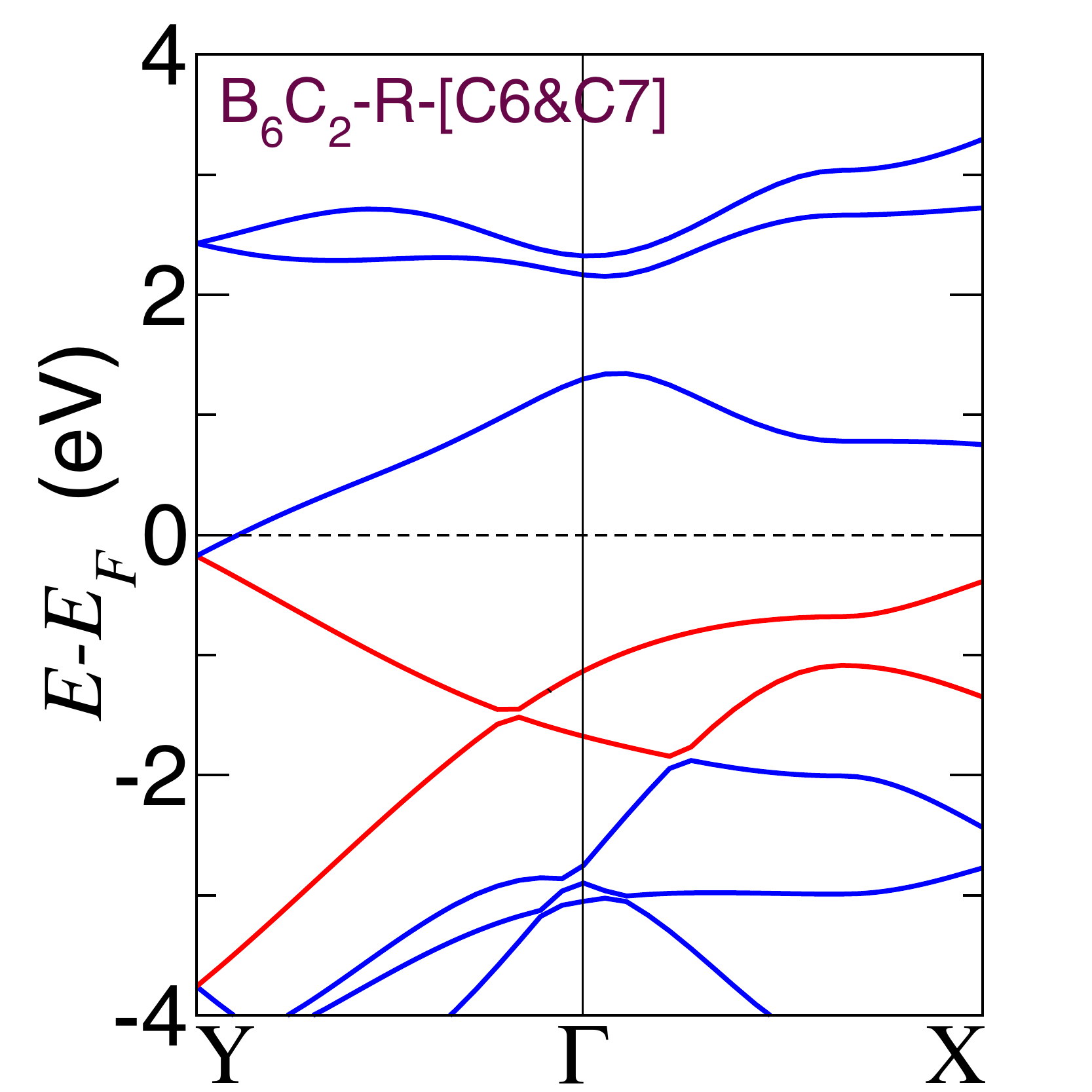} }
  \subfloat[]{   \includegraphics[width=0.22\textwidth]{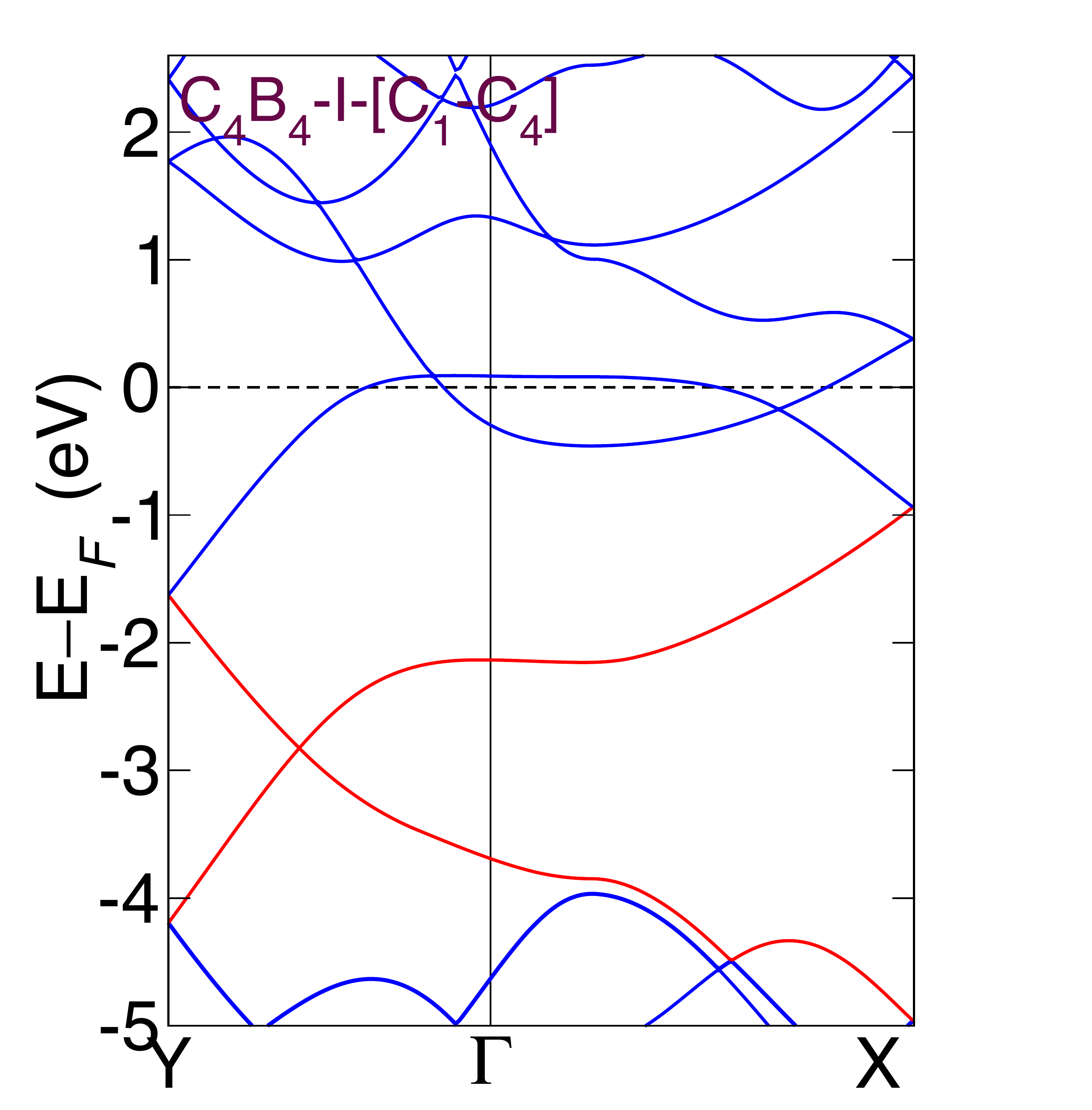} }
  \caption{Crystal structure and DFT-PBE band structure for system in which
  (a,e) a single B5 atoms in ridge site is replaced by C atom, B$_{7}$C$_{1}$-R-[C5],
  (b,f) a single B2 atom in inner site is replaced by C atom,  B$_{7}$C$_{1}$-I-[C2],
  (c,g) a dimer of B atoms (B5 and B6) in ridge sites is replaced by C atoms B$_{6}$C$_{2}$-R-[C5$\&$C6],
  (d,h) a dimer of B atoms (B2 and B3) in ridge sites is replaced by C atoms B$_{6}$C$_{2}$-I-[C2$\&$C3],
  (i,m) a dimer of B atoms (B7 and B8) in ridge sites is replaced by C atoms B$_{6}$C$_{2}$-R-[C7$\&$C8],
  (j,n) a dimer of B atoms (B1 and B4) in ridge sites is replaced by C atoms B$_{6}$C$_{2}$-I-[C1$\&$C4],
  (k,o) B$_6$ and B$_7$ atoms in ridge sites are replaced by C atoms B$_{6}$C$_{2}$-R-[C6$\&$C7], and (l,p) all of B atoms in inner sites are replaced by C atoms B$_{4}$C$_{4}$-I-[C1-C4], .
  }
\label{fig-sub-6}
\end{figure}

The crystal structure of pure $8Pmmn$ borophene is presented in Fig.~\ref{fig-sub-1}(a).
The basic unit cell is rectangular and contains eight B atoms.
For DFT calculation we use pseudopotential Quantum Espresso code~\cite{Espresso} based on plane wave basis set within the GGA in the Perdew-Burke-Ernzerhof (PBE) parameterization~\cite{Perdew}. Simulation of borophene rectangular unit cells is based on the slab model having a 25 ${\AA}$ vacuum separating slabs. We also consider monolayer of $8Pmmn$ structure, where some of the B atoms are substituted by C atoms in the form of B$_{8-x}$C$_{x}$ ($x$=0, 1, 2). 
The obtained structural properties after the ionic relaxations such as lattice parameters, $x$, $y$, and $z$ component of the B and C atoms in crystal coordinates are shown in Table \,\ref{table:5} for pure borophene (B$_8$) and C-doped systems B$_{8-x}$C$_{x}$.
We have depicted the crystal structure for situation where a single B atom in the
ridge (inner) sites is replaced by C atom, denoted by B$_{7}$C$_{1}$-R-[C5] (B$_{7}$C$_{1}$-I-[C2])
in Fig.~\ref{fig-sub-6}(a) (Fig.~\ref{fig-sub-6}(b)).
An interesting situation where a dimer of B atoms in ridge (inner) sites is replaced by a dimer of C atoms are denoted by B$_{6}$C$_{2}$-R-[C5$\&$C6] and B$_{6}$C$_{2}$-R-[C7$\&$C8] (B$_{6}$C$_{2}$-I-[C2$\&$C3] and B$_{6}$C$_{2}$-R-[C1$\&$C4]) and are shown in Fig.~\ref{fig-sub-6}(c) and Fig.~\ref{fig-sub-6}(d) (Fig.~\ref{fig-sub-6}(i) and Fig.~\ref{fig-sub-6}(j)).
For completeness all inner B atoms are replaced by C atoms which are denoted by 
B$_{4}$C$_{4}$-R-[C1-C4].
The uniform $k$-point grids of 24$\times$24$\times$1  are used for the self-consistent field
calculations of all systems. The Kinetic energy cut-offs for the wavefunctions and the charge
density are $850$ and $8500$ eV, respectively.
For each systems, the Broyden-Fletcher-Goldfarb-Shanno quasi-Newton algorithm is
used to relax the internal coordinates of the B and C atoms and possible distortions with convergence threshold on forces for ionic minimization as small as 10$^{-4}$ eV$/{\AA}$.

\begin{table}[H]
\centering
\caption{Optimized lattice parameters (in ${\AA}$) and atomic position for a system in which:
(1) pure borophene (B$_8$)
(2) a single B5 atoms in ridge sites is replaced by C atom,  B$_{7}$C$_{1}$-R-[C5],
(3) a single B2 atoms in inner sites is replaced by C atom,  B$_{7}$C$_{1}$-I-[C2],
(4) a dimer of B atoms (B5 and B6) in ridge sites is replaced by C atoms B$_{6}$C$_{2}$-R-[C5$\&$C6],
(5) a dimer of B atoms (B7 and B8) in ridge sites is replaced by C atoms B$_{6}$C$_{2}$-R-[C7$\&$C8],
(6) a dimer of B atoms (B1 and B4) in inner sites is replaced by C atoms B$_{6}$C$_{2}$-I-[C1$\&$C4],
(7) a dimer of B atoms (B2 and B3) in inner sites is replaced by C atoms B$_{6}$C$_{2}$-I-[C2$\&$C3],
(8) B$_6$ and B$_7$ atoms in ridge sites are replaced by C atoms B$_{4}$C$_{4}$-R-[C6$\&$C7], and
(9) all of B atoms in inner sites are replaced by C atoms B$_{4}$C$_{4}$-I-[C1-C4].} \label{table:5}
\begin{tabular}{cccc|cccc|cccccc}
\hline
&(1)&   B$_{8}$ [borophene] &&&&  (2) B$_{7}$C$_{1}$-R-[C5]   &&&(3)&  B$_{7}$C$_{1}$-I-[C2] & \\
\hline
& $a$=4.5210 & $b$=3.2620 & $c$=26.0 && $a$=4.5686 & $b$=3.2297 & $c$=26.0 && $a$=4.5191 & $b$=3.2714 & $c$=26.0 & \\
\hline
\hline
 atom & $x$ & $y$ &  $z$ & atom & $x$ & $y$ & $z$ & atom & $x$ & $y$ & $z$ &  \\
\hline
B$_{1}$ & -0.1850 &   0.5000 & -0.4841 & B$_{1}$ & -0.1864 & 0.5120  & -0.4831 & B$_{1}$ & -0.1757  & 0.5000 & -0.4872 \\
B$_{2}$ & -0.3150  & 0.0000  &  0.4841 & B$_{2}$ & -0.3169 & 0.0333  & 0.4880  & C$_{2}$ & -0.2991  & 0.0000 & 0.4832 \\
B$_{3}$ & 0.3150 &  0.0000  & 0.4843 & B$_{3}$ & 0.3169 & 0.033   &0.4810   & B$_{3}$ & 0.3385 & 0.0000 & 0.4854 \\
B$_{4}$ & 0.1850 &  0.5000 & -0.4843 & B$_{4}$ & 0.1864 & 0.5120  & -0.4841 & B$_{4}$ & 0.1982 & 0.5000 & -0.4852  \\
B$_{5}$ &  0.5000 &  0.2476 & -0.4578 & C$_{5}$ & 0.5000 & 0.2091  & -0.4527 & B$_{5}$ & 0.5000  & 0.2470 & -0.4420 \\
B$_{6}$ & 0.5000 & -0.2476 & -0.4578 & B$_{6}$ & 0.5000 & -0.2922 & -0.4537 & B$_{6}$ & 0.5000  & -0.2470 & -0.4450  \\
B$_{7}$ & 0.0000  & 0.2532 &   0.4578 & B$_{7}$ & 0.0000 & 0.2451  & 0.4574  & B$_{7}$ & 0.0000  & 0.2470 & 0.4460  \\
B$_{8}$ &  0.0000 & -0.2532 &  0.4578 & B$_{8}$ & 0.0000 & -0.2559 & 0.4594  & B$_{8}$ & 0.0000  & -0.2470 & 0.4440  \\

\hline
\end{tabular}
\begin{tabular}{cccc|cccc|cccccc}
\hline
&(4)&   B$_{6}$C$_{2}$-R-[C5$\&$C6] &&&(5) &  B$_{6}$C$_{2}$-R-[C7$\&$C8] &&&(6)&   B$_{6}$C$_{2}$-I-[C1$\&$C4] & \\
\hline
& $a$=4.2809 & $b$=3.5399 & $c$=26.0 && $a$=4.3721 & $b$=3.5053 & $c$=26.0 && $a$=4.4132 & $b$=3.1501 & $c$=27.0 & \\
\hline
\hline
 atom & $x$ & $y$ & $z$ & atom & $x$ & $y$ & $z$ & atom & $x$ & $y$ & $z$ & \\
\hline
    B$_{1}$      &   -0.2116    &    0.5000    &    0.5326 & B$_{1}$ & -0.2095 & 0.5000 & -0.4713 & C$_{1}$ & -0.1924 & 0.5000 & -0.4863 \\
   B$_{2}$       &   -0.3228    &    0.0000    &    0.4716 & B$_{2}$ & -0.3218 & 0.0000 & 0.4743 & B$_{2}$ & -0.3133 & 0.0000 & 0.4866 \\
B$_{3}$       &    0.3228    &    0.0000    &    0.4716 & B$_{3}$ & 0.3218 & 0.0000 & 0.4723 & B$_{3}$ &  0.3133 & 0.0000 & 0.4866  \\
  B$_{4}$       &    0.2116    &    0.5000    &    0.5326 & B$_{4}$ & 0.2095 & 0.5000 & -0.4703  & C$_{4}$ & 0.1924 & 0.5000 & -0.4863 \\
  C$_{5}$       &    0.5000    &    0.8049    &    0.5347 & B$_{5}$ & 0.5000 & 0.1970 & -0.4613 & B$_{5}$ & 0.5000 & 0.2528 & -0.4475  \\
C$_{6}$       &    0.5000    &    0.1950    &    0.5347 & B$_{6}$ & 0.5000 & -0.1970 & -0.4611 & B$_{6}$ & 0.5000 & -0.2528 & -0.4485   \\
B$_{7}$       &    0.0000    &    0.7399    &    0.4609 & C$_{7}$ & 0.0000 & 0.2721 & 0.4619 & B$_{7}$ & 0.0000 & 0.2457 & 0.4452  \\
B$_{8}$       &    0.0000    &    0.2600    &    0.4609 & C$_{8}$ & 0.0000 & -0.2721 & 0.4619 & B$_{8}$ &  0.0000 & -0.2457 & 0.4482 \\
\hline
\end{tabular}
\begin{tabular}{cccc|cccc|cccccc}
\hline
&(7) &  B$_{6}$C$_{2}$-I- [C2$\&$C3] &&&(8) &  B$_{6}$C$_{2}$-R-[C6$\&$C7] &&&(9)&   B$_{4}$C$_{4}$-I-[C1-C4] & \\
\hline
& $a$=4.1139 & $b$=3.1369 & $c$=26.0 && $a$=4.4473 & $b$=3.2116 & $c$=26.0 && $a$=4.2279 & $b$=3.2260 & $c$=27.0 & \\
\hline
\hline
 atom & $x$ & $y$ &  $z$ & atom & $x$ & $y$ & $z$ & atom & $x$ & $y$ & $z$ &  \\
\hline
B$_{1}$ & -0.2061 & 0.5000 & 0.5104 & B$_{1}$ & -0.1961 & 0.5000  & 0.5226 & C$_{1}$     &   -0.1966    &   0.5000   &   -0.4849 \\
C$_{2}$ & -0.3167& 0.0000 & 0.4744 & B$_{2}$ & -0.3038 & 0.0000  &  0.4774 & C$_{2}$     &   -0.3033    &  -0.0000   &    0.4849 \\
C$_{3}$ & 0.3167 & 0.0000 & 0.4744 & B$_{3}$ &  0.3038 & 0.0000  &  0.4774 & C$_{3}$    &  0.3033    &  -0.0000   &    0.4849 \\
B$_{4}$ & 0.2061 & 0.5000 & 0.5104 & B$_{4}$ &  0.1961 & 0.5000   &0.5226   & C$_{4}$    &  0.1966    &   0.5000   &   -0.4849 \\
B$_{5}$ & 0.5000 & 0.7383 & 0.5630 & B$_{5}$ & 0.5000 & 0.8024  & 0.5375  & B$_{5}$    &  0.5000    &  0.2571    &  -0.4423  \\
B$_{6}$ & 0.5000 & 0.2616 & 0.5630 & C$_{6}$ & 0.5000 & 0.3041 & 0.5545  & B$_{6}$    &  0.5000     &  -0.2571   &   -0.4423  \\
B$_{7}$ & 0.0000 & 0.7455 & 0.4520 & C$_{7}$ & 0.0000 & 0.8041  &  0.4454 &   B$_{7}$    &  0.0000    &   0.2429   &    0.4424 \\
B$_{8}$ & -0.0000 & 0.2544 & 0.4520 & B$_{8}$ & 0.0000 & 0.3023 &  0.4624 & B$_{8}$     &   0.0000    &  -0.2429   &    0.4424  \\
\hline
\end{tabular}
\end{table}

As we have shown in the main text and the previous sections of SM, the
formation of the Dirac nodes is controlled by those atoms residing at the inner sites, while
the location and tilting properties are controlled by the atoms residing at the ridge sites.
With the aim of finding new materials with tilted Dirac cone, as well as, in order to confirm our analytical model,
in this section we investigate whether and to what degree the location and tilt of Dirac-cone would change if boron atoms
replaced by another atoms like carbon. We establish that the coarse grained honeycomb lattice of
tilted Dirac fermions presented in the main text, is fully supported by {\em ab initio} calculations.

For instance, for the case of substitution of single C atom in ridge sites B$_7$C$_1$-R-[C5], we obtain that lattice parameters and position of atoms are not too different from the pure B$_8$ (borophene) and, as a consequence, the shape of
band structure including location and tilt of Dirac-cone does not change much (see second part of Table~\ref{table:5} and Fig.~\ref{fig-sub-6}(e)).
On the other hand, the situation is different in the case of substitution of single C atom in inner sites, B$_7$C$_1$-I-[C2]
due to the breaking of $\tilde{C}_{2x}$ and $\tilde{C}_{2y}$ symmetries (it breaks C$_{2z}$ and inversion symmetries when we consider effective hexagonal lattice). So, single C atoms in hexagonal lattice gap out the tilted Dirac cone bands, as presented in Fig.~\ref{fig-sub-6}(f).
This observation confirms that the atoms at the inner (honeycomb-like) sites are responsible for the formation of the parent Dirac-cone.

\begin{figure} [H]
  \centering
  \includegraphics[width=0.42\textwidth]{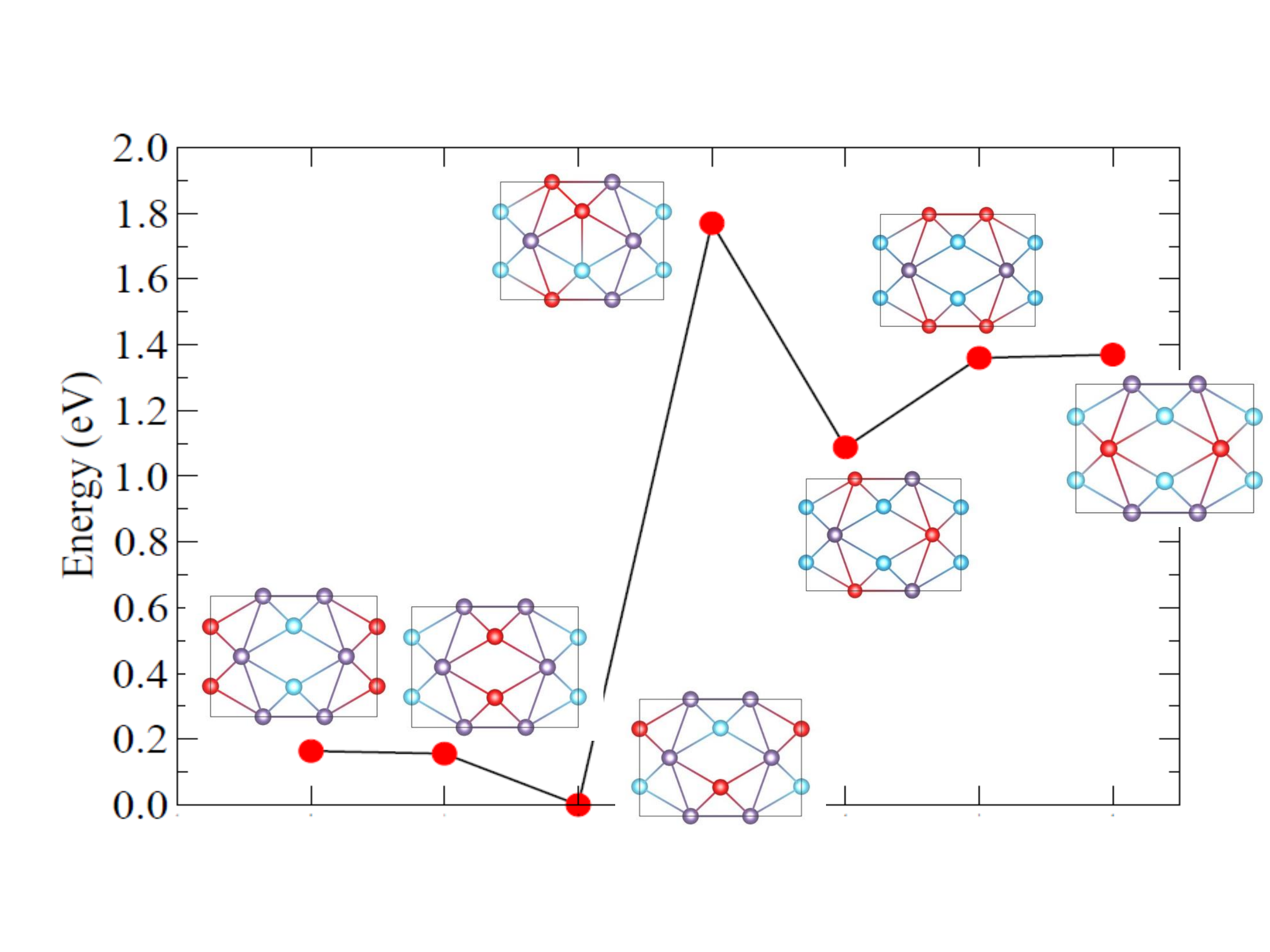}
  \caption{Total energy for different configuration of C-doped borophene based on DFT calculation.}
\label{fig-sub-x}
\end{figure}

In the following, we will consider the situation in which a dimer of B atoms in the ridge (inner) sites is replaced by C atoms B$_{6}$C$_{2}$-R (B$_{6}$C$_{2}$-I).
In the case of B$_{6}$C$_{2}$-R-[C5$\&$C6], Table~\ref{table:5} indicates that atoms in ridge sites get closer to the inner atoms in hexagonal lattice.
As a result, the ratio between the effective hopping energy differences and the first neighbor hopping $t$, namely $(t^p-t^x)/(t-2t^x)$ and $(\tilde{t}-\bar{t})/t$
(that according to our model in section~\ref{tbmode.sec} of SM, control location and tilting of the Dirac cone, respectively) are expected to be larger than pure borophene.
The movement of Dirac cones can be qualitatively seen in panel (g) and (m) of Fig.~\ref{fig-sub-6} where a dimer of C atoms are replacing
the corresponding B atoms on the ridge sites. A weaker movement can be seen in panel (a) as only one C has been replaced in the ridge site, and hence a weaker shift in the ratio $(t^p-t^x)/(t-2t^x)$.
Fig.~\ref{fig-sub-6}(m) for B$_{6}$C$_{2}$-R-[C7$\&$C8] indicates that more or less the same story hold when we replace B$_7$ and B$_8$ by C atoms. There is another configuration of two C atoms in the ridge sites B$_{6}$C$_{2}$-R-[C6$\&$C7] (see Fig.~\ref{fig-sub-6}(k)) which is energetically favorable as shown in Fig.~\ref{fig-sub-x}, but it does not have significant value of tilt.
Now let us consider the opposite case, displayed in panel (d) and (h) of Fig.~\ref{fig-sub-6} where  the C dimers replace B dimers on the inner sites, namely B$_{6}$C$_{2}$-I-[C2$\&$C3] and B$_{6}$C$_{2}$-I-[C1$\&$C4].
In these cases the ridge atoms move away from the $xy$ plane (see Table~\ref{table:5}), thereby giving rise to opposite effect of panel (c). Therefore in panel (d) the Dirac nodes move away from each other
and tilting of the Dirac cone is smaller than pure borophene.
This behavior is reinforced in B$_{4}$C$_{4}$-R-[C1-C4] when we replace all inner B atoms by C atoms as the shape of the bands is similar to that of graphene bands (Fig.~\ref{fig-sub-6}(p)).

So summarize, in terms of the effectiveness of the C substitution in changing the tilt, the most effective 
substitutions are B$_{6}$C$_{2}$-R-[C5$\&$C6] and B$_{6}$C$_{2}$-I-[C2$\&$C3]. Next priority is the single C-doped.
This is because similar to pure borophene they are the C-doped materials in which Dirac cone are dominantly formed by low-energy isolated sets of $p_z$ bands.
So, in the following we only present the results for pure borophene B$_8$, B$_{6}$C$_{2}$-R-[C5$\&$C6], and B$_{6}$C$_{2}$-I-[C2$\&$C3].
The ridge positions in general offer more stable positions for the substitution of C atoms (see Fig.~\ref{fig-sub-x}). Among the compositions with two C atoms, panel (k) in Fig.~\ref{fig-sub-6}
has the lowest energy. Then panel (c) has a comparable but slightly higher energy. Among panels (a), (b), panel (a) being ridge site has lower energy.

\begin{figure}[H]
\centering
\subfloat[]{ \includegraphics[width=0.33\textwidth]{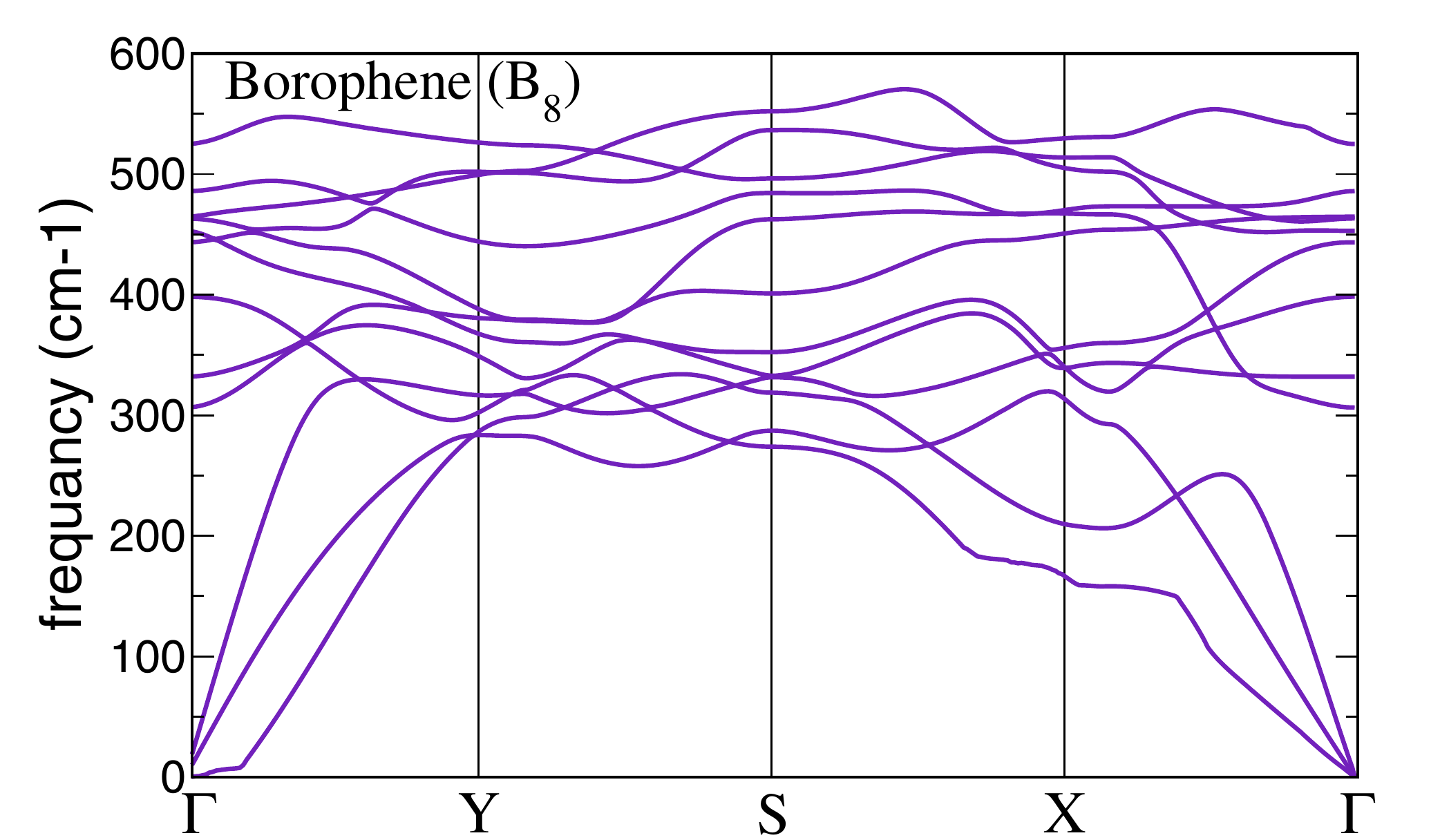}}
  \subfloat[]{    \includegraphics[width=0.33\textwidth]{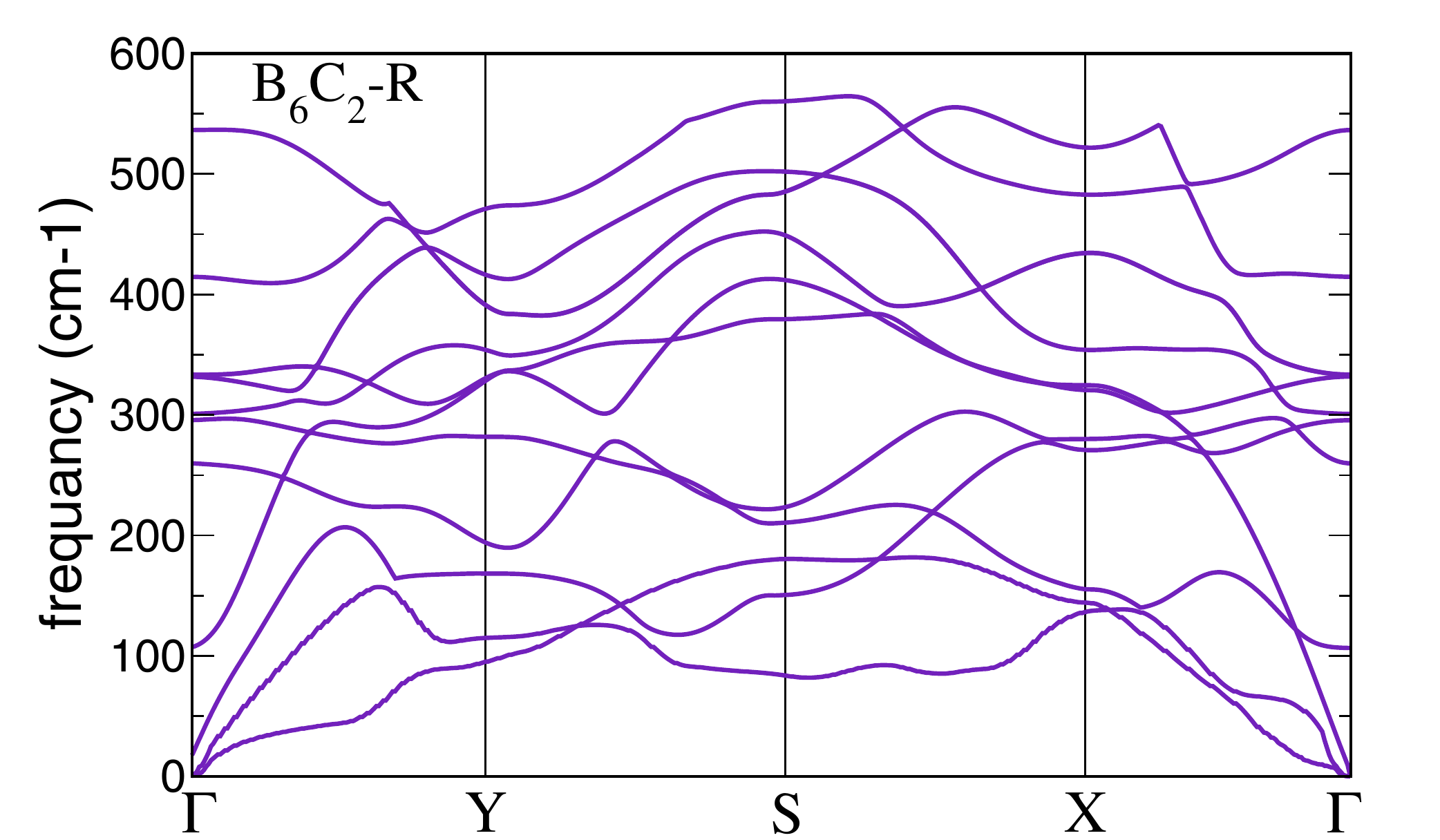}}
  \subfloat[]{   \includegraphics[width=0.33\textwidth]{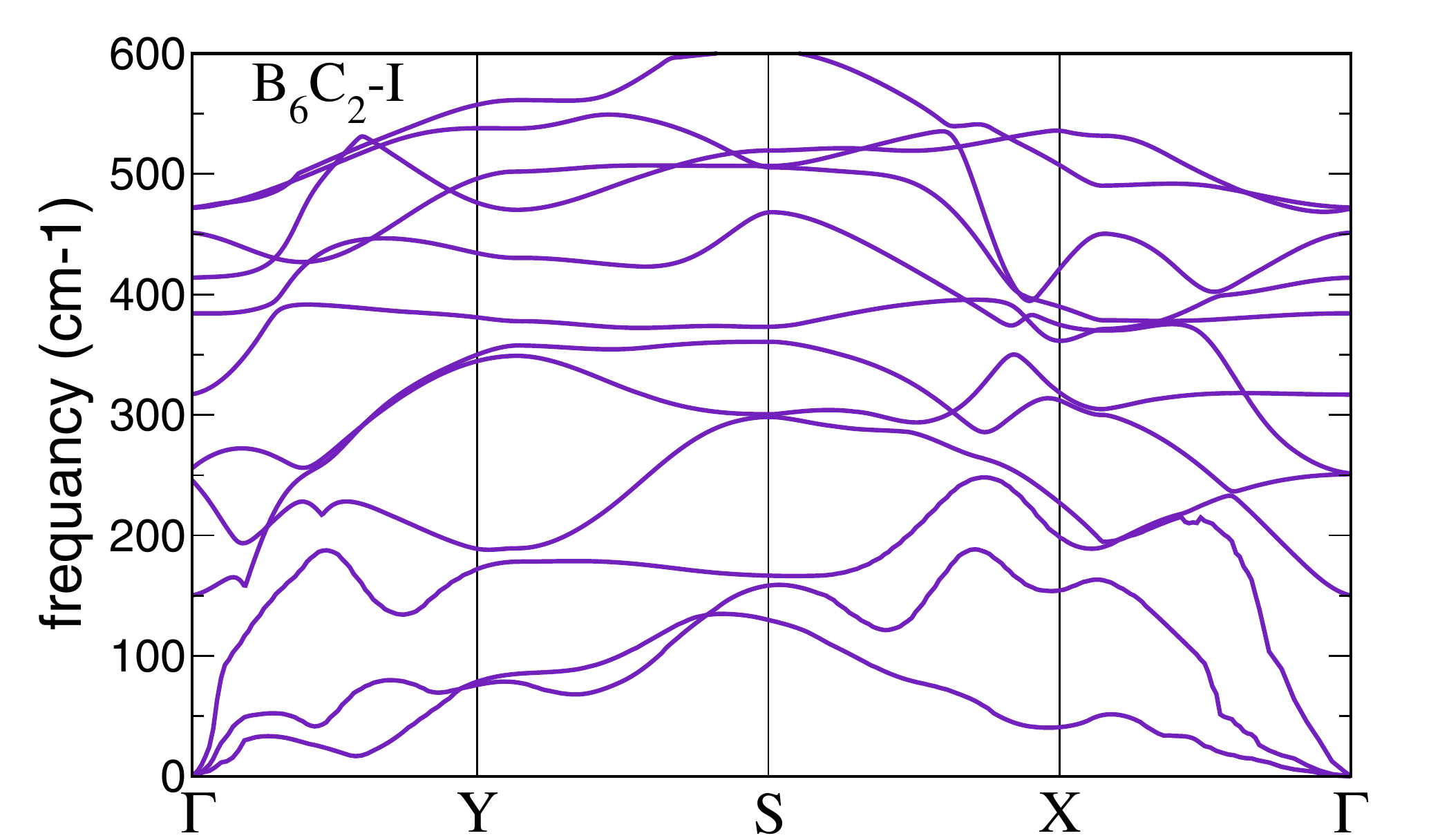} }
  \caption{The phonon dispersion of (a) pure $8Pmmn$ borophene B$_{8}$, and a system in which a dimer of B atoms residing in
  (b) ridge sites and (c) inner sites, is replaced by a dimer of C atoms. }
\label{fig-sub-7}
\end{figure}

Due to high flexibility, the range of strain at which $8Pmmn$ borophene remains stable is very
much higher than that of other 2D materials such as silicene~\cite{Roman}, MoS$_{2}$~\cite{Wang} and black phosphorene~\cite{Li},
and yet,
slightly larger than graphene~\cite{Liu} and h-BN~\cite{Wu}. We expect it can also be able to withstand strong strain generated by carbon substitutions. Nevertheless, we need to confirm whether the $8Pmmn$ borophene remains stable by the chemical substitutions with C atoms.
Study of phonon dispersion provides a way to investigate the dynamical stability of crystal structures. The calculated phonon dispersions of pure $8Pmmn$ borophene and two systems in which a dimer of C atoms is substituted into borophene (B$_{6}$C$_{2}$-R-[C5$\&$C6] and B$_{6}$C$_{2}$-I-[C2$\&$C3]) are shown in Fig.~\ref{fig-sub-7}.
In this method, the negative value of imaginary phonon frequencies is an indication of
dynamical instability. As can be seen in Fig.~\ref{fig-sub-7}, there are no negative imaginary frequencies, and therefore all three systems studies here are stable with respect to substitution of C atoms for B atoms.

\section{{\em Ab initio} hopping matrices elements}
\label{tij.sec}
The parameters of our coarse-grained $8Pmmn$ structure are obtained from atomic scale hopping parameters $t_{ij}$
depicted in panel (a) of Fig.~\ref{fig-sub-3}.
Based on these $t_{ij}$ parameters, as explained in section~\ref{rg.sec}, the effective hopping parameters
of our effective model defined on the parent honeycomb structure are calculated.
The accurate determination of the effective parameters therefore depends on accurate {\em ab initio} calculation
of atomic hoppping amplitudes $t_{ij}$ that can be extracted from the corresponding Wannier functions~\cite{Mostofi,Freimuth,Marzari}.

The honeycomb lattice tight-binding model presented in our paper to describe the formation of tilt in $8Pmmn$ structure
is rather generic, and the hopping parameters can span a wide range of values. Nevertheless for a specific material
they are fixed numbers.
In contrast to graphene where hopping energy to sites further away than nearest neighbours are much smaller than the
first neighbor hopping, our renormalized hopping scenario predict large effective hopping mediated by B$_R$ atoms in borophene.
Even the graphene fits within our renormalized hopping picture, as in the case of pure graphene, there are no
ridge sites, and hence all $t_{ij}$ parameters other than the nearest neighbor hoppings are nearly zero.
In contrast, in the $8Pmmn$ structure, they are numerically on the scale of electron volts,
giving rise to effective hopping parameters of the same order.
So it is sufficient to calculate the atomic $t_{ij}$ parameters.

\begin{figure}[H]
\centering
\subfloat[]{ \includegraphics[width=0.22\textwidth]{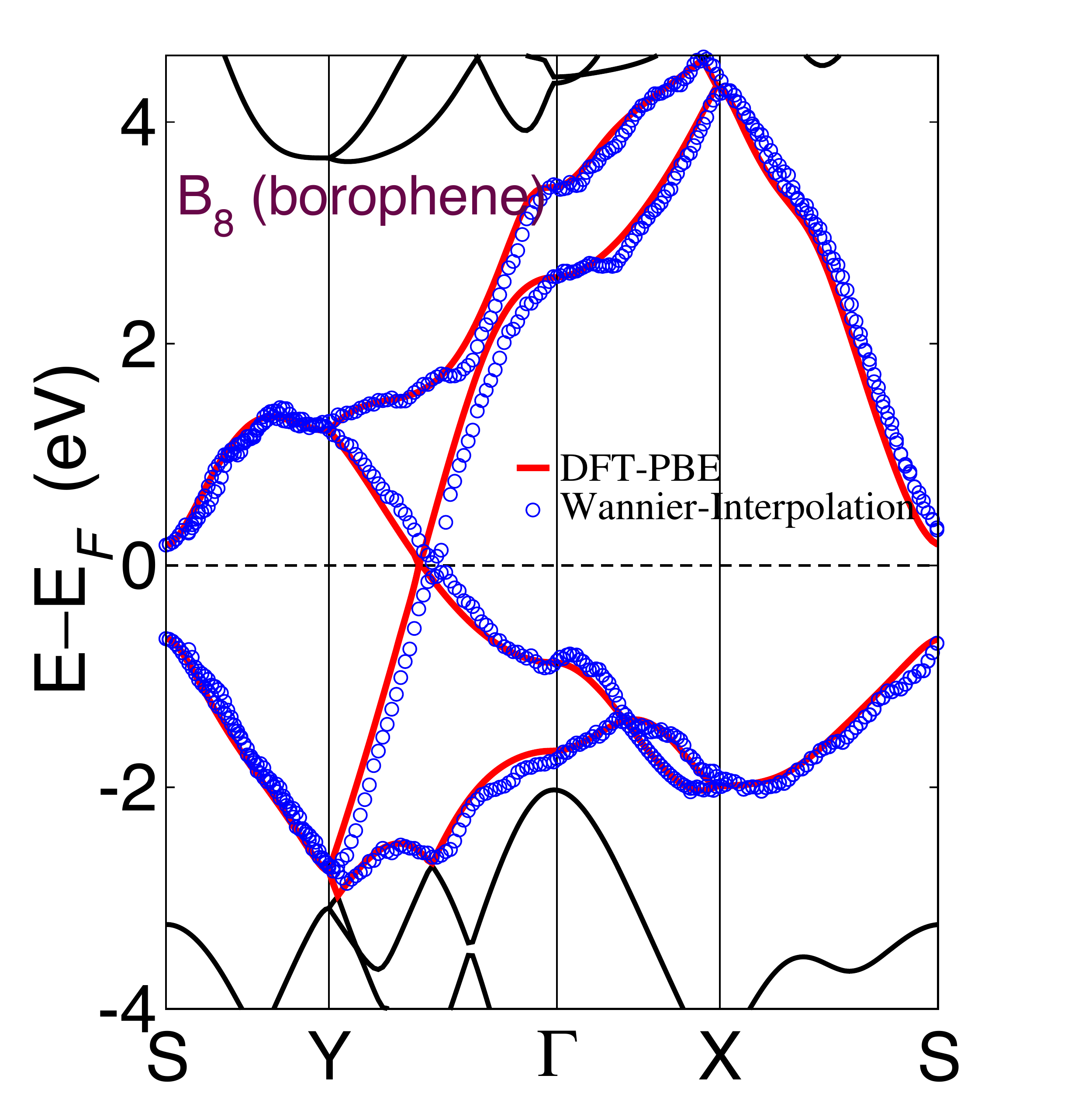}}
  \subfloat[]{    \includegraphics[width=0.22\textwidth]{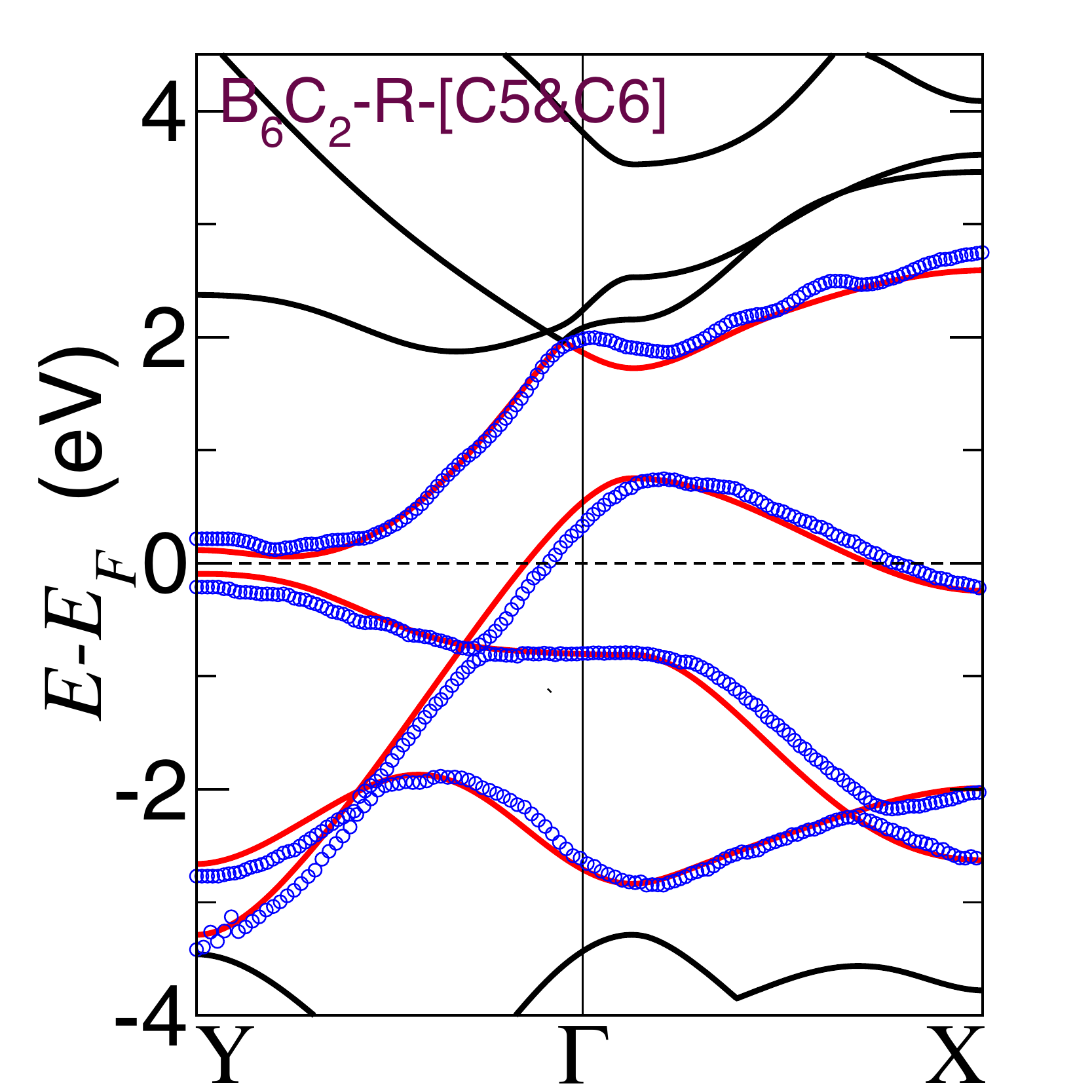}}
  \subfloat[]{   \includegraphics[width=0.22\textwidth]{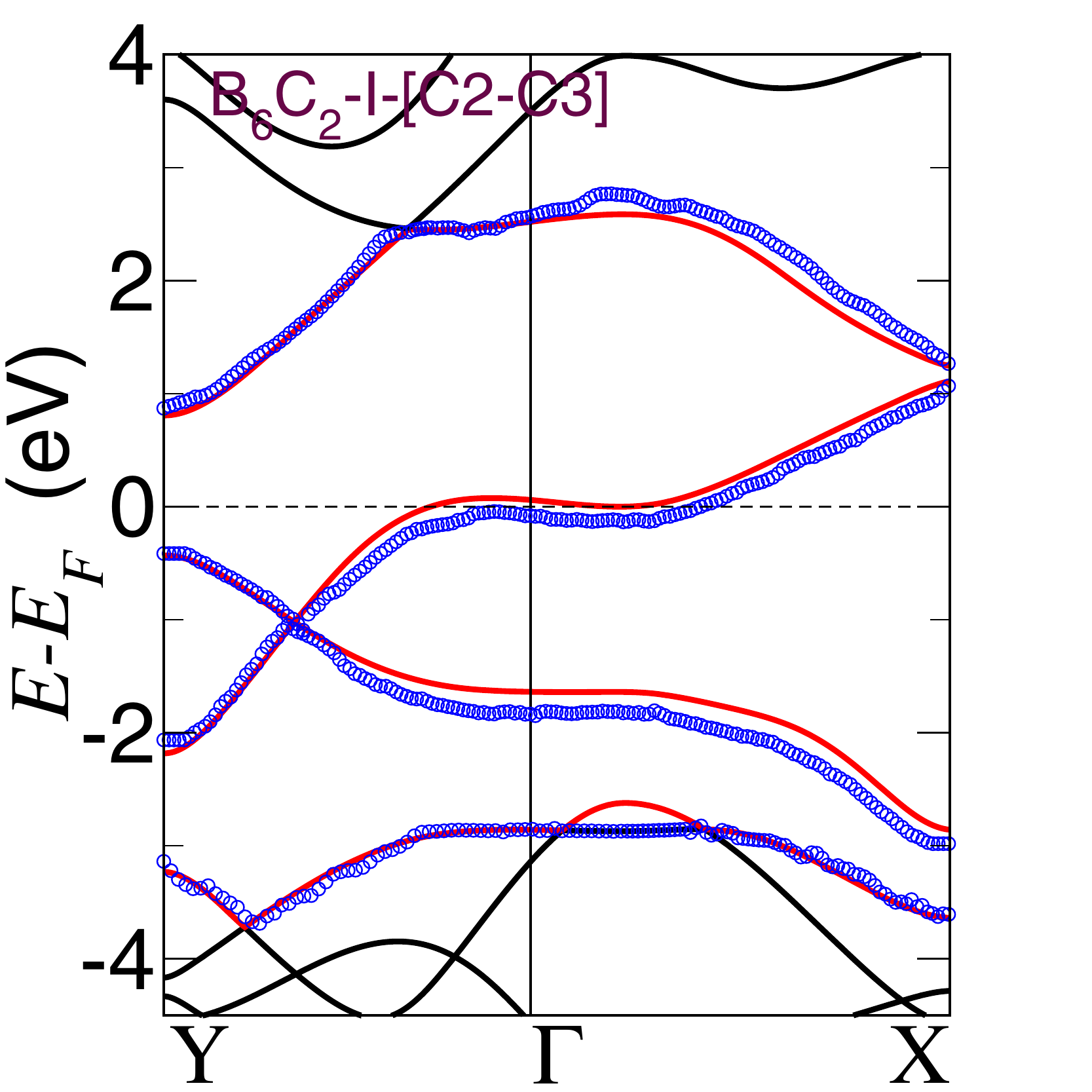} }
  \caption{Comparison of the DFT-PBE
band structures with the corresponding Wannier-interpolated
band structures obtained with $p_z$ Wannier orbitals of B$_I$ atoms together with  $p_z$ and $p_x$ Wannier orbitals
of B$_R$ atoms for (a) pure $8Pmmn$ borophene B$_{8}$ (b) B$_{6}$C$_{2}$-R-[C5$\&$C6], and (c) B$_{6}$C$_{2}$-I-[C2$\&$C3].}
\label{fig-sub-8}
\end{figure}

We calculate the hopping energies for all considered systems using Wannier functions. The maximally localized Wannier functions (MLWFs) are constructed with the WANNIER90
library~\cite{Mostofi,Freimuth}. It is worth noting that graphene show well-isolated $\pi$ bands at the Fermi energy, which induces a
simple single band model with $p_z$  states. In borophene, our projected band-structure indicate that the $p_x$ states of B$_R$ atoms
are not completely isolated from the $p_z$ states. To verify the adequacy of the above orbitals and hence the validity of calculated Wannier functions, in
Fig.~\ref{fig-sub-8} for three B$_{8}$, B$_{6}$C$_{2}$-R-[C5$\&$C6], and B$_{6}$C$_{2}$-I-[C2$\&$C3] systems,
we compare the DFT-PBE band structure (solid red) with the corresponding Wannier-interpolated bands (blue) obtained with $p_z$ Wannier orbitals on the B$_I$ site and the $p_x$ and $p_z$ Wannier orbitals on the B$_R$-site. To avoid complexity, only four bands with $p_z$ character are shown.
 As seen from the band structures, the overall
agreement between original and Wannier-interpolated bands
is remarkably good. Small deviations appear for states far from the
Fermi energy, which given our picture based on the renormalization, are irrelevant for the low-energy physics.

Now that we have established in Fig.~\ref{fig-sub-8} that the $p_z$ orbitals of inner sites and $p_x$ and $p_z$ orbitals of
ridge sites are relevant degrees of freedom, we are ready to calculate the $t_{ij}$ parameters by inclusion of these
orbitals.
The results of \emph{ab initio} hopping matrix elements for pure borophene, B$_{7}$C$_{1}$-I-[C2], B$_{7}$C$_{1}$-R-[C5], B$_{6}$C$_{2}$-R-[C5$\&$C6], and B$_{6}$C$_{2}$-I-[C2$\&$C3] are presented in the left partition of Table~\ref{table:6}.
These values are then used to derive the parameters of the effective model in the right partition of the same table.
When a dimer of B$_{R}$ atoms occupying the ridge sites is replaced by a C dimer B$_{6}$C$_{2}$-R-[C5$\&$C6], we obtain comparatively large matrix elements (except $t_{6,5}$). This is due to the fact that the ridge atoms
come closer to the $xy$ plane upon replacement by C atoms as depicted in Fig.~5(a) of main text.
So, the ridge atoms mediate stronger hoppings in B$_{6}$C$_{2}$-R-[C5$\&$C6] system, meaning that they can better connect
further away inner sites on the effective honeycomb model. As a consequence, the tilt of Dirac-cone
 is considerably larger than the corresponding values in pristine borophene. The situation is reversed when B$_{I}$ atoms of
 the inner sites are replaced by a dimer of C atoms B$_{6}$C$_{2}$-I-[C2$\&$C3]. In this case,
 the ridge atoms move away from the plane of inner atoms, as a results, the hopping parameters are smaller than corresponding ones in pristine borophene. In this situation, the system tends to reduce the tilt of Dirac cone.

\begin{table}[H]
\centering
\caption{(left) \emph{ab initio} atomic hopping matrix elements (in $eV$) for borophene and C-doped borophene obtained
from Wannier functions. Conventions for labeling of the hopping matrix elements are given in Fig.~\ref{fig-sub-3}(a).
(right) Renormalized parameters of Fig.~\ref{fig-sub-3}(c). $\zeta_y$ is the tilt parameter calculated from our
analytical model. $k_D/k_Y$ quantifies the location of Dirac node extracted from {\em ab initio} data. } \label{table:6}
\begin{ruledtabular}
\begin{tabular}{ccccccccc|ccccccc}
system&$t_{36}$&$t_{65}$&$t_{53}$&$t_{81}$&$t_{38}$&$t_{37}$&$t_{72}$&$t_{78}$&$t$&$t^{p}$&$t^{x}$&$\tilde{t}$&$\bar{t}$&$\zeta_y$&$k_D/k_{Y}$\\
\hline
pure borophene $B_8$           &  2.09  & -2.66  & 2.09 & 2.09  & -1.87  & -1.87   & -1.87&-2.54& -2.21 & -2.36  & -1.07  & -1.99 & -2.51& 0.46 & 0.48\\
B$_{7}$C$_{1}$-I-[C2]&  2.21  & -2.59  & 2.32 & 2.35  & -1.52  & -1.61   & -1.47&-2.35& -2.32 & -1.87  & -1.19  & -1.88 & -2.40& ---- & ---- \\
B$_{7}$C$_{1}$-R-[C5]&  2.23  & -2.73  & 2.25 & 2.14  & -1.89  & -1.74   & -1.85&-2.13& -2.30 & -2.03  & -1.09  & -2.11 & -2.69& 0.56 & 0.42 \\
B$_{6}$C$_{2}$-I-[C2$\&$C3]&  1.93  & -2.52  & 1.96 & 1.92  & -1.52  & -1.52   & -1.55&-2.34&-2.37  & -1.75  & -0.95  & -1.62 & -2.05& 0.36 & 0.66 \\
B$_{6}$C$_{2}$-R-[C5$\&$C6]&  2.14  & -2.33  & 2.12 & 2.15  & -2.23  & -2.21   & -2.20&-2.43&-2.05  & -2.49  & -1.09  & -2.24 & -2.85& 0.59 & 0.32 \\
\end{tabular}
\end{ruledtabular}
\end{table}

\begin{figure}[H]
  \centering
  \subfloat[]{ \includegraphics[width=0.32\textwidth]{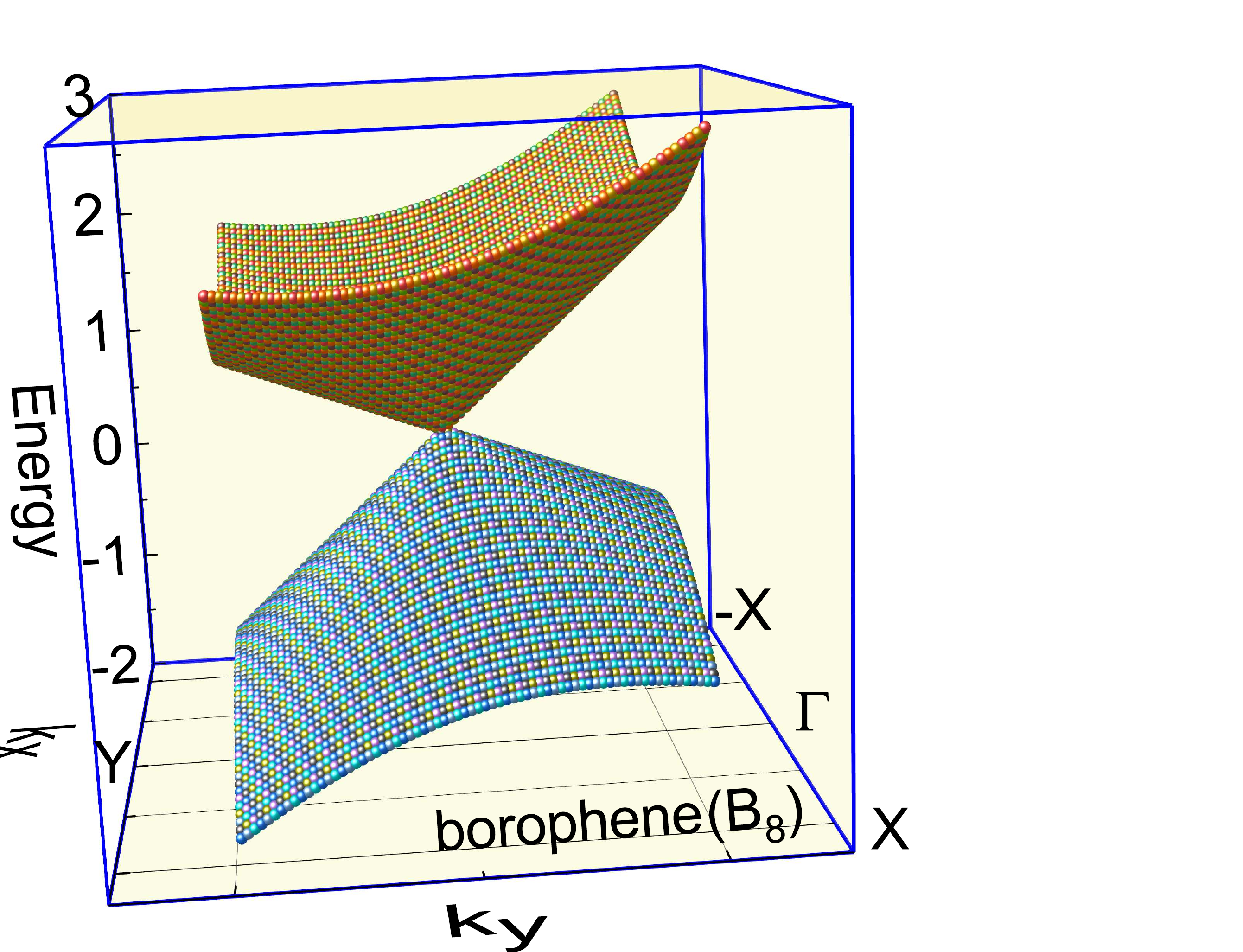}}
  \subfloat[]{ \includegraphics[width=0.32\textwidth]{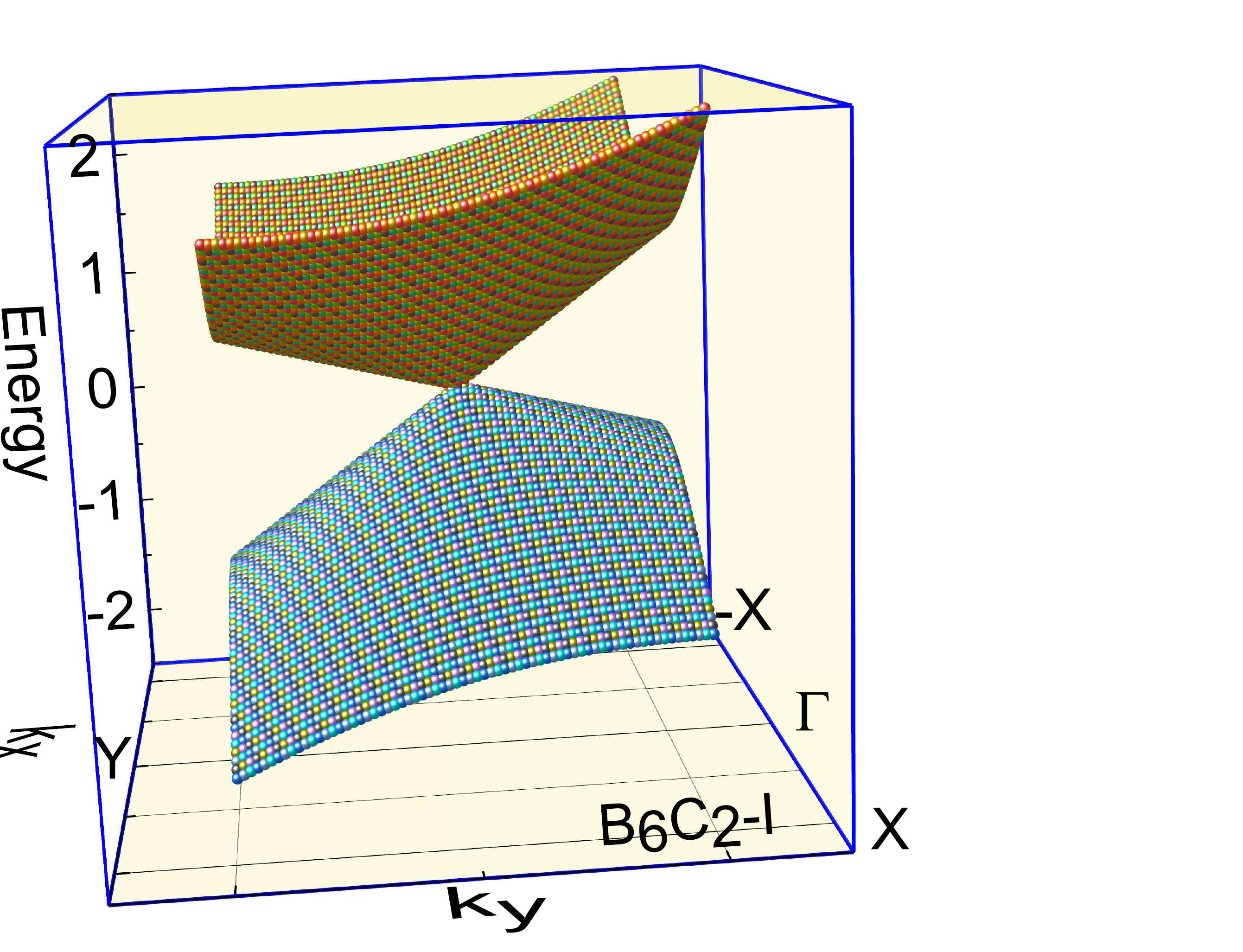}}
  \subfloat[]{ \includegraphics[width=0.32\textwidth]{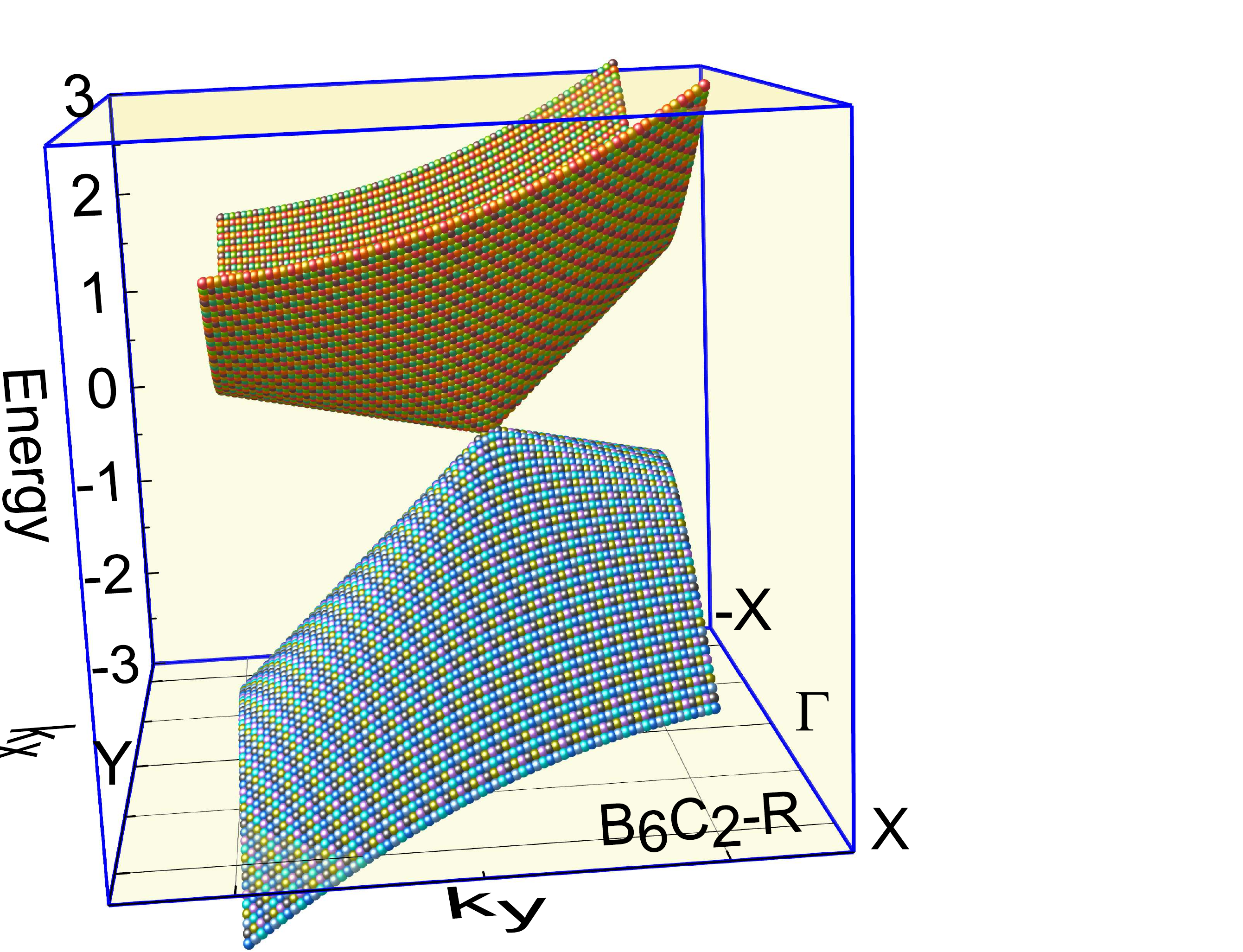}}
  \caption{3D plot band structures obtained based on our effective tight-binding model using \emph{ab initio} hopping matrix elements
  for (a) pure B$_{8}$ (b) B$_{6}$C$_{2}$-I-[C2$\&$C3], and (c) B$_{6}$C$_{2}$-R-[C5$\&$C6].}
\label{fig-sub-9}
\end{figure}

Now that we have a picture of the formation of the Dirac cone, and an analytical model
to produce it, we use the model parameters obtained in Table~\ref{table:6} to construct
a picture of tilted Dirac cone as a function of $(k_x,k_y)$ in pure and C-doped borophene.
The solution of the tight-binding model leads to the $\zeta_x=0,~~\zeta_y=\pm 2(\tilde t-\bar t)/t$,
where the further neighbor effective hoppings $\tilde t$ and $\bar t$ of the model are calculate in the right
partition of Tab.~\ref{table:6}.
As shown in Table. \,\ref{table:6}, placing C atoms in the ridge sites generates larger $(\tilde t-\bar t)/t$ that
ultimately increases the tilt from the $\zeta_y=0.46$ of the pristine borophene to $\zeta_y=0.59$ in B$_{6}$C$_{2}$-R-[C5$\&$C6].
Employing Eqs.~\eqref{shifted.eqn} and  \eqref{equ2} the energy dispersion for pristine borophene, B$_{6}$C$_{2}$-I-[C2$\&$C3],
and B$_{6}$C$_{2}$-R-[C5$\&$C6] are reconstructed in Fig.~\ref{fig-sub-9} using the \emph{ab initio} effective hopping amplitude reported in
Tab.~\ref{table:6}.

{\bf Accuracy of tilt parameters:}
The ultimate outcome of our model is to understand the mechanism of the formation of the tilt.
Hence the value of tilt parameters are important. Let us close this supplementary material by
a comparison between the values of the tilt parameter directly extracted from the {\em ab initio} data,
and the tilt parameter predicted by Eq.~\eqref{tilt.eqn} of our model.

We extract the tilt directly from the DFT data by fitting the slopes $m_R$ and $m_L$ of the {\em ab initio}
dispersion relations along the $\Gamma Y$ direction as,
\begin{equation*}
\begin{aligned}
&\zeta^{\rm DFT}_y=\frac{m_{R}-m_{L}}{m_{R}+m_{L}}.
  \end{aligned}
\end{equation*}
Using the direct DFT data, we find that the tilting parameter value $\zeta^{\rm DFT}_y$ are $0.49$, $0.47$, and $0.66$
in pristine borophene, B$_{6}$C$_{2}$-I, and B$_{6}$C$_{2}$-R-[C5$\&$C6] respectively.
This is in qualitative agreement with the values obtained from our model, Eq.~\eqref{tilt.eqn}
and the trend in tilt values upon placing the carbon atoms in ridge/inner sites are correctly reproduced in
our model (see Tab.~\ref{table:6}).  The physical picture emerging from both DFT data and our model
is that the substitutions of B$_{R}$ atoms by C dimers where
the two C atoms belong to the ridge sites can increase the tilt of Dirac cone.
Moreover, the location of tilted Dirac-cone is very sensitive to the position of C atoms on the $8Pmmn$ lattice.
This location has been quantified by the parameter $k_D/k_{Y}$ to measure the distance from Dirac point to
$\Gamma$ and the results are reported in Tab.~\ref{table:6}.
If C atoms are replaced in the ridge sites, Dirac-node moves closer to $\Gamma$ point.

To conclude, the tilted Dirac cone dispersion in $8Pmmmn$ borophene can
be well described by our effective tight-binding model.
This effective model is not only important for understanding the origin of formation of the tilt in the Dirac cone
in borophene, but it also increases considerably the predictive power to find new compounds possessing tilted Dirac cone.
Furthermore, out concrete model paves the path for many other studies including the effects of interactions/disorder/symmetry breaking, etc
in a physically motivated model of tilted Dirac cone in $8Pmmn$ space group.


\end{document}